\newif\ifdouble
\newif\ifsingle
\newif\ifchange
\doubletrue

\ifdouble
  \documentclass[sigconf, usenames, dvipsnames]{acmart}
\else
  \documentclass[manuscript]{acmart}
\fi

\usepackage{dblfloatfix} 
\usepackage{booktabs} 
\usepackage{arydshln}
\usepackage{subfig}
\newcommand{\change}[1]{\textcolor{purple}{#1}}
\renewcommand{\change}[1]{#1}

\AtBeginDocument{%
  \providecommand\BibTeX{{%
    \normalfont B\kern-0.5em{\scshape i\kern-0.25em b}\kern-0.8em\TeX}}}

\copyrightyear{2022}
\acmYear{2022}
\setcopyright{acmcopyright}\acmConference[UIST '22]{The 35th Annual ACM Symposium on User Interface Software and Technology}{October 29-November 2, 2022}{Bend, OR, USA}
\acmBooktitle{The 35th Annual ACM Symposium on User Interface Software and Technology (UIST '22), October 29-November 2, 2022, Bend, OR, USA}
\acmPrice{15.00}
\acmDOI{10.1145/3526113.3545626}
\acmISBN{978-1-4503-9320-1/22/10}

\newcommand{\system}{Sketched Reality}

\begin{document}
\pagenumbering{arabic}
\pagestyle{plain}
\title{\system{}: \change{Sketching Bi-Directional Interactions Between Virtual and Physical Worlds with AR and Actuated Tangible UI}}

\author{Hiroki Kaimoto}
\affiliation{%
  \institution{The University of Tokyo}
  \city{Tokyo}
  \country{Japan}}
\affiliation{%
  \institution{University of Calgary}
  \city{Calgary}
  \country{Canada}}  
\email{hkaimoto@xlab.iii.u-tokyo.ac.jp}
\authornote{Both authors contributed equally to the paper}

\author{Kyzyl Monteiro}
\affiliation{%
  \institution{IIIT-Delhi}
  \city{New Delhi}
  \country{India}}
\affiliation{%
  \institution{University of Calgary}
  \city{Calgary}
  \country{Canada}}  
\email{kyzyl17296@iiitd.ac.in}
\authornotemark[1]

\author{Mehrad Faridan}
\affiliation{%
  \institution{University of Calgary}
  \city{Calgary}
  \country{Canada}}
\email{mehrad.faridan1@ucalgary.ca}

\author{Jiatong Li}
\affiliation{%
  \institution{University of Chicago}
  \city{Chicago}
  \country{U.S.A.}}
\email{jtlee@uchicago.edu}

\author{Samin Farajian}
\affiliation{%
  \institution{University of Calgary}
  \city{Calgary}
  \country{Canada}}
\email{samin.farajian@ucalgary.ca }

\author{Yasuaki Kakehi}
\affiliation{%
  \institution{The University of Tokyo}
  \city{Tokyo}
  \country{Japan}}
\email{kakehi@iii.u-tokyo.ac.jp}

\author{Ken Nakagaki}
\affiliation{%
  \institution{University of Chicago}
  \city{Chicago}
  \country{U.S.A.}}
\email{knakagaki@uchicago.edu}

\author{Ryo Suzuki}
\affiliation{%
  \institution{University of Calgary}
  \city{Calgary}
  \country{Canada}}
\email{ryo.suzuki@ucalgary.ca}

\renewcommand{\shortauthors}{Kaimoto, et al.}

\begin{abstract}
This paper introduces Sketched Reality, an approach that combines AR sketching and actuated tangible user interfaces (TUI) for \textbf{\textit{bi-directional sketching interaction}}. Bi-directional sketching enables virtual sketches and physical objects to \textit{``affect''} each other through physical actuation and digital computation. In the existing AR sketching, the relationship between virtual and physical worlds is only one-directional --- while physical interaction can affect virtual sketches, virtual sketches have no return effect on the physical objects or environment. In contrast, bi-directional sketching interaction allows the seamless coupling between sketches and actuated TUIs. In this paper, we employ tabletop-size small robots (Sony Toio) and an iPad-based AR sketching tool to demonstrate the concept.
In our system, virtual sketches drawn and simulated on an iPad (e.g., lines, walls, pendulums, and springs) can move, actuate, collide, and constrain physical Toio robots, as if virtual sketches and the physical objects exist in the same space through seamless coupling between AR and robot motion. This paper contributes a set of novel interactions and a design space of bi-directional AR sketching. We demonstrate a series of potential applications, such as tangible physics education, explorable mechanism, tangible gaming for children, and in-situ robot programming via sketching. 
\end{abstract}



\begin{teaserfigure}
\centering
\includegraphics[width=\columnwidth]{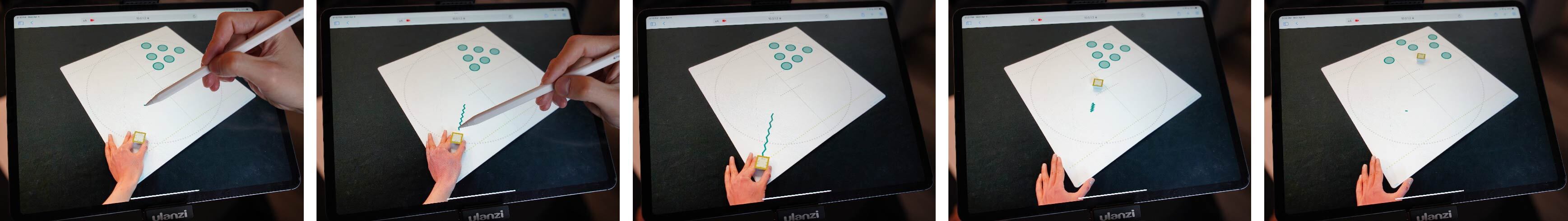}
\caption{\change{Sketched Reality explores bi-directional interactions between AR-based virtual sketches and actuated tangible UIs.
When the user sketches a virtual spring (green lines), the user can start pulling the spring with a physical robot (yellow box), so that the sketched virtual spring \textit{affects} the physical robot by applying the force like a slingshot game (virtual $\rightarrow$ physical).
When the robot hits the virtual sketched circles, then the robot also \textit{affects} the virtual objects, as if the physical robot collides with virtual sketched circles like a billiard board game (physical $\rightarrow$ virtual).}
}
\label{fig:top}
\end{teaserfigure}


\maketitle

\section{Introduction}
With the advent of augmented and mixed reality (AR/MR) technology, many Human-Computer Interaction (HCI) researchers recently started exploring AR sketching as a means of interactive authoring and exploration of AR objects.
In contrast to screen-based sketching~\cite{laviola2006mathpad2, scott2013physink}, AR sketching allows us to embed virtual sketches in the real world, which helps combine digital and physical worlds in more immersive and improvisational ways for designing~\cite{arora:2018:symbiosissketch}, annotating~\cite{vuforia-chalk}, and collaborating~\cite{gasques2019you} in the real world.

\begin{figure}[t]
\centering
\includegraphics[width=1\linewidth]{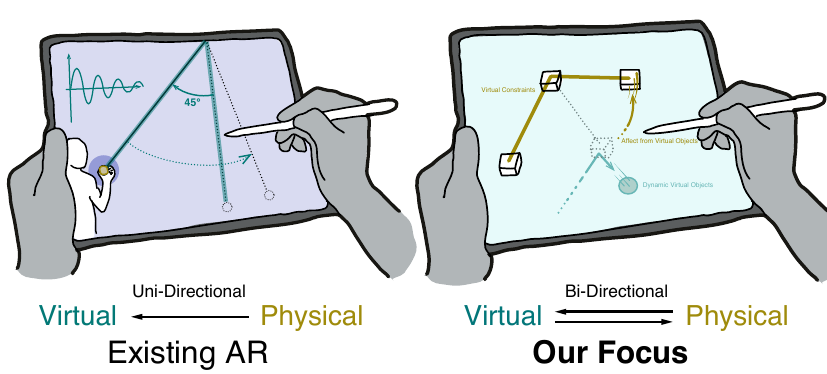}
\caption{While, in conventional AR (such as in \cite{suzuki2020realitysketch}), the virtual graphical information is affected by physical objects and actions, Sketched Reality explores how virtual can also affect physical objects by incorporating actuation with AR sketches.}
\label{fig:related-work}
\end{figure}
\change{However, in the existing AR sketching tools, or in the existing AR interfaces in general, the coupling between AR sketches and the physical world is only \textbf{\textit{one directional}} ---
meaning that virtual sketches have \textbf{\textit{no effect}} on the physical environment. (Figure~\ref{fig:related-work} left).}

\change{This paper introduces Sketched Reality, an approach that combines AR sketching and actuated tangible user interfaces (TUI) for \textbf{\textit{bi-directional sketching interaction}} (Figure~\ref{fig:related-work} right).
In our paper, \textit{bi-directional sketching interaction} refers to the interaction in-between virtual sketches and actuated physical objects that affect each other through physical actuation and digital computation.}


To demonstrate this concept, we employ tabletop-size mobile robots (Sony Toio) and an iPad-based AR sketching interface to seamlessly couple physical robot and virtual interactive sketches. 
With our system, the user can sketch and embed lines onto a tabletop surface using an iPad and WebXR (A-Frame and 8th Wall).
Similar to conventional AR sketching~\cite{suzuki2020realitysketch}, these sketched lines are dynamic so that they can interact with tabletop robots that can represent an actuated tangible object.
By dynamically calculating the collision and simultaneously moving these robots, it is possible to make illusions, as if virtual sketches and the physical robots affect each other. 

This paper presents a set of design space of such a bi-directional sketching interaction in the following four categories: 1) \textit{boundary constraints} (e.g., contour, walls), 2) \textit{geometric constraints} (e.g., constant length, angle), 3) \textit{applied force} (e.g., gravity, friction, attraction force), and collision (e.g., hitting objects).
In total, we formulate the eight different interactions between virtual sketches and physical robots.
By combining these elements, the user can quickly and easily create interactive effects on-demand in an improvisational and tangible manner, without any pre-defined programs or configurations. 
For example, sketched lines in AR can dynamically create a \textit{virtual constraint} that can actually enclose and move actuated tangible objects.
Or a sketched virtual material can provide a \textit{force} to a physical object so that the user can shoot like a slingshot.
We demonstrate the potential of this approach through various use cases and application scenarios, such as tangible physics education for children, explorable mechanism, tangible gaming, and in-situ robot programming and actuated TUI via sketching.
We believe this paper opens up a new way of interactions and exploration for AR and tangible user interfaces. 

\change{The core novelty of this paper lies in the first exploration of the \textit{concept} of bi-directional sketching interaction and its comprehensive \textit{design space}.}
More specifically, this paper makes the following contributions: 
\begin{enumerate}
\item A concept and design space of Sketched Reality, allowing AR sketches to bi-directionally interact with actuated TUIs. 
\item An implementation of the system that uses a small tabletop robot (Sony Toio) as actuated objects and mobile AR (iPad and WebXR) as AR sketching interfaces. 
\item Interaction techniques and application scenarios enabled by our system, which includes tangible education, interactive games, and in-situ robot programming and control via sketching.
\end{enumerate}


\section{Related Work}



\subsection{Augmented Reality Sketching Tools} 
In recent years, HCI researchers have explored various augmented and virtual reality (AR/VR) sketching interfaces. 
For example, Just a Line~\cite{just-a-line}, TiltBrush~\cite{tiltbrush},
Gravity Sketch~\cite{gravity-sketch}, and Vuforia Chalk AR~\cite{vuforia-chalk} are commercially available sketching tools that support various 2D/3D sketching in AR/VR environments. 
In particular, one of the unique benefits of AR sketching over screen-based sketching is that the sketched elements can be embedded, situated, and contextualized in the real world, which allows the user interaction tightly coupled with the real environment.
For example, Vuforia Chalk AR~\cite{vuforia-chalk} allows the user to directly annotate a physical object via sketching so that the sketches can be embedded in the associated physical location, which is useful for many application scenarios (e.g., real-time remote assistant between experts and field technicians in a factory or construction site.) 
By combining a tablet as a sketching canvas, SymbiosisSketch~\cite{arora:2018:symbiosissketch}, VRSketchIn~\cite{drey2020vrsketchin}, PintAR~\cite{gasques2019you}) leverage a physical surface to serve as a geometric constraint for sketching in an immersive environment. Alternatively, SweepCanvas~\cite{li:2017:sweepcanvas} and SketchingWithHands~\cite{kim2016sketchingwithhands} use a physical object as a reference to create 3D shapes.
These AR sketching tools enable a variety of applications, including education~\cite{rajaram2022papertrail}, entertainment~\cite{arora:2018:symbiosissketch}, design~\cite{yu2021cassie}, and collaboration~\cite{gasques2019you}.

While many of these tools only focus on embedding \textit{static} sketches, more recent work has started exploring how to embed \textit{dynamic} sketches to further blend virtual sketches with the real physical environment. 
For example, RealitySketch~\cite{suzuki2020realitysketch} explores \textit{embedded and responsive} AR sketching, in which sketched objects can dynamically animate and interact with physical motion in the real world.
To push the boundary of this AR sketching and tangible interaction research domain, we explore this novel idea and demonstrates this concept by combining augmented reality sketching with actuated tangible user interfaces. 

\subsection{Actuated Tangible User Interfaces}
The original motivation for actuated tangible user interfaces was driven by the vision of coupling bits and atoms~\cite{poupyrev2007actuation}.
While traditional tangible interfaces can couple visual representation with physical interaction, these traditional TUI systems often face a challenge of digital-physical discrepancy --- the physical manipulation can change the digital representation, but the digital computation cannot change the physical
representation of passive and static objects. 
Such seamless coupling between virtual and physical worlds is not possible without \textit{actuating} physical environment (e.g., changing physical environments, corresponding to the changes in the digital world).
To address this limitation, HCI researchers have explored the research concept of actuated tangible user interfaces~\cite{poupyrev2007actuation} and shape-changing user interfaces~\cite{rasmussen2012shape, coelho2011shape, alexander2018grand} since the early 2000s~\cite{pangaro2002actuated}.
In actuated tangible user interfaces, physical objects are not merely augmented with digital overlays but are themselves dynamic and self-reconfigurable, so that they can change their physical properties to reflect the state of the underlying computation. 
These works have been greatly explored through different techniques such as magnetic actuation~\cite{patten2007mechanical}, ultrasonic waves~\cite{marshall2012ultra}, magnetic levitation~\cite{lee2011zeron}, and wheeled and vibrating robots~\cite{nowacka2013touchbugs}.
In recent years, a growing body of research started exploring tabletop mobile robots as actuated tangible objects. 
For example, Zooids~\cite{le2016zooids, kim2020user} introduces a swarm user interface, one of the class of actuated tangible interfaces which leverage tabletop swarm robots.
Similarly, many different systems have also been proposed such as ShapeBots~\cite{suzuki2019shapebots}, HERMITS~\cite{nakagaki2020hermits}, HapticBots~\cite{suzuki2021hapticbots}, Rolling Pixels~\cite{lee2020rolling}, some of which also leverage the same commercially available Sony Toio robots~\cite{nakagaki2020hermits, kim2019swarmhaptics}.

While most of these actuated tangible interfaces or swarm user interfaces are designed to interact based on \textit{pre-programmed} behavior, other researchers have also explored the way to enable the user to program and design the motion on demand. 
For example, Topobo~\cite{raffle2004topobo}, MorphIO~\cite{nakayama2019morphio}, and Animastage~\cite{nakagaki2017animastage} allow the user to program physical behaviors with \textit{direct manipulation}, as opposed to programming on a computer screen, so that the user can quickly control the actuation in improvisational ways.
Most closely related to our work, Reactile~\cite{suzuki2018reactile} also leverages projection-based AR sketching to create virtual objects, which can be later bound and programmed for the dynamic motion of swarm user interfaces.
However, Reactile only explores a small subset o bi-directional interaction (i.e., only bounded geometric parameters between virtual sketches and physical objects). 
In contrast, this paper provides a more holistic view of bi-directional sketching interaction, providing eight different categories of the \textit{bi-directional virtual-physical interaction} of AR sketches and actuated tangible user interfaces. 
We show that by combining these novel sets of design space and interaction, we can achieve a more expressive and richer set of interactions and applications, that were previously unexplored in the literature.

\subsection{Augmented Reality and Robotics}
Augmented reality and robotics is a growing area over the last decades~\cite{makhataeva2020augmented}.
While the majority of augmented reality interfaces for robotics are considered AR-enhanced Human-Robot Interaction (AR-HRI)~\cite{walker2018communicating, suzuki2022augmented}, there are also a number of prior works that explore augmented reality for robotic and actuated tangible interfaces~\cite{suzuki2022augmented}.
For example, exTouch~\cite{kasahara2013extouch} and Laser control Robot~\cite{ishii2009designing} use AR as a control and manipulation for robots. 
AR provides rich visual feedback and affordances, thus these AR interfaces provide a more intuitive understanding of how the robots or actuated interfaces should work. 
For example, AR is used to change the color, appearance, and added information to the robot~\cite{sugimoto2005augmented}. 
Additionally, systems that allow human-robot interaction via sketching have been explored in multiple prior systems~\cite{boniardi2016autonomous, shah2010robust, setalaphruk2003robot}. These prior works allow users to easily control physically actuated locomotive robots to navigate in space. Some other works have explored such sketch-based user interaction with augmented reality setup, letting users directly draw instruction over a real-physical environment~\cite{ishii2009designing, mueller2012interactive}.

As the primary goals of these prior works were to control robots' behavior naturally, AR sketches (or reality overlaid virtual digital information) affect the robot in a single direction, but not another way around. 
In such a way, unlike conventional human-robot instructional interaction, we explore the combination of AR sketches and actuated tangible user interfaces, that enhance bi-directional interaction for users to affect digital information and physical objects interactively via in-situ embedded sketching.

As we can see, most of the existing works use AR to only \textit{augment} the physical interface, but the \textit{interaction} between AR objects and robots is not widely explored in the literature~\cite{suzuki2022augmented}, except for a few examples.
For example, Kobito~\cite{aoki2005kobito} is one of the earliest explorations of synchronous coupling between AR and physical objects, which provides an illusion of virtual brownies pushing a physical cube.
By taking inspiration from the prior art, this paper explores how we can support such interaction through improvisational AR sketching so that the user can create these effects more quickly and intuitively. 






\begin{figure*}[t]
\centering
\includegraphics[width=\linewidth]{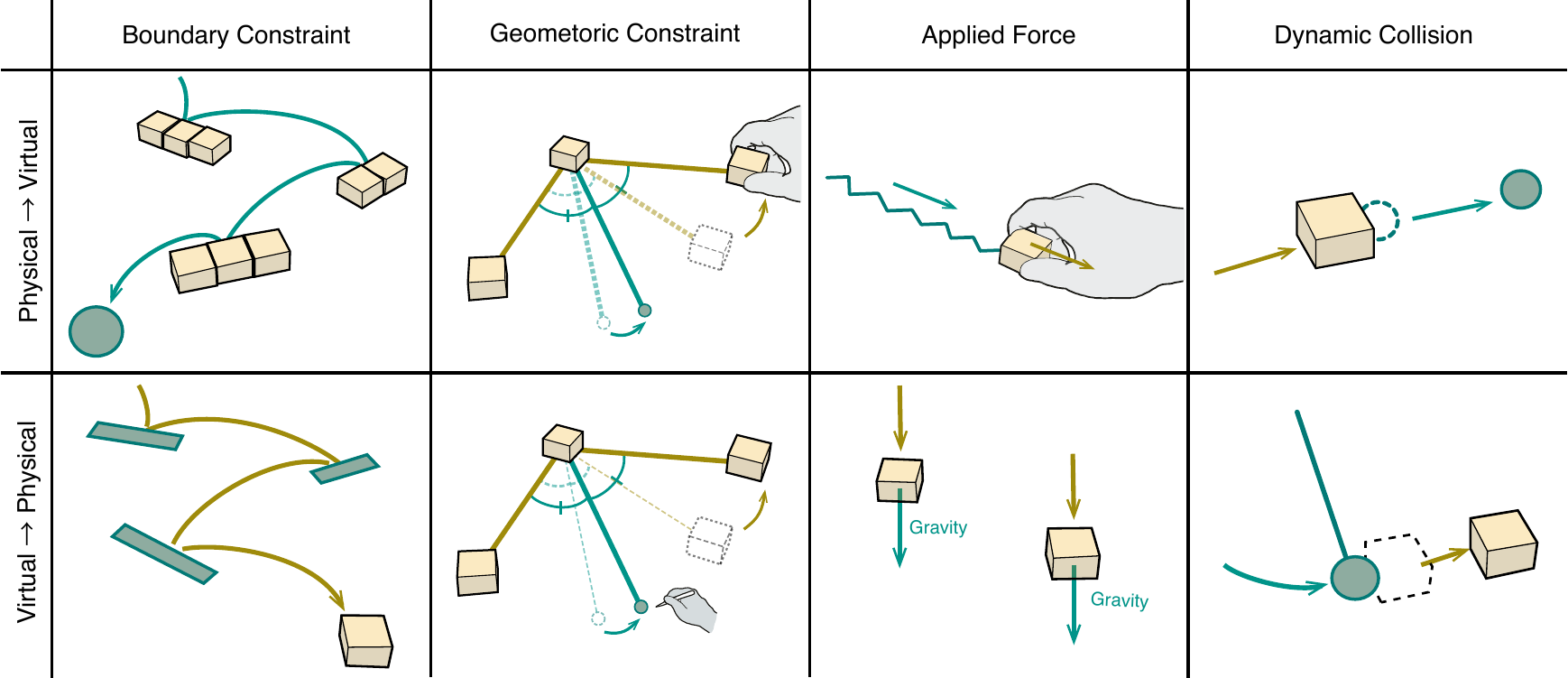}
\caption{Design space of Sketched Reality: The horizontal axis shows four basic categories of our design space: 1) boundary constraints (left), 2) geometric constraints (middle left), 3) applied force (middle right), and 4) dynamic collision (right). The vertical axis shows descriptions for each element about  1) physical → virtual interaction: physical movement/interaction affects the behavior of the virtual objects (top line), and 2) virtual → physical interaction: virtual movement/interaction affects the behavior of the physical objects (bottom line).}
\label{fig:designspace}
\end{figure*}

\section{Sketched Reality}

\subsection{Concept}
\subsubsection{Definition}
This section introduces the concept of Sketched Reality. 
Sketched Reality is an approach to combining AR sketching and actuated tangible user interfaces (TUI) for \textit{bi-directional sketching interaction}.
A bi-directional AR sketching interaction allows the virtual sketches to move, actuate, collide, and constrain physical objects through synchronized coupling between physical actuation and virtual phenomenon (Figure~\ref{fig:related-work} right).
This enables us to further blend digital and physical worlds~\cite{ishii2012radical, suzuki2022augmented}.

\subsubsection{Requirements and Scope}
For Sketched Reality, the physical world needs to be synchronized with the virtual sketched animation and movement. 
To achieve this goal, we need to incorporate the physical actuation and reconfiguration through, for example, actuated tangible user interfaces~\cite{poupyrev2007actuation}, swarm user interfaces~\cite{le2016zooids}, shape-changing user interfaces~\cite{rasmussen2012shape}, and reconfigurable environments~\cite{cavallo2019dataspace}.
The sketched interaction with all of these different configurations of the system is vast and goes beyond the scope of this paper.
Therefore, in this paper, we specifically focus on bi-directional sketching interactions with \textit{tabletop-scale small robots}, that can act as actuated tangible tokens and objects on a horizontal surface. 


\subsection{Design Space of Bi-Directional Interaction}
In this section, we explore the design space of bi-directional interaction to understand how virtual and physical elements can affect each other. 
We have identified four basic categories: 1) boundary constraints, 2) geometric constraints, 3) applied force, and 4) dynamic collision (Figure \ref{fig:designspace}). 
For each element, we describe 1) \textit{physical $\rightarrow$ virtual interaction}: physical movement/interaction affects the behavior of the virtual objects, and 2) \textit{virtual $\rightarrow$ physical interaction}: virtual movement/interaction affects the behavior of the physical objects. 

\subsubsection{Boundary Constraint}
Boundary constraints refer to a virtual or physical constraint based on a static (stationary object), such as a wall or contour. 
The boundary constraint allows interaction between a dynamic and static object --- a dynamic object can freely move in the physical or virtual space, whereas a static object stays at a certain position.
This also applies to both virtual and physical worlds. 
For example, the physical boundary constraints also do the same thing for a virtual dynamic object (Figure~\ref{fig:designspace} left top), and the virtual boundary constraints can prevent and reflect the physical object  (Figure~\ref{fig:designspace} left bottom). 

\subsubsection{Geometric Constraint}
Second, geometric constraints refer to a fixed positional relationship between virtual and physical objects.
For example, consider the situation where the user applies the geometric constraint as a middle angle between two lines, like a bisector line of a triangle or two lines.
In physical to virtual interaction, when the user moves the endpoint of one line, the bisector line also moves accordingly to maintain an equal angle between two lines (Figure~\ref{fig:designspace} middle left top). 
On the other hand, in virtual to physical interaction, the position of a physical robot can dynamically change based on the maintained geometric constraint. 
For example, in a similar situation, when the user moves the angle of the two lines by virtually moving one end of the line, then the position of the robot moves accordingly to maintain the equal angle of geometric constraint (Figure~\ref{fig:designspace} middle left bottom). 
The geometric constraints can be various positional relationships such as length and angle, which can govern how virtual or physical objects behave through the connected and bound parameters. 


\subsubsection{Applied Force}
Through bi-directional interaction, the user can also apply force to both virtual and physical objects. 
For example, in physical to virtual interaction, the user can apply an extension force to a virtual spring by deforming a virtual spring object  (Figure~\ref{fig:designspace} middle right top).
In a similar way, the physical interaction can deform soft materials such as clothes or elastic string sketched in the virtual world. 
On the other hand, in virtual to physical interaction, the user can apply force to a physical object.
For example, the user can apply gravity force to the physical robot, so that the robot starts moving in a certain direction with a gravity force (Figure~\ref{fig:designspace} middle right bottom).
Such virtual force can take different forms such as magnetic force or friction.


\subsubsection{Collision}
Finally, dynamic collision refers to an interaction between virtual and physical objects, in which these objects collide with each other.
In contrast to the boundary constraint, which focuses on the collision between \textit{dynamic} and \textit{static (stationary)} objects, the dynamic collision focuses on the interaction of multiple \textit{dynamic} objects between virtual and physical worlds. 
For example, in physical to virtual interaction, a physical dynamic object hits a virtual dynamic object, which creates a movement and collision of the virtual object. 
In the same way, in virtual to physical interaction, a virtual object hits a physical object, then it moves the physical object. 

\section{Demonstrating Sketched Reality}

\subsection{System Overview}
To demonstrate the concept of Sketched Reality, we develop a system that combines an AR sketching interface and actuated tangible interfaces, based on tabletop-size small robots. 
More specifically, our system employs iPad and WebXR for AR sketching and Sony Toio mobile robots~\footnote{\url{https://www.sony.com/en/SonyInfo/design/stories/toio/}} for actuated tangible objects.

\begin{figure}[h!]
\centering
\includegraphics[width=1\linewidth]{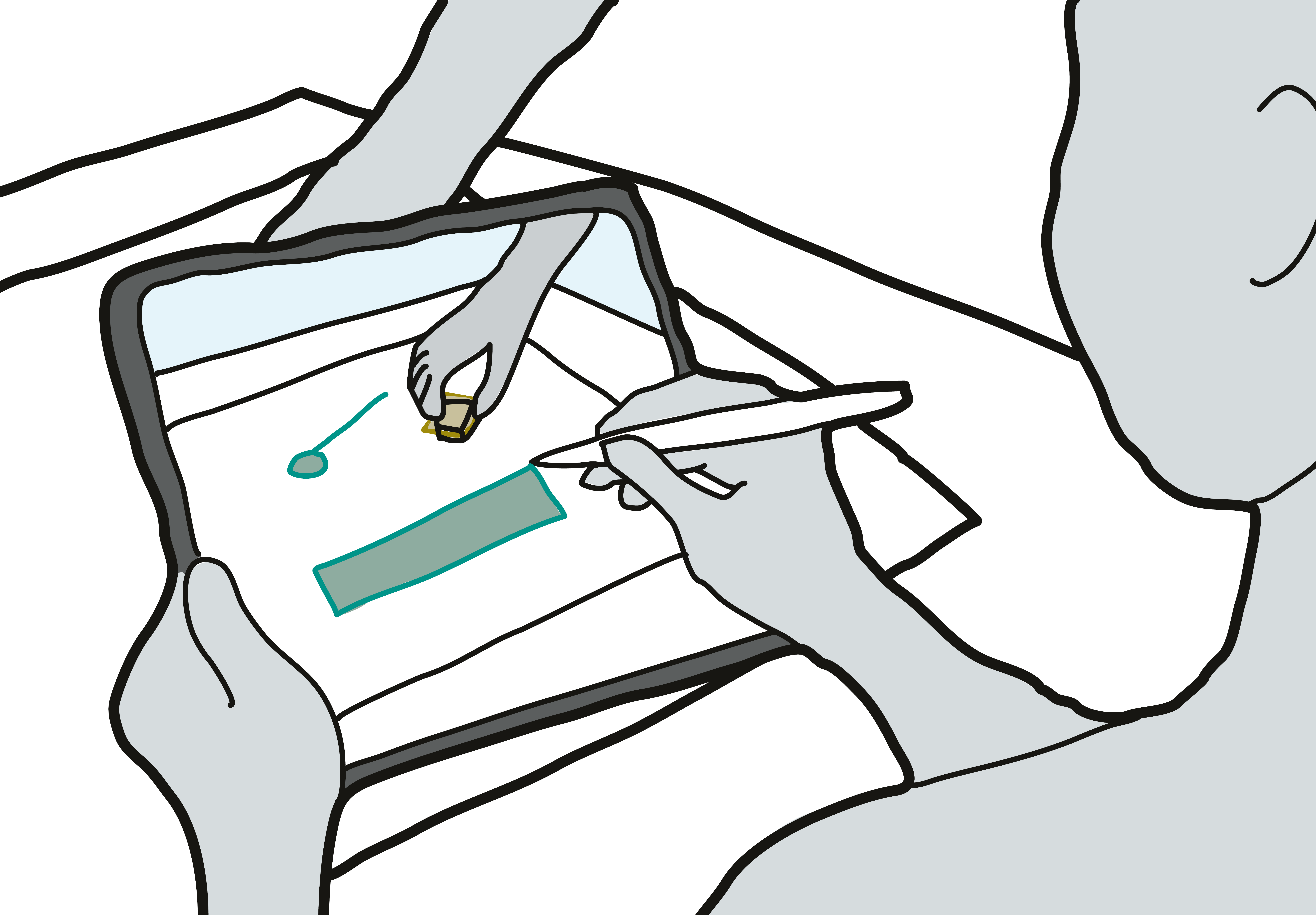}
\caption{Basic setup of our system. The user draws virtual sketches through the iPad screen and Apple Pencil. After the simple calibration, the embedded 2D canvas onto a table surface in AR can be synchronized with the Sony Toio coordination.
The green lines and objects represent the virtually sketched objects whereas a yellow cube object represents the physical Toio robot.}
\label{fig:setup}
\end{figure}

\subsubsection{Interaction Workflow and Techniques}
In our system, the user can \textit{actuate} Sony Toio robots through AR sketching interactions. 
As we discussed in the Design Space section, there are several approaches, such as colliding virtual and physical robots, constraining robots' motion with virtual lines, and applying force to the robots through virtual property or environmental force.
Similar to RealitySketch~\cite{suzuki2020realitysketch}, the user can interact and manipulate the tangible Toio robots to modify the virtual sketched behaviors, but also modify or simulate the virtual sketches to change the behavior of tangible Toio robots through virtual geometric constraints or collision.

In our Sketched Reality system, the system allows the following interaction workflow for bi-directional sketching interaction:
\begin{enumerate}
\item The user first selects a horizontal or vertical surface in the real world, then calibrates the position and coordination of the embedded 2D canvas to synchronize the coordination of the mobile robots. 
\item The user sketches a line in an AR screen using the iPad touch screen and Apple Pencil.
\item The system recognizes a sketched line and creates a virtual body (closed shape like a circle or rectangle) or constraint (open shape such as lines or spring) based on the drawn shape.
\item The user can apply property or force to both virtual sketched objects and physical robots in AR by selecting dynamic (affect gravity force) or static object (stationary object which does not affect gravity force).
\item The user can interact with virtual or physical objects through virtually manipulating objects in AR scene or tangibly manipulating physical objects in the real world.
\end{enumerate}
In the following section, we will describe each step in more detail. 

\subsection{Select a Surface and Calibrate the Position}
\subsubsection{Surface Detection}
Sketched Reality system leverages mobile augmented reality (A-Frame and 8th Wall on iPad) to embed sketches onto a real-world surface.  
The user sketches with a pen on a touchscreen, where the sketched elements are overlaid onto a camera view of the real world.
All of the sketched elements are 2D, so the user first needs to define the embedded 2D canvas by selecting a horizontal or vertical surface in the real world.
Our system employs WebXR (8th Wall~\cite{8th-wall}) for the surface detection so that the system allows the user to sketch on a detected surface. 

\subsubsection{Calibration between AR Canvas and Physical Robot's Coordination}
\begin{figure}[h!]
\includegraphics[width=0.32\linewidth]{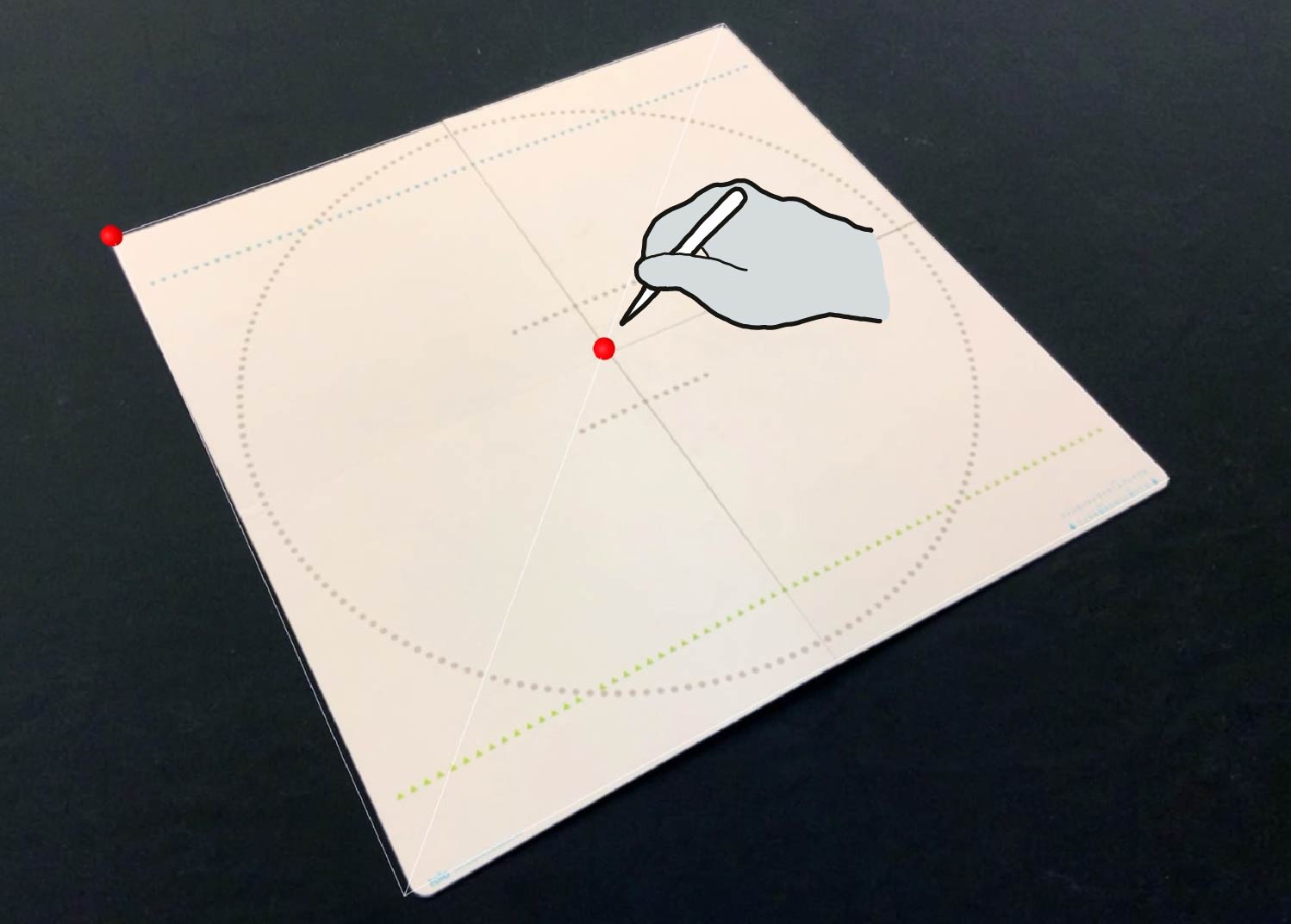}
\includegraphics[width=0.32\linewidth]{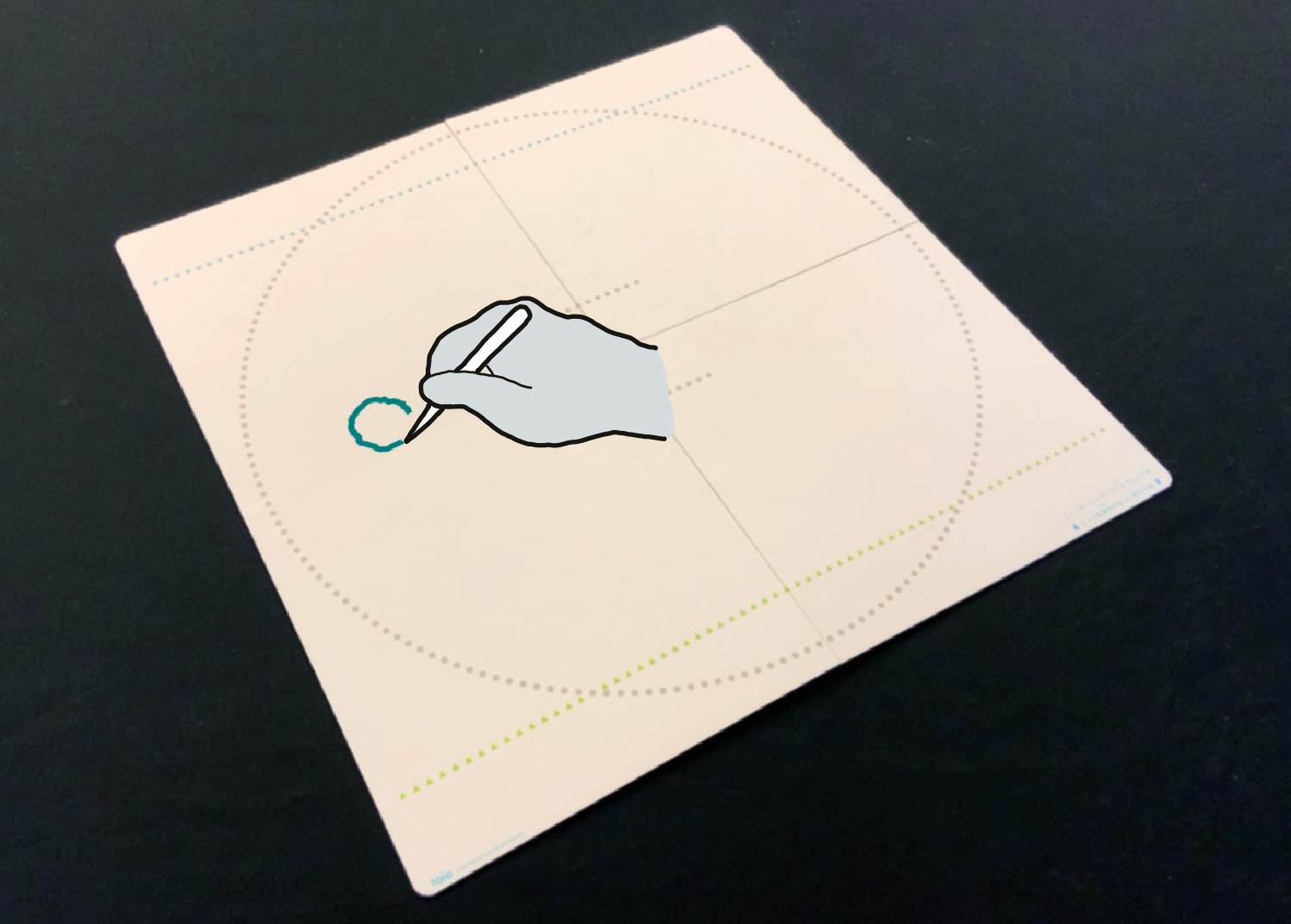}
\includegraphics[width=0.32\linewidth]{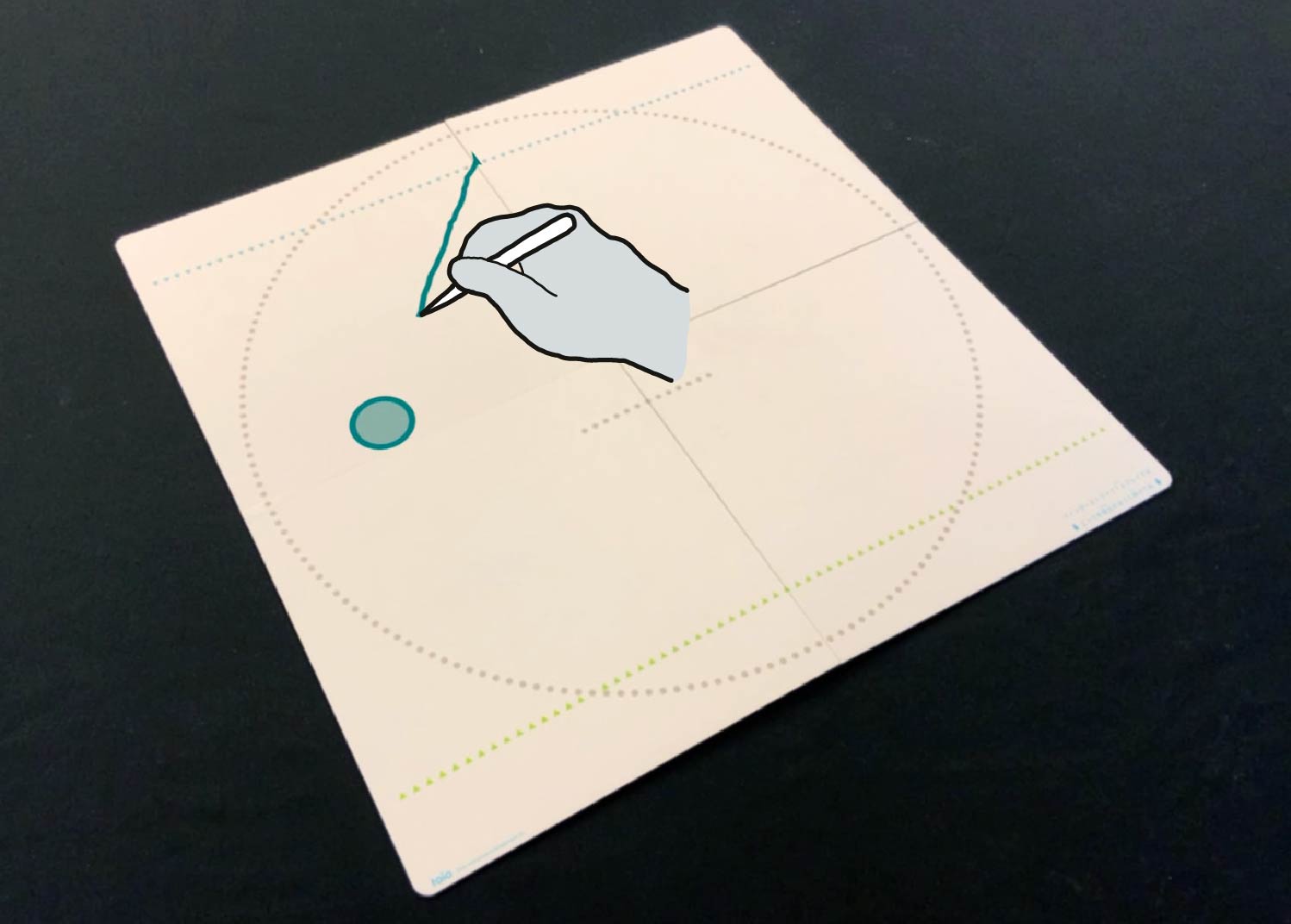}

\caption{Position Calibration and Sketching to Create Circle Shapes: By tapping the screen and selecting the surface with two points, the user can calibrate the position between the AR canvas and the physical robot's coordination(left). Then, the user can create the 2D object in the AR scene by sketching a closed line. When the user draws a closed circle, the user can create a 2D ball shape of the body on a screen(middle). The user also can create an attached constraint with a single line by drawing to the sketched 2D body(right).}
\label{fig:system-step-1}
\end{figure}

In Sketched Reality, the virtual and physical worlds need to be seamlessly coupled and synchronized with each other. 
Therefore, the coordination of the virtual and physical worlds is an important requirement for the system. 
To do so, once the system detects the surface, then the system allows the user to calibrate the position of the embedded 2D canvas in AR and Toio robot coordination. 
To calibrate the position, the user simply taps the center and left top corner of the physical square mat (Figure~\ref{fig:system-step-1} left).
Once the user taps two points, then the AR view shows the two red dots to confirm the position. 
When the calibration step is done, then the user can start drawing on a physical mat. 

The main reason we chose the Sony Toio robot is its sophisticated and easily deployable tracking system. 
For tracking and localization of the robot, Toio has a built-in look-down camera at the base of the robot to track the position and orientation on a mat by identifying unique printed dot patterns, similar to the Anoto marker~\footnote{\url{https://en.wikipedia.org/wiki/Anoto}}.
The built-in camera reads and identifies the current position of the robot, enabling easy 2D tracking of the robots with no external hardware. 
Therefore, we can precisely track the location of multiple Toio robots on a physical square mat (Figure~\ref{fig:system-step-1} left).

We leverage this built-in position and orientation tracking capability for synchronization between AR and the physical world.
Based on the above simple calibration process, the system can match the coordinate systems between the Toio mat and the AR sketching canvas.

\subsection{Sketch to Create Shapes and Constraints}
\subsubsection{Sketch a Line and Recognize a Shape}
In Sketched Reality system, the user can create a virtual object with a simple line drawing. 
In our system, we have two virtual object categories: 1) body and 2) constraint.
For constraint, it also has several variations such as fixed-line constraint, spring, or geometric relationship.
The user can create these virtual objects by sketching them on a screen. 
In general, the closed line can be categorized as a body and the open line can be categorized as a constraint. 


\subsubsection{Create a Body with a Closed Line}
\begin{figure}[h!]
\centering
\includegraphics[width=0.32\linewidth]{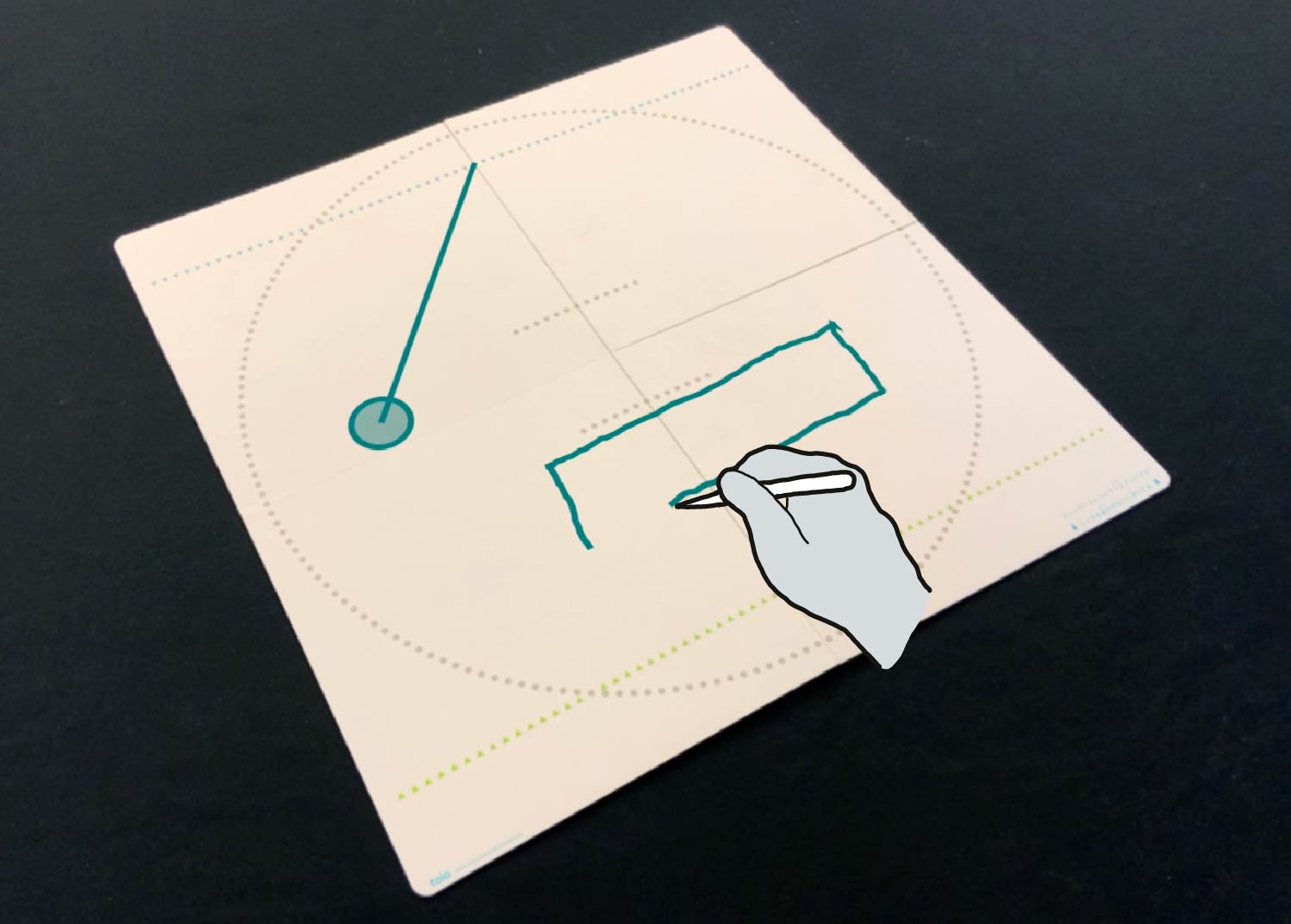}
\includegraphics[width=0.32\linewidth]{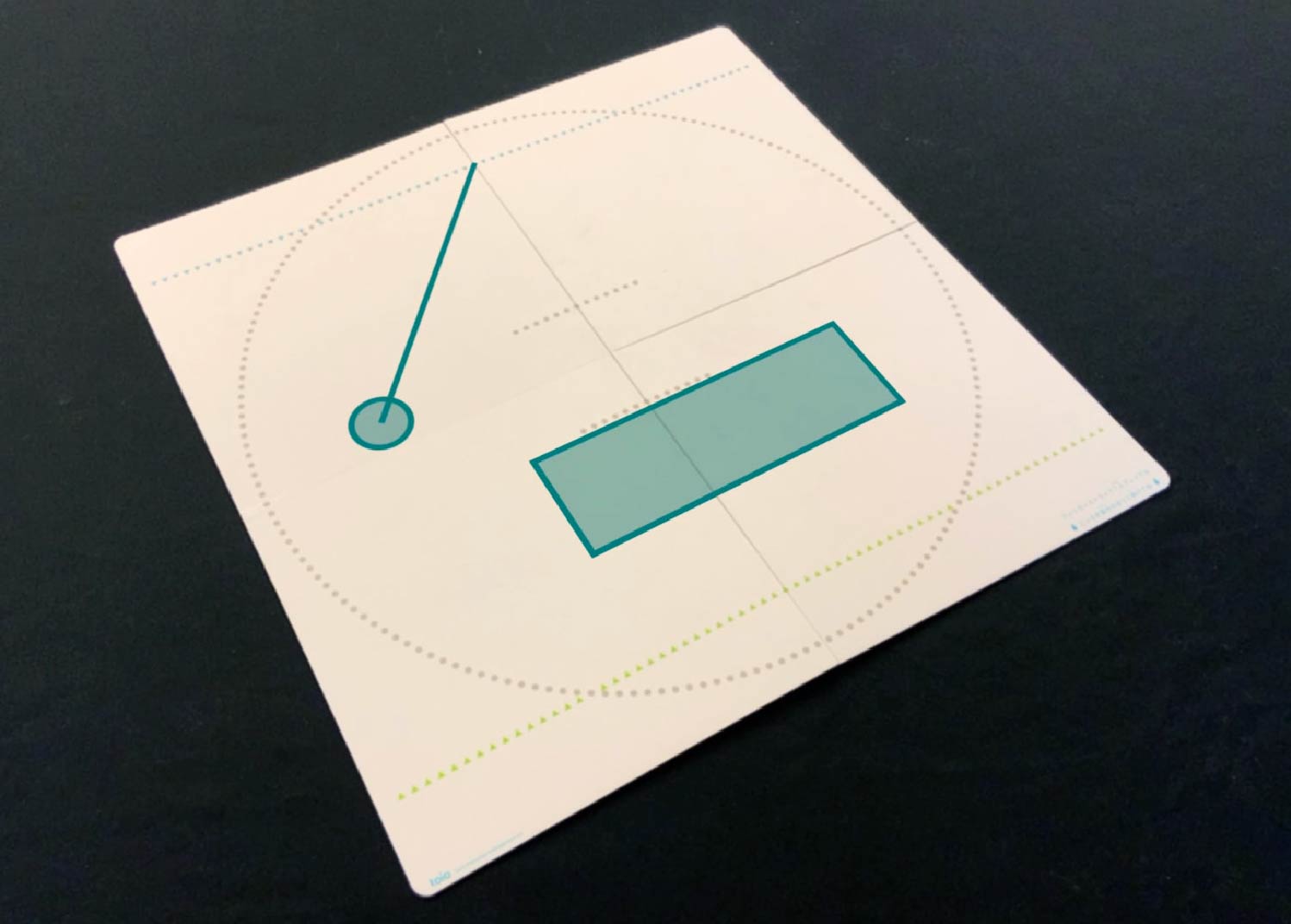}
\includegraphics[width=0.32\linewidth]{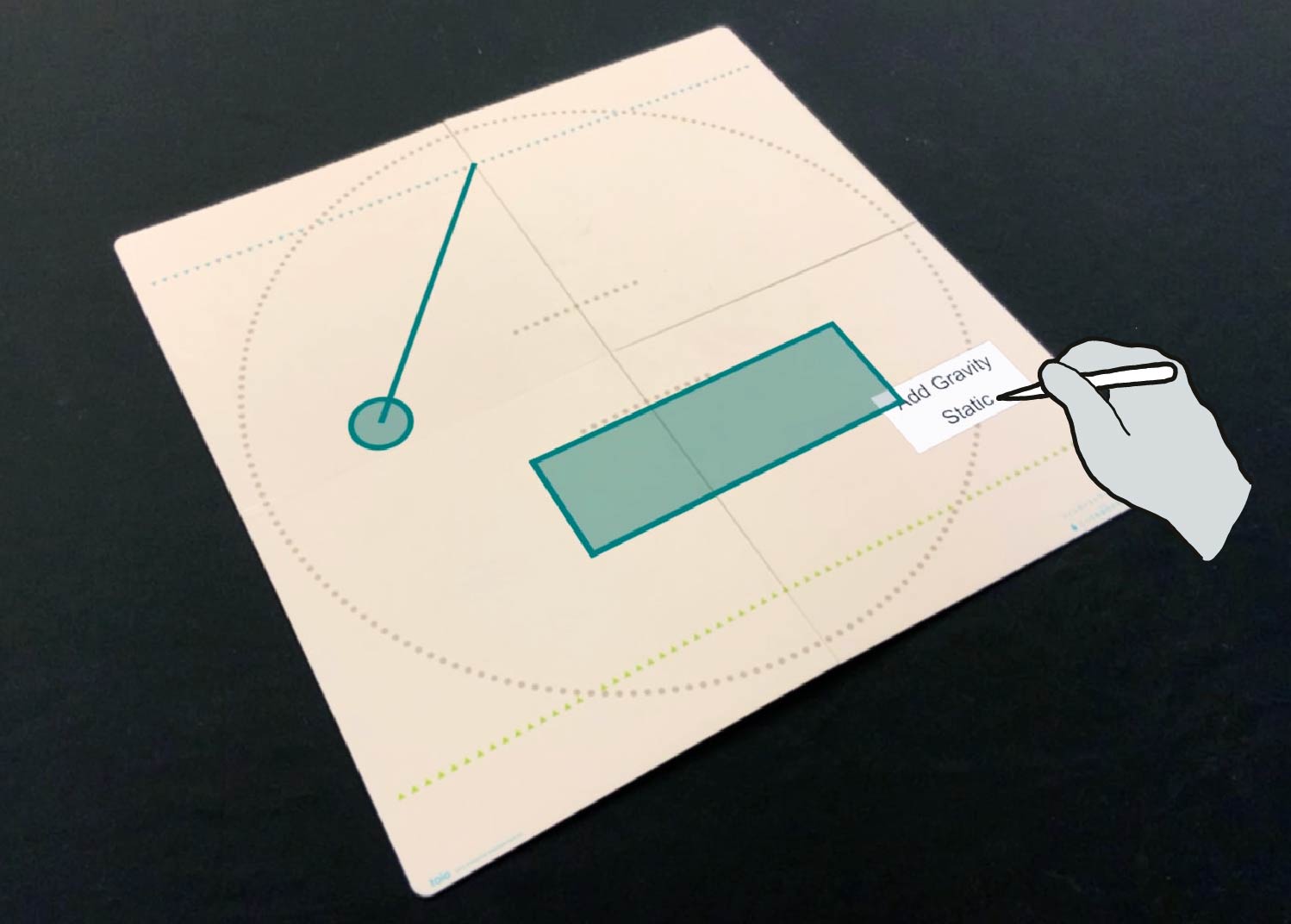}
\caption{Sketching to Create Rectangle Shapes: The user can create a 2D rectangle shape of the body by sketching a closed rectangle-shaped line (left, middle). Then, the user applies static property to the sketched virtual rectangle from the AR menu (right).}
\label{fig:system-step-2}
\end{figure}

First, the system detects the body or constraint based on the shape. 
By default, the system can basically recognize two different categories: 1) body and 2) constraint. 
The body refers to the virtual object such as circles or rectangles, whereas the constraint refers to the virtual relationship between bodies, such as a constant line or spring.
For example, in Figure~\ref{fig:system-step-1}, the user draws a circle shape, then the system recognizes it as a virtual circular shape body, which is embedded in the AR view. 
Once the body is generated by sketches, the system shows the 2D object in the AR scene.
In a similar manner, Figure~\ref{fig:system-step-2} left and middle illustrates the situation where the user draws a rectangle shape and the system recognizes and embeds the rectangle body in the AR scene.
By default, this 2D object is a floating body, which enables the collision with virtual or physical objects but does not affect gravity force. 


\subsubsection{Create a Constraint with a Non-Closed Line}
Similarly, the user can also create a constraint by drawing a non-closed line. 
By default, the system supports two different constraints: 1) a static line constraint, and 2) a spring constraint. 
The system recognizes these two constraints based on the sketched line shape.
Similar to RealitySketch~\cite{suzuki2020realitysketch}, if one end of the line is attached to a body (either virtual or physical), it creates an attached constraint in which when the body moves, the attached end of the constraint also moves and follows. 
For example, Figure~\ref{fig:system-step-1} right and Figure~\ref{fig:system-step-2} left show the user draws a line to the virtual sketched circle in AR.
In this way, the user can create a fixed-length geometric constraint between the point on the surface and the attached virtual object. 
The end of these line constraints can not only be attached to the virtual sketched objects, but also attached to the physical robot.


\subsection{Apply Force or Property to Virtual or Physical Object}
\subsubsection{Applying Force or Property to Virtual Objects}
Once the user creates the virtual sketched object, the user can also change the property of the body or constraint through the AR interface. 
For the body, the user can choose the property of either 1) dynamic body or 2) static body. 
The dynamic body refers to the object that can move on a canvas, being affected by the gravitational force, such as a bouncing ball and block that falls down.
On the other hand, a static body refers to a stationary object, such as a wall, boundary constraints, or slope that does not move.
For example, the dynamic body can move based on the collision, while the static body does not move.
Virtual objects with different properties can achieve different virtual-physical interactions, such as boundary constraints (static vs dynamic objects) or collisions (dynamic vs dynamic objects).
To specify the property, the user can simply tap the sketched object, then the AR interface shows the menu of \textit{``Add Gravity''} or \textit{``Static''}.
For example, Figure~\ref{fig:system-step-2} right illustrates the user applying the static body property to the sketched rectangle shape.
Once the user selects the menu, then the property of the selected virtual object will change. 
In a similar way, the user can also change other properties of the constraint. 
For example, the user can change the elasticity of the static line constraint or spring constraint by tapping and selecting the menu attached to these sketched constraints. 
Similar to RealitySketch~\cite{suzuki2020realitysketch}, the user can also apply the geometric constraint, such as the parameter of the constraint (e.g., length or angle of the constraint).


\begin{figure}[h!]
\includegraphics[width=0.32\linewidth]{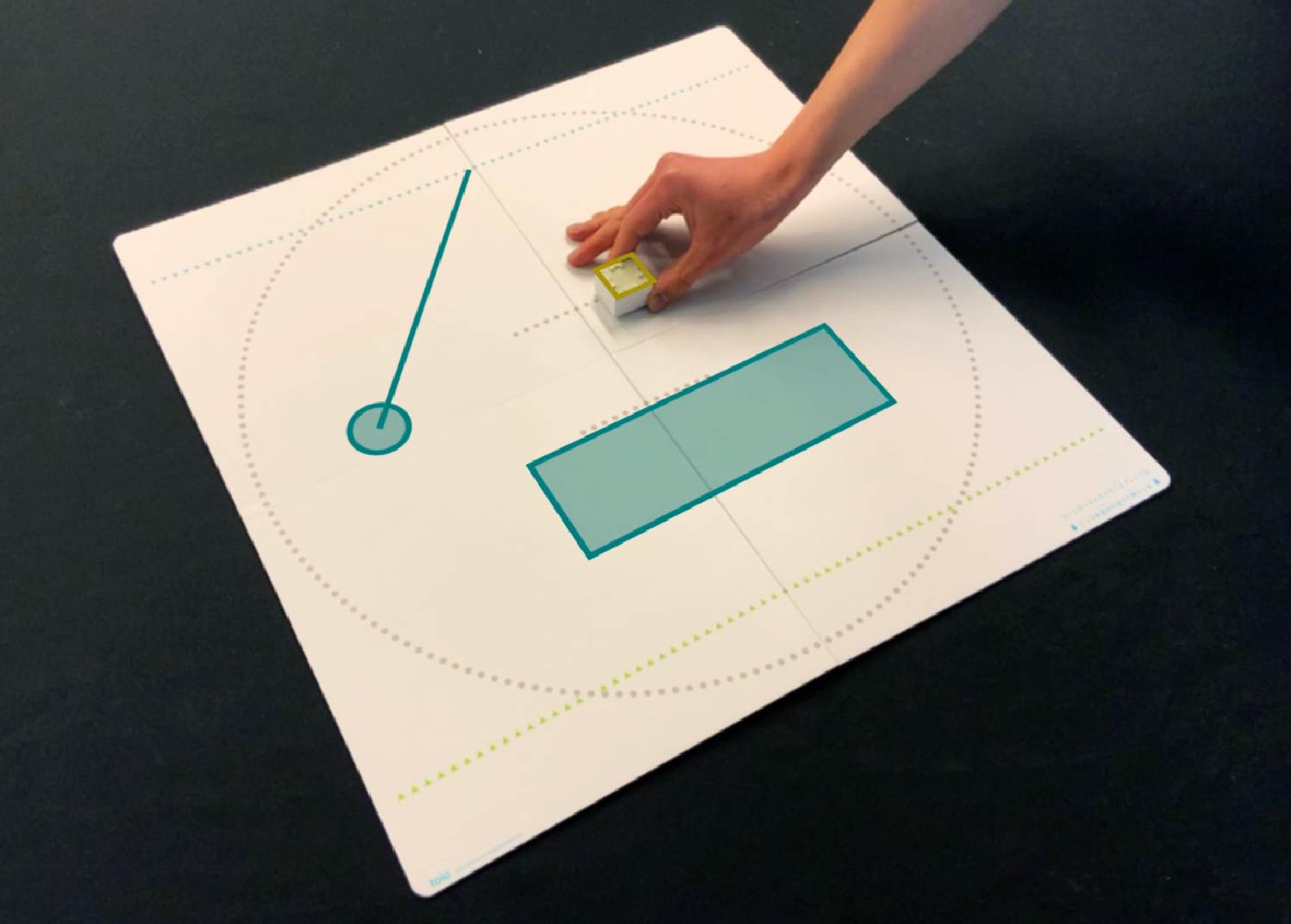}
\includegraphics[width=0.32\linewidth]{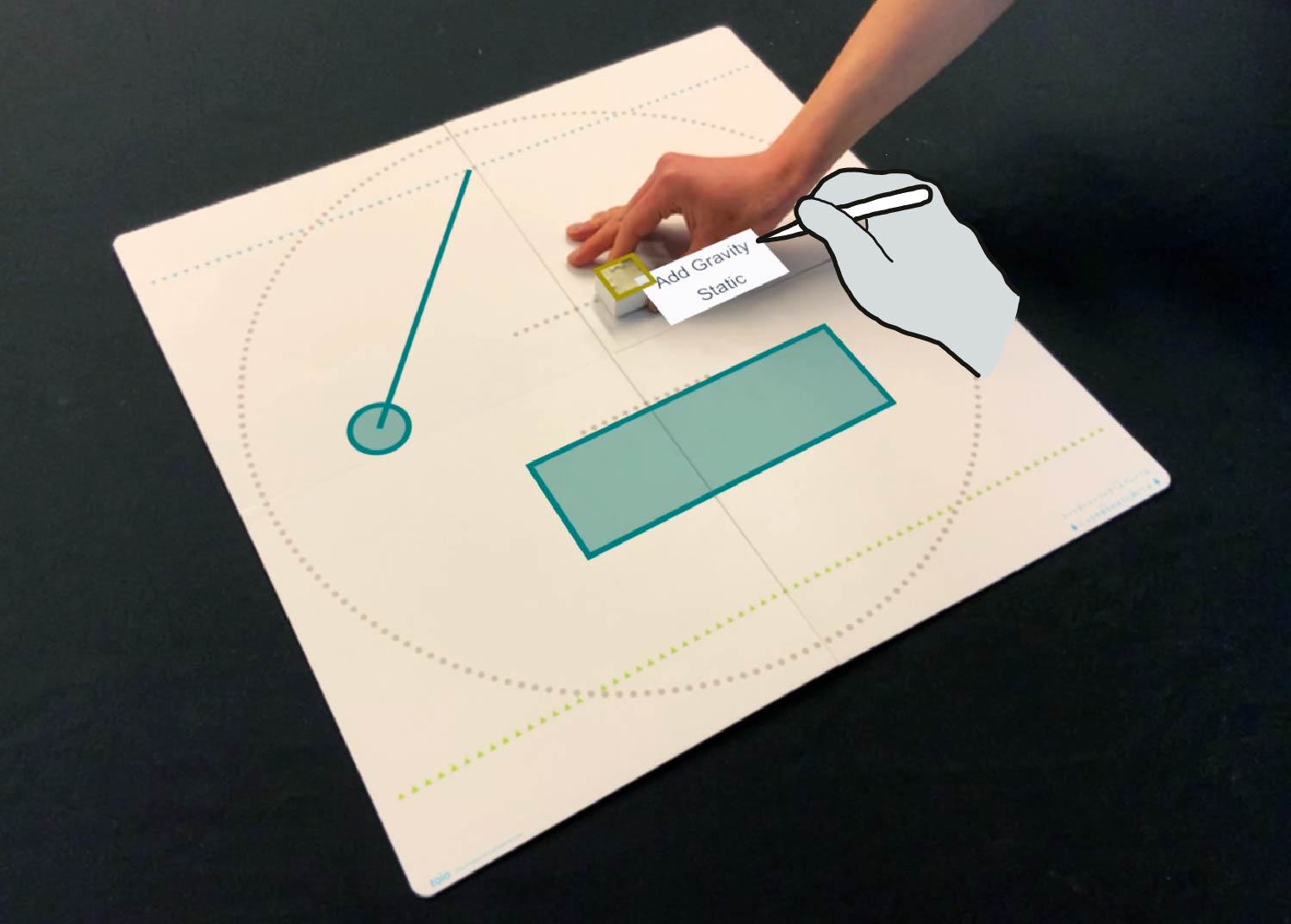}
\includegraphics[width=0.32\linewidth]{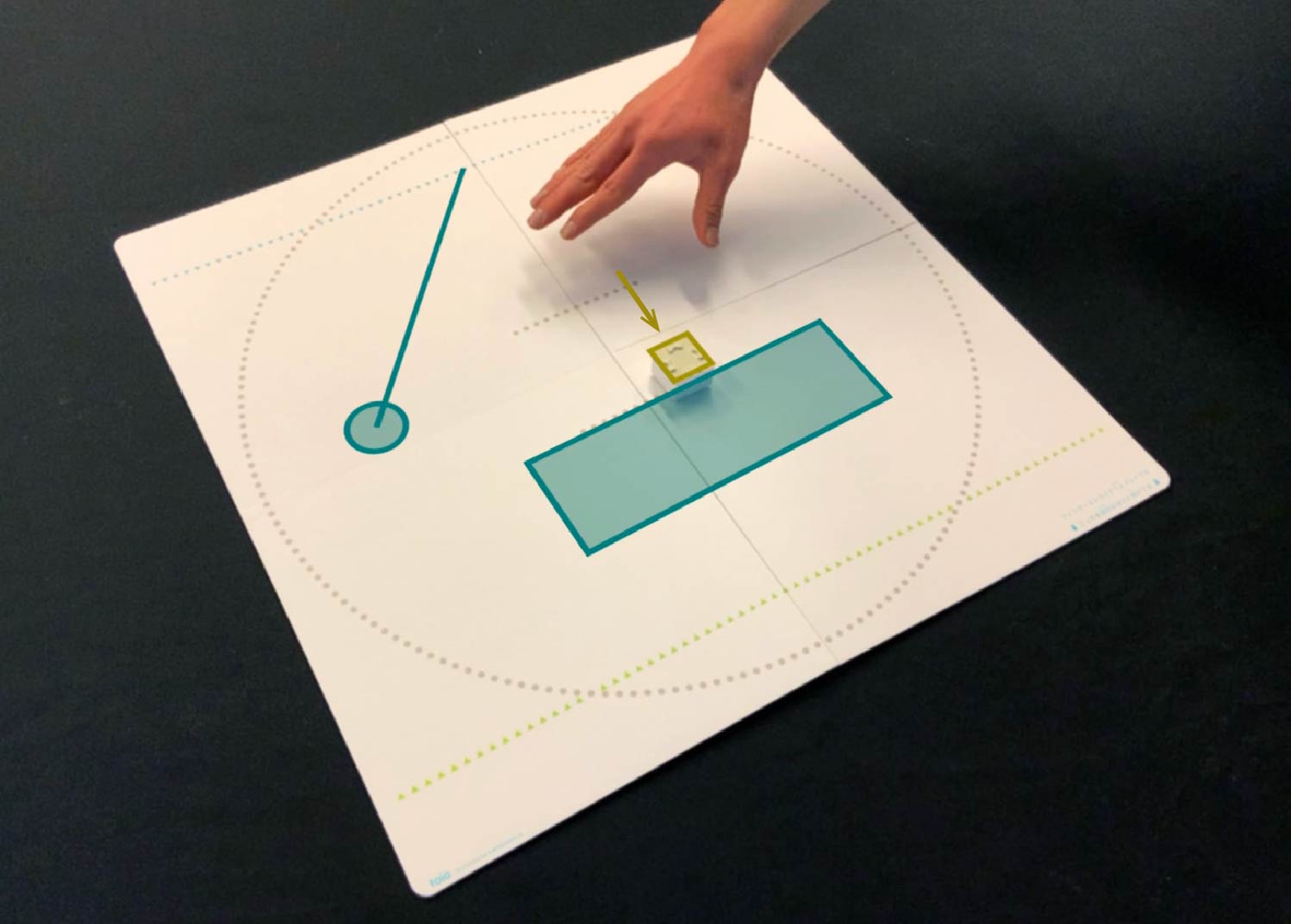}

\caption{Apply Force or Property to Physical Robots: When the user places a physical robot on the mat (left), the user can see the menu of \textit{``Add Gravity''} or \textit{``Static''} (middle). The physical robot can collide with the virtual rectangle object by applying gravity (right).}
\label{fig:system-step-3}
\end{figure}

\subsubsection{Applying Force or Property to Physical Robots}
In our system, physical robots also behave as dynamic or static bodies. 
Therefore, the user can also change the property or apply force to the physical robots, in the same way, we do with virtual sketched objects. 
This functionality helps the robots interact with virtual objects through bi-directional interaction.
When the user places a physical robot on the mat, then the system starts tracking the position of the robot and shows a yellow overlaid shape on top of the robot in AR view. 
In a similar manner, the user can tap the robot in the AR interface to show up in the menu of \textit{``Add Gravity''} or \textit{``Static''}.
For example, if the user specifies the robot to be dynamic and effect by gravity, the robot can fall down in the direction of gravity.
Furthermore, Figure~\ref{fig:system-step-3} shows that the user selects the added gravity to apply the dynamic body property to a robot, then the robot starts falling down. 
Since the user applies the static object property to the virtual rectangle shape in Figure~\ref{fig:system-step-2}, the robot hits the ground and stays on top of the static body ground.

\begin{figure}[h!]
\includegraphics[width=0.32\linewidth]{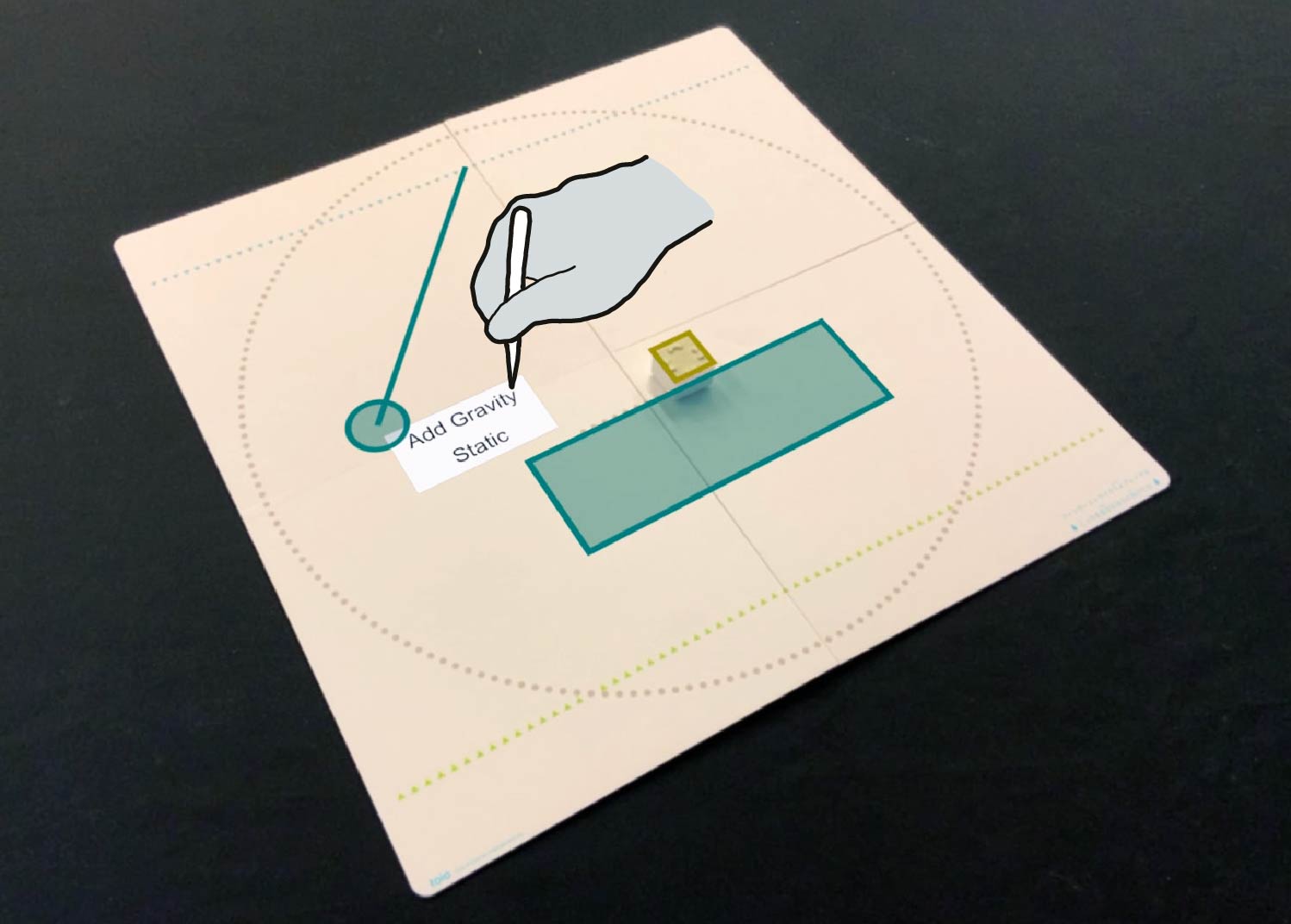}
\includegraphics[width=0.32\linewidth]{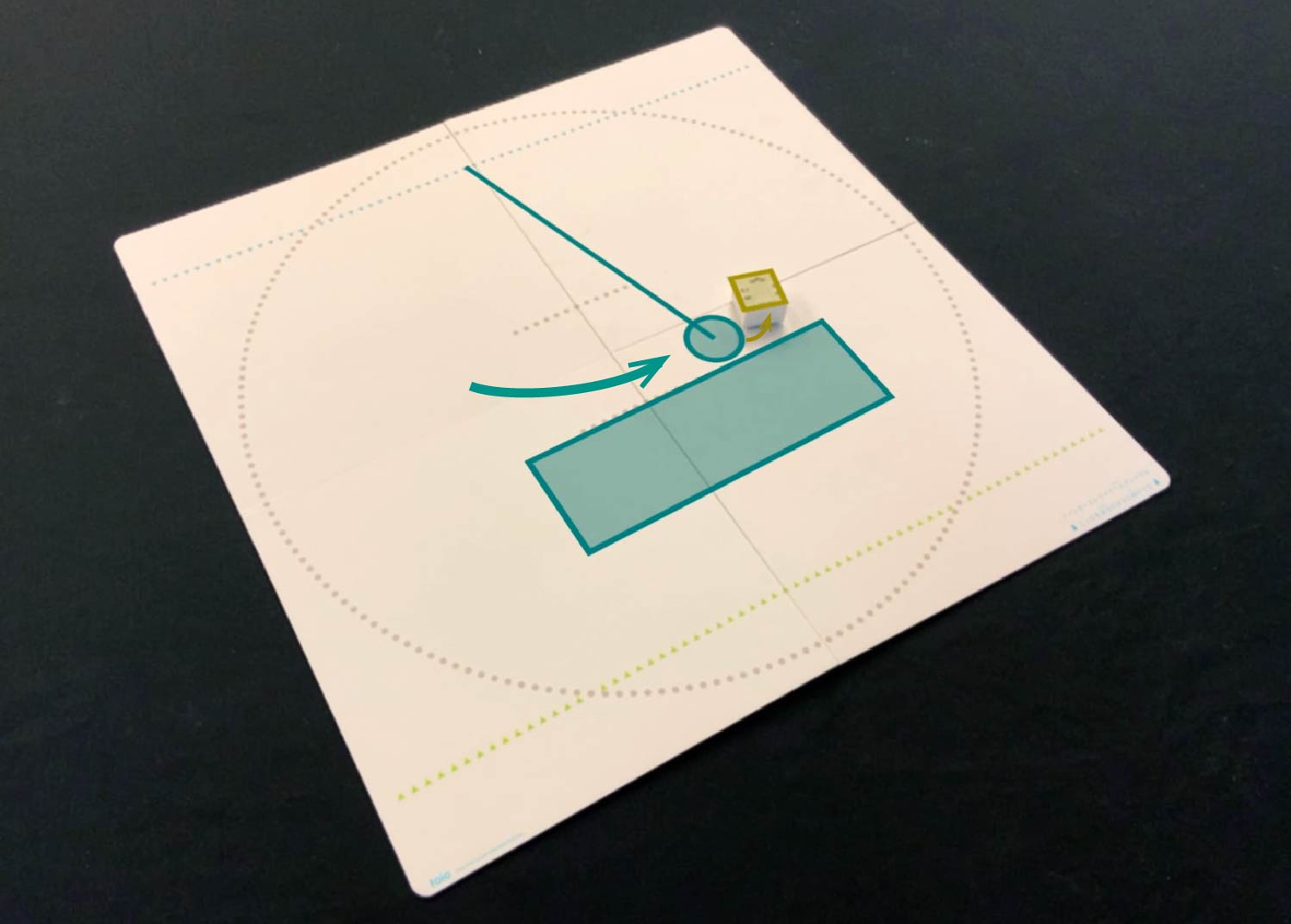}
\includegraphics[width=0.32\linewidth]{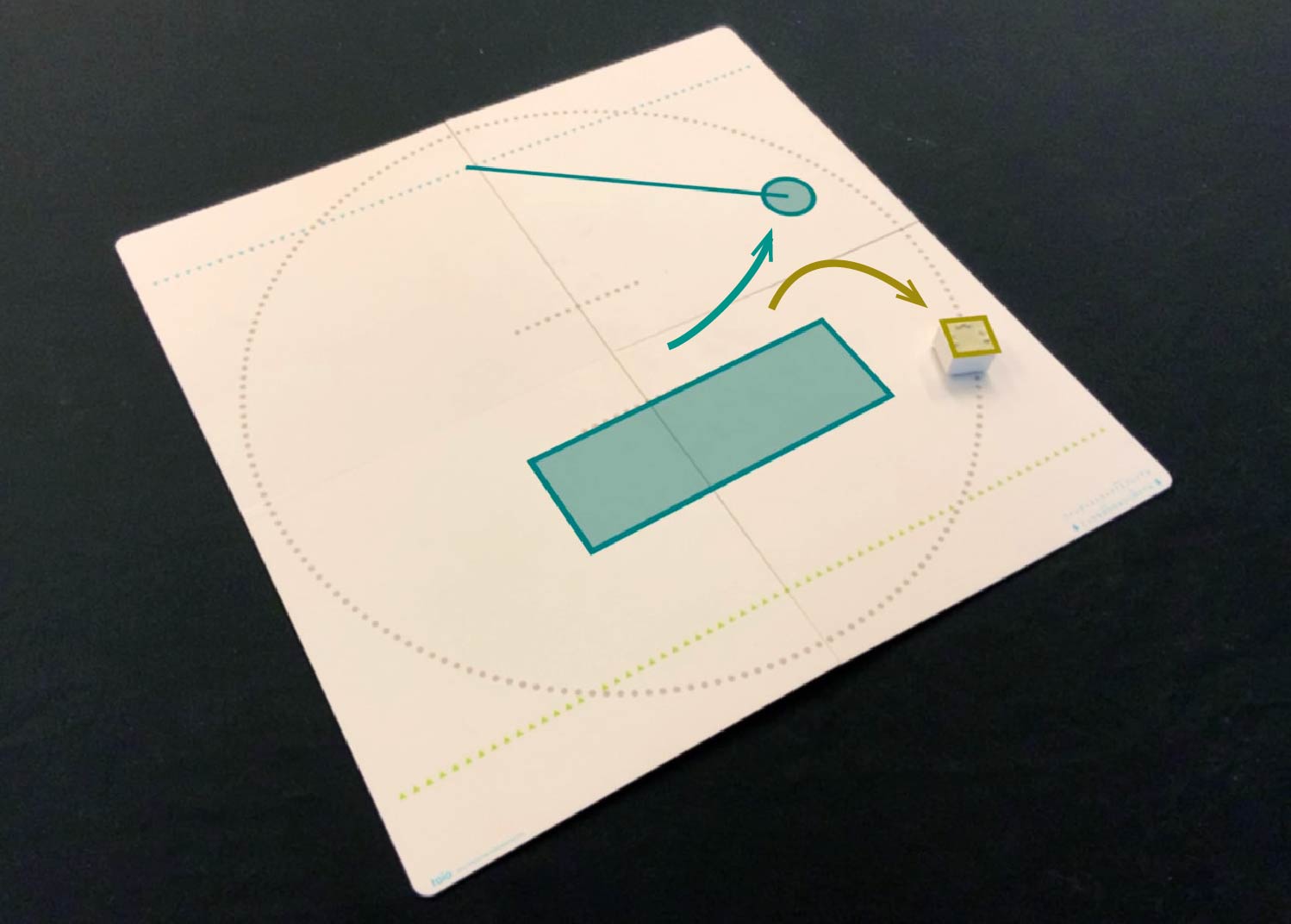}
\caption{Interact between Virtual and Physical Objects: When the user applies gravity to the virtual sketched pendulum created with a circle-shaped body and an attached constraint (right), the virtual pendulum can affect the physical robot through collision, geometry constraints, boundary constraints, and applied force (middle). After the collision with the virtual pendulum, the physical object falls from the virtual rectangle scaffold through applied gravity force (left).}
\label{fig:system-step-4}
\end{figure}

By applying force or property to virtual or physical objects, then these objects start interacting with each other. 
For example, Figure~\ref{fig:system-step-4} shows that when the user applies gravity to a virtual pendulum ball, then it starts swinging to hit the robot to collide with each other so that the physical robot moves and starts falling down from the static ground.
The user can also change the shape of the physical robot by tapping and drawing a physical robot. 
For example, when the user taps the physical robot and starts drawing, then it changes the virtual shape of the physical robot (e.g., change the yellow shape to a circular shape), which can change how to behave when interacting with virtual objects.

\subsection{Interaction between Virtual and Physical Objects}
\subsubsection{Manipulating Virtual Objects}
Once sketching and property change are done, the user can start interacting with the virtual objects. 
The dynamic body object can be basically manipulated with pen or touch interaction, so that the user can drag the object to a different position to interact with virtual or physical objects.
These virtual interactions can affect the physical robots through collision, geometry constraints, boundary constraints, and applied force.

\subsubsection{Manipulating Physical Robots}
In the same way, the user can also manipulate and interact with physical robots through tangible and embodied interaction.
With this interaction, the user can also affect and move the virtual objects in a seamless manner. 
This interaction allows a couple of different interactions for actuated TUI.
First, it can support users in manipulating and controlling the actuated TUI and robot motion. 
Second, it can also support users to give another cycle to interact, observe, and interpret tangible physics models to explore and examine models. 
This allows users to 1) observe how the improvisational sketched models affect the motion of actuated TUIs based on the sketched relationships, 2) tangibly interact with the virtual physical models to perceive and explore the behaviors, 3) interpret to further understand how certain sketched relationship or physic model affects the behaviors. 
After users interact and interpret, they can go back to sketch mode to modify and improvise the models. In such a way, it can support iterative and explorative understandings. 

\begin{figure*}[h]
\centering
\includegraphics[width=1\linewidth]{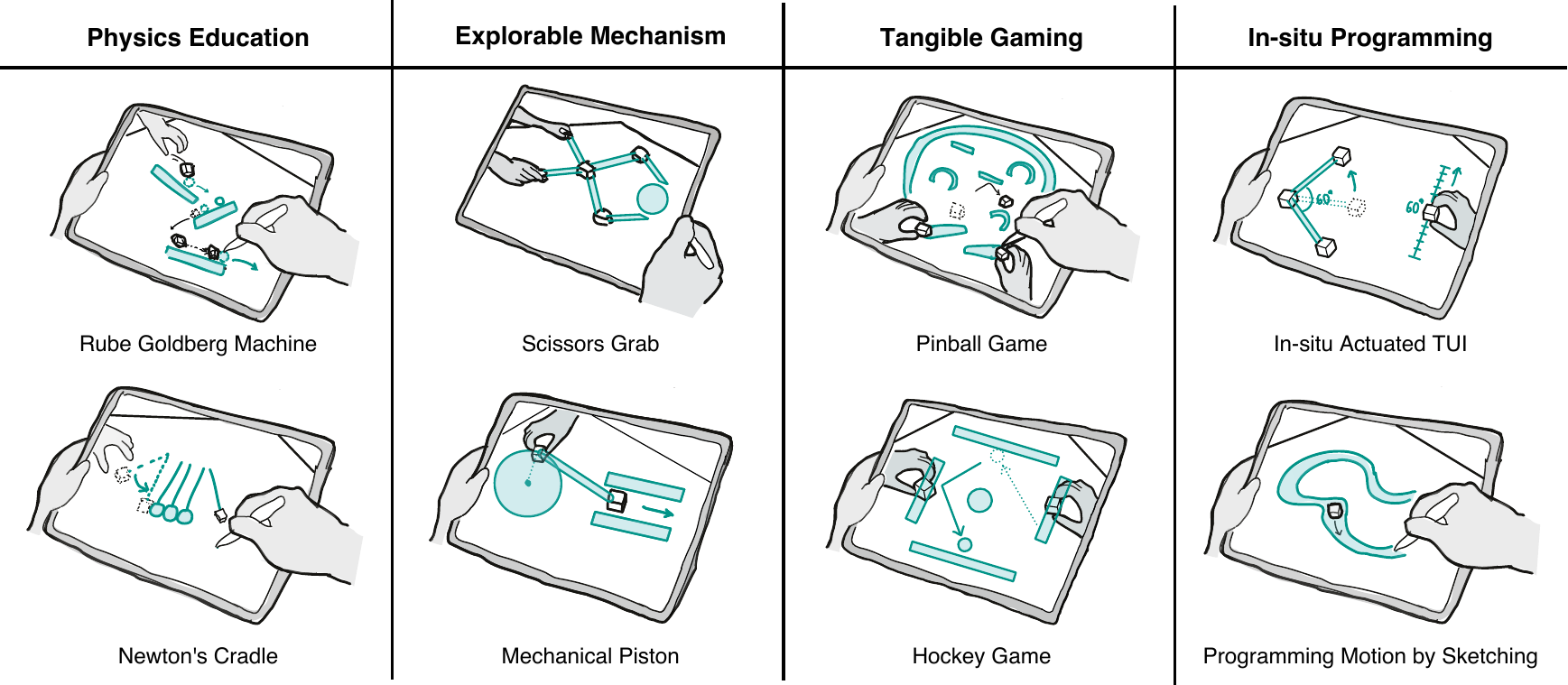}
\caption{Application Sketches: Sketches highlighting the capability of Sketched Reality through Physical Education (left), Explorable Mechanism (middle left), Tangible Gaming (middle right), In-situ Programming (right).}
\label{fig:applicatino-sketches}
\end{figure*}

\section{Implementation}


\subsection{Mobile Robot Tracking and Control}
We use Sony Toio, small tabletop-size robots, for actuated physical objects of our system.
The system can track each Toio robot's position based on the pattern-printed tracking mat.
Each robot has 3.2 cm x 3.2 cm x 2.5 cm and can travel at the speed of 24 cm/sec horizontally.
The tracking mat covers 55 cm × 55 cm on top of the table and can track the position with a 1 mm error. 
To control the Toio robot, we need a Node.js server to run the control script, based on the open-source library of toio.js~\footnote{\url{https://github.com/toio/toio.js/}}.
The Node.js server running on MacBook Pro 16-inch (Intel i7) can communicate with Toio robots through Bluetooth communication. 
One MacBook Pro can communicate and control up to 7 Toio robots simultaneously.
For the controlling algorithm, we adapt to the open-source library of the existing Toio-based research project~\cite{nakagaki2020hermits}~\footnote{\url{https://github.com/mitmedialab/HERMITS_UIST20}} and rewrite the algorithm for Node.js based on toio.js.

\subsection{AR Sketching Interface}
For AR Sketching, we implement the interface based on WebXR, using A-Frame, Three.js, and 8th Wall.
In A-Frame and Three.js, the HTML canvas can be embedded as an interactive 2D surface on top of the plane geometry.
We show this canvas texture onto a square-shaped plane geometry, which can match with the Toio tracking mat. 
To detect the touch and pen interaction, we use ray casting from the camera to get the intersection between the ray and the canvas plane geometry. 
Then, by calculating the position within the 2D canvas based on UV mapping of the touched point, we convert the 3D touch position into the 2D coordinate of the canvas. 
To dynamically render 2D shapes, we employ Konva.js for HTML Canvas drawing and manipulation and React.js for the JavaScript framework.
Konva.js only supports the rendering of the virtual shape, thus we also use Matter.js for the 2D physics engine.
By computing the position of each shape at each frame, we can animate the virtual sketched shape in Konva.js based on the physics simulation. 
In a similar manner, the system can also show the yellow object based on the robot's position.
For the basic shape recognition, we use the JavaScript version of \$1 Unistroke Recognizer, to detect the shape of a circle, rectangle, static constraint line, and spring.

\subsection{Server and Communication}
To synchronize between AR sketching and Toio robots, the system needs to communicate between the client-side web browser and Node.js server. 
To this end, the system uses the WebSocket protocol to 1) send the command to the next position of the Toio robot from the browser to the Node.js server, and 2) send the current position of each Toio robot from the Node.js server to the browser. 
In this way, the system allows synchronous communication between the AR interface and Toio control. 
The system runs a Node.js server on MacBook Pro (2021 16-inch Intel i7 CPU, 16GB RAM), which can communicate and control all of the robots.

\section{Applications}


The bi-directional interaction between AR sketch and actuated user interface opens up a broad set of applications, allowing the user to interact with digital information in improvisational, responsive, reality-embedded, and tangible ways. This section introduces multiple application areas and examples within each area to highlight the capability of Sketched Reality (Figure \ref{fig:applicatino-sketches}).

\subsection{Physics Education}
Firstly, Sketched Reality can be used for physics education. While physics is commonly difficult to be taught only with equations and textbooks, through Sketched Reality, the combination of tangible objects together with overlaid sketches that define the abstract properties between objects can be useful for children and students to tangibly learn the concepts. 

\begin{figure}[h]
\includegraphics[width=0.32\linewidth]{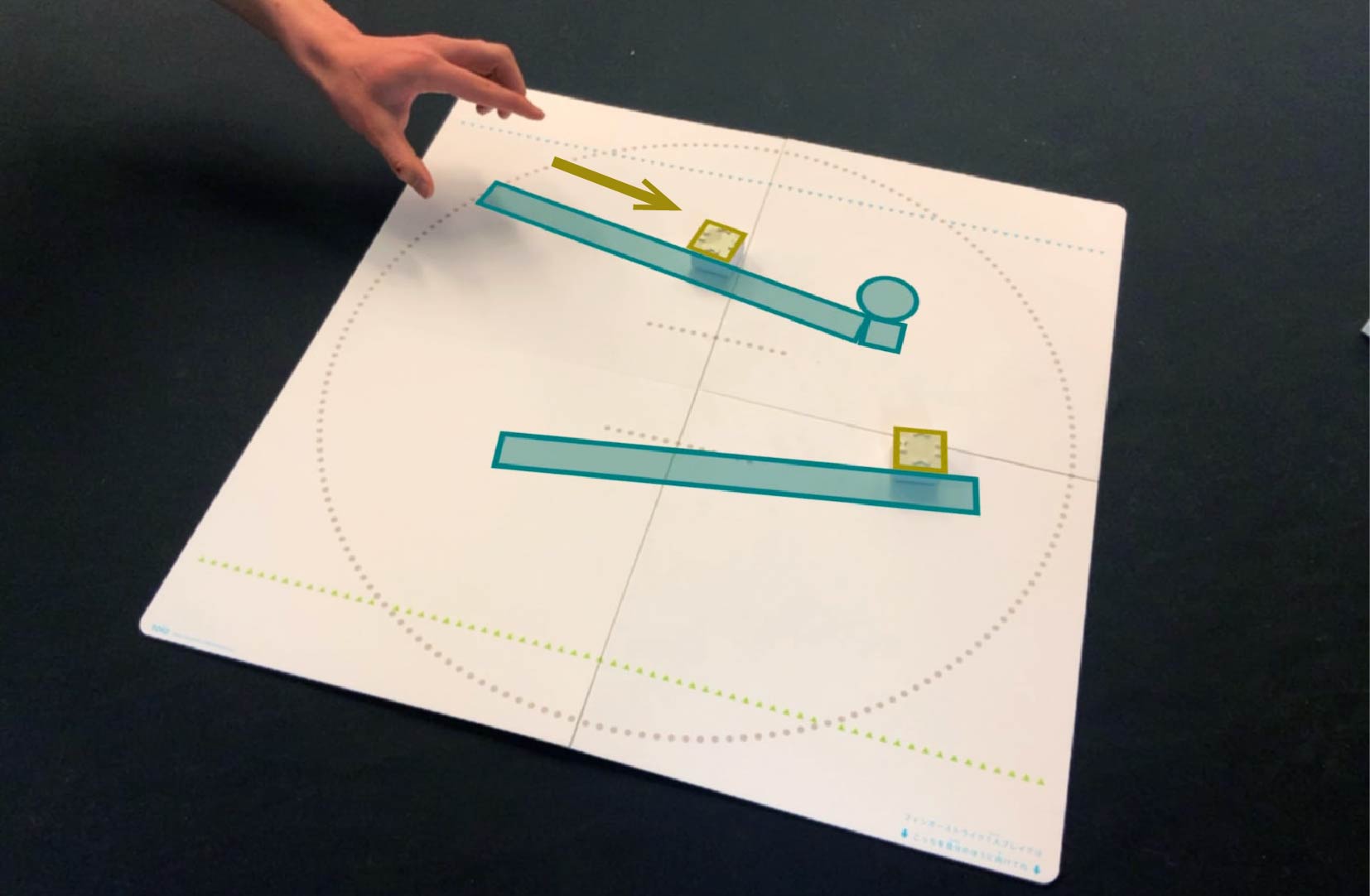}
\includegraphics[width=0.32\linewidth]{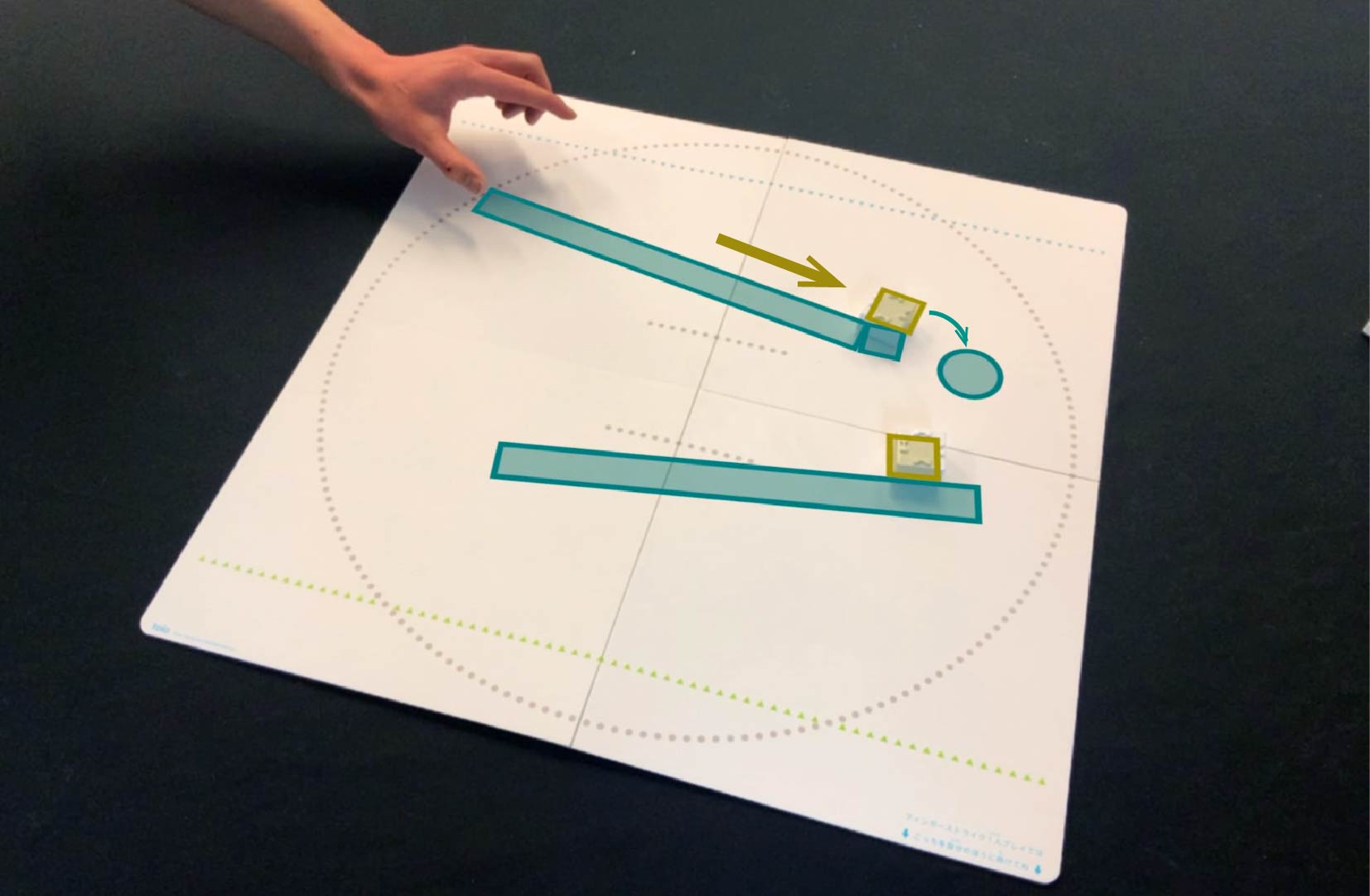}
\includegraphics[width=0.32\linewidth]{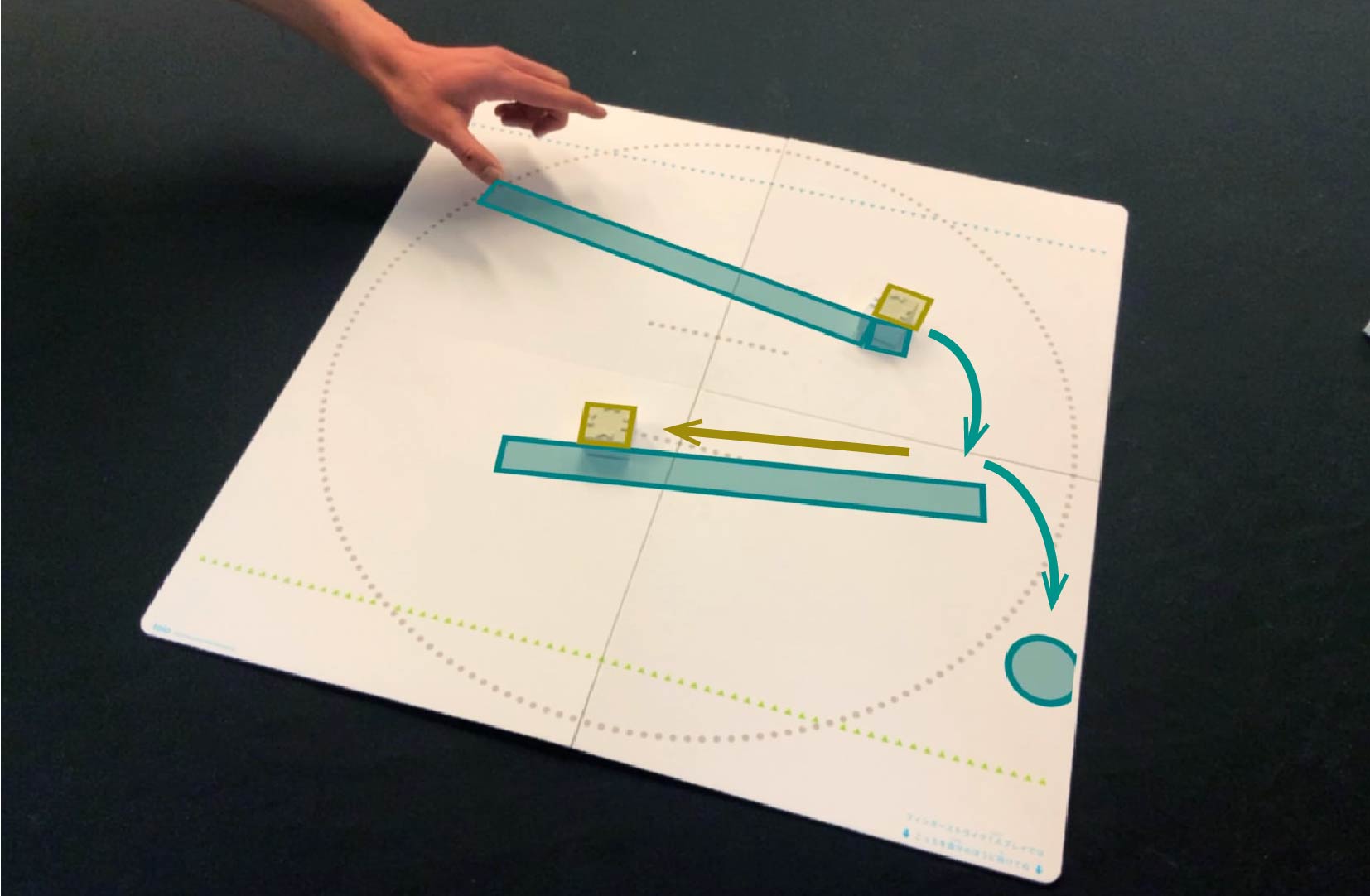}
\caption{Physics Education Application - Rube Goldberg Machine: a user releases a physical Toio robot from the above (left), then the physical robot, with virtual gravity, rolls down the virtual slope then hits a virtual ball (middle). The virtual ball, in turn, affects another physical Toio robot, that is triggered to move (right).}
\label{fig:rube_goldberg_machine}
\end{figure}

\begin{figure}[h]
\includegraphics[width=0.32\linewidth]{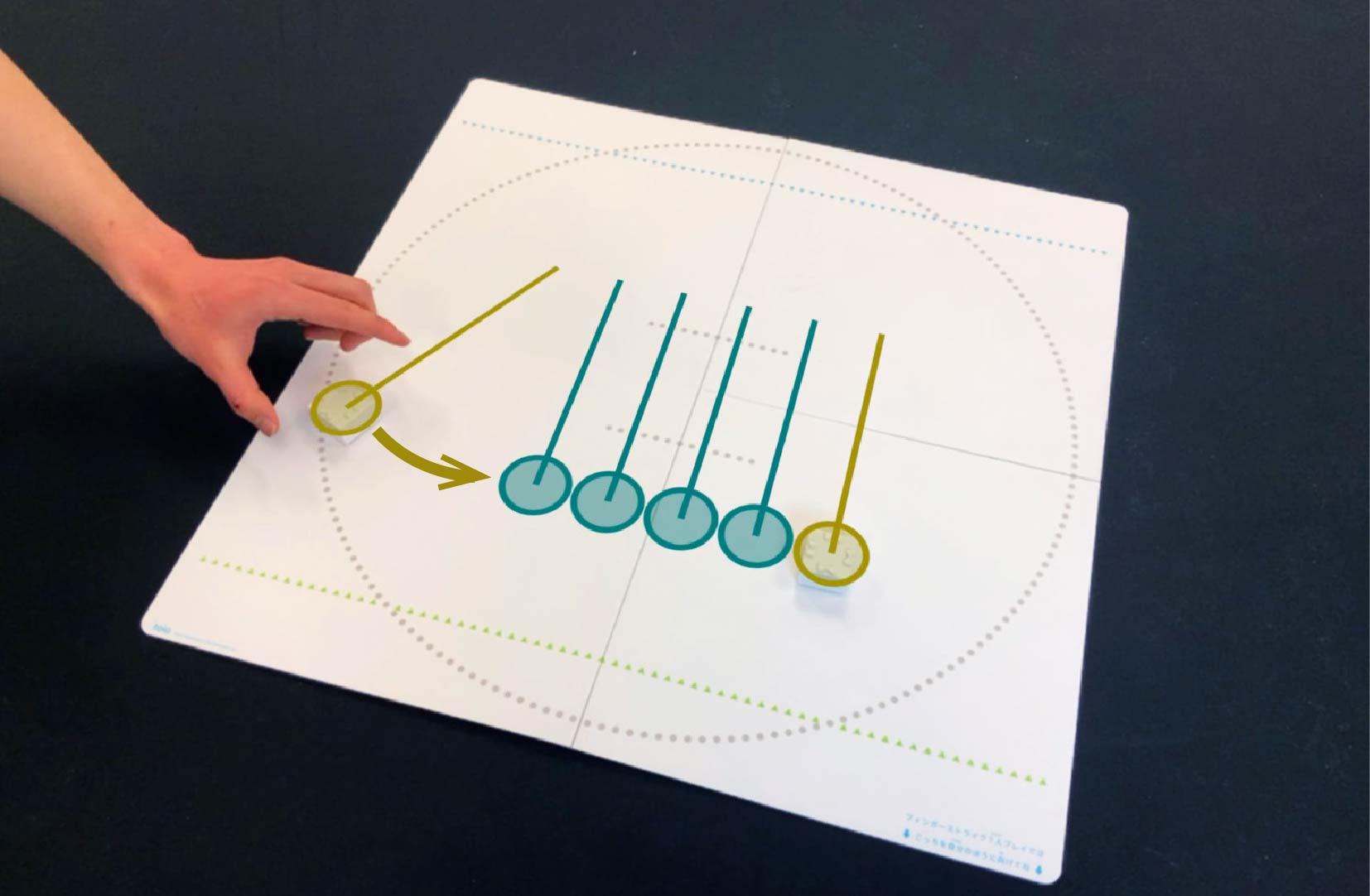}
\includegraphics[width=0.32\linewidth]{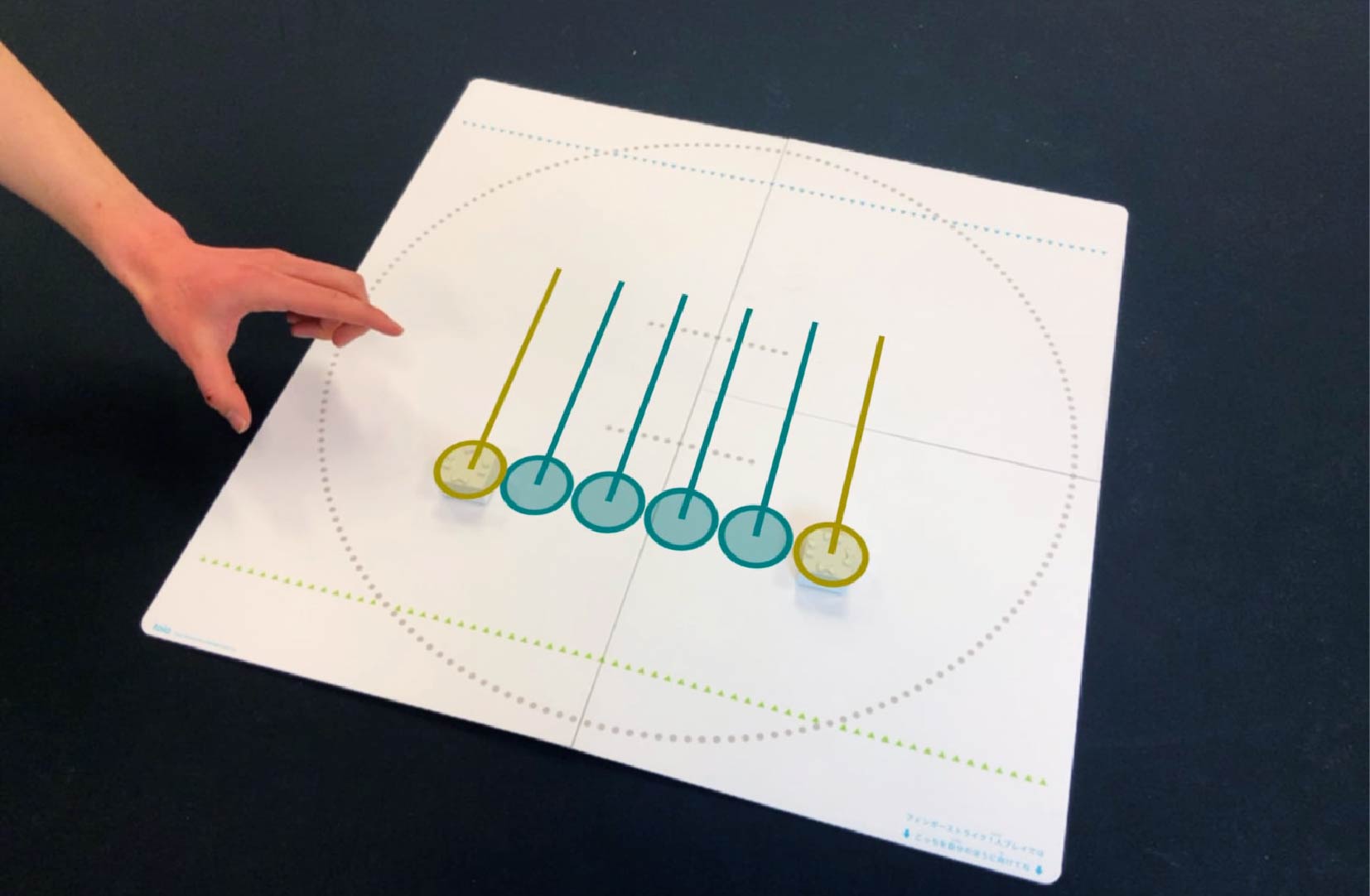}
\includegraphics[width=0.32\linewidth]{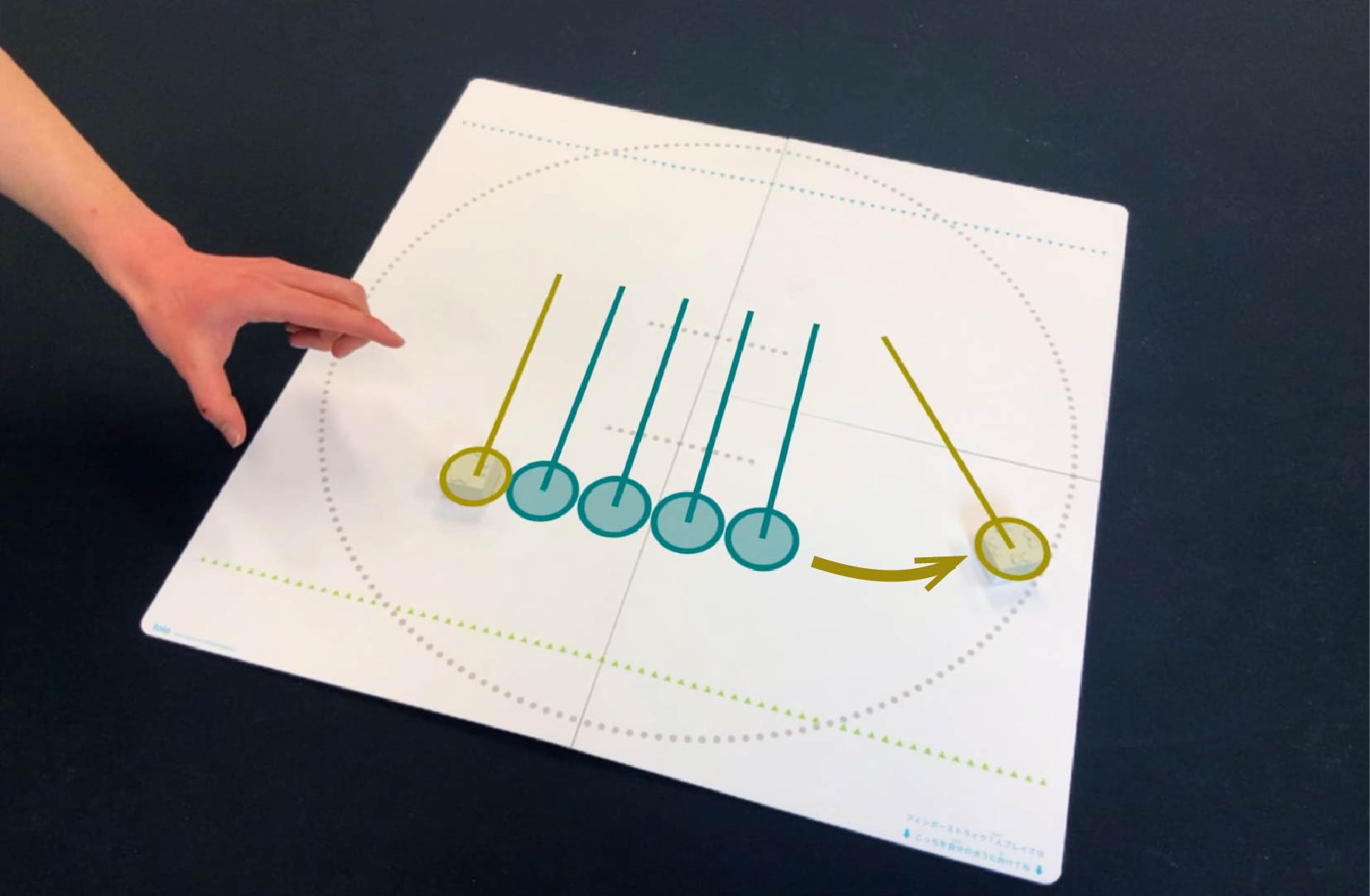}
\caption{Physics Education Application - Newton's Cradle: users can sketch a set of virtual strings and balls that reacts to a physical robot's movement triggered by users. Once a physical robot, representing a ball, is released by a user (left), the kinematic force is propagated through the virtual balls after the collision with the virtual ball (middle), then the motion is transmitted to another physical robot (right).}
\label{fig:newton's_cradle}
\end{figure}

As shown in Figure \ref{fig:rube_goldberg_machine} and \ref{fig:newton's_cradle}, users can sketch different objects and strings in AR that interacts with actuated TUIs. By constructing their own physics setups that can be interacted tangibly, users could learn basic physics knowledge.  In these setups, users can iteratively adjust different parameters of the physics sandbox through sketches.

The sketch can be done by students/learners, for them to exploratively and interactively learn physics behaviors, while it can also be done by teachers in classrooms for them to instantly construct physics simulations to describe and demonstrate different physics concepts intangible manner. Experiential and dynamic aspects give tangibility to users. Through actuation, the actuated TUIs can represent physics properties in a dynamic and haptic way -- e.g. users can feel the magnitude of gravity or mass via haptic feedback from the actuated tangibles \cite{kim2019swarmhaptics}.

\subsection{Explorable Mechanism}

\begin{figure}[h]
\includegraphics[width=0.32\linewidth]{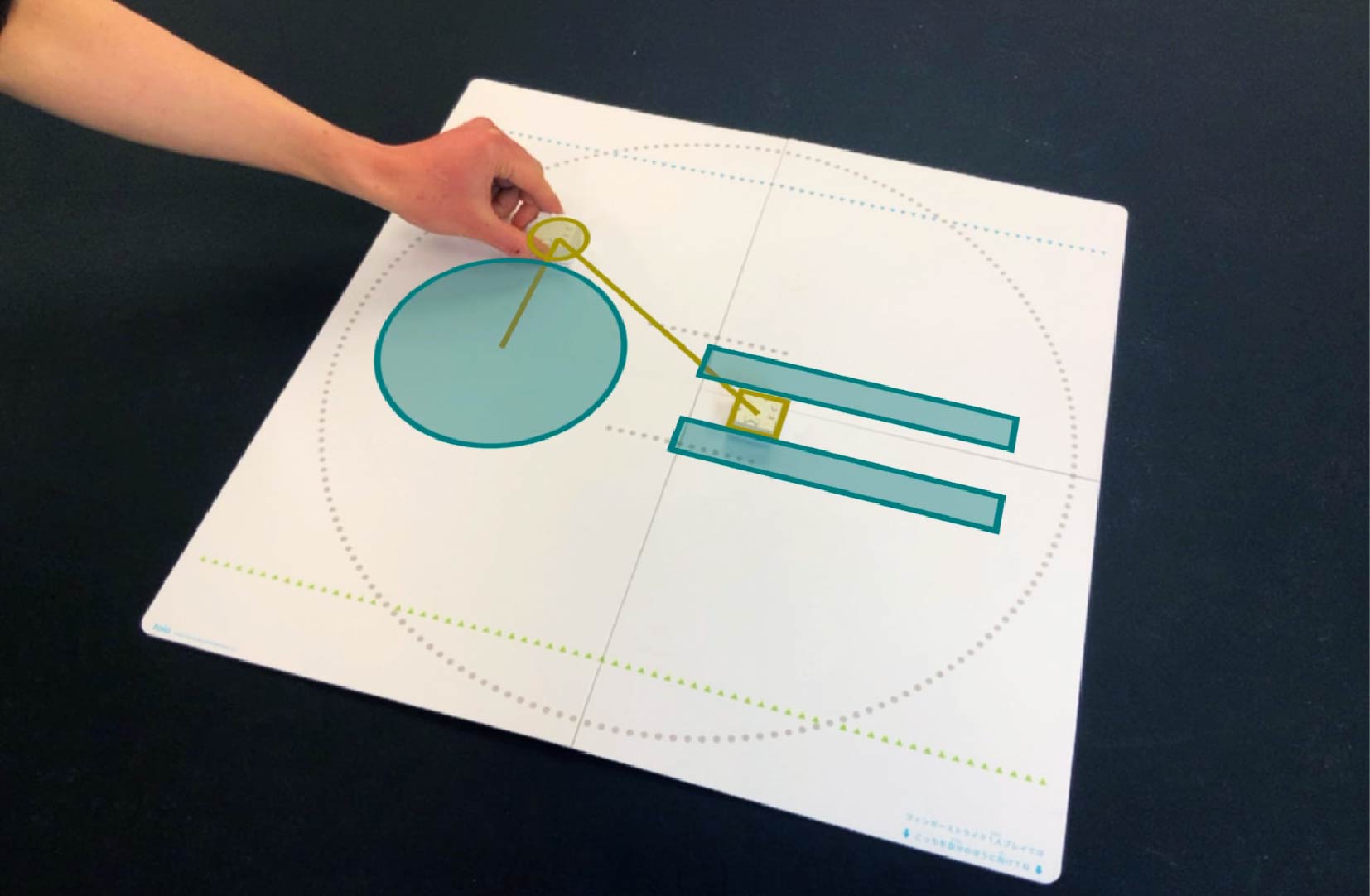}
\includegraphics[width=0.32\linewidth]{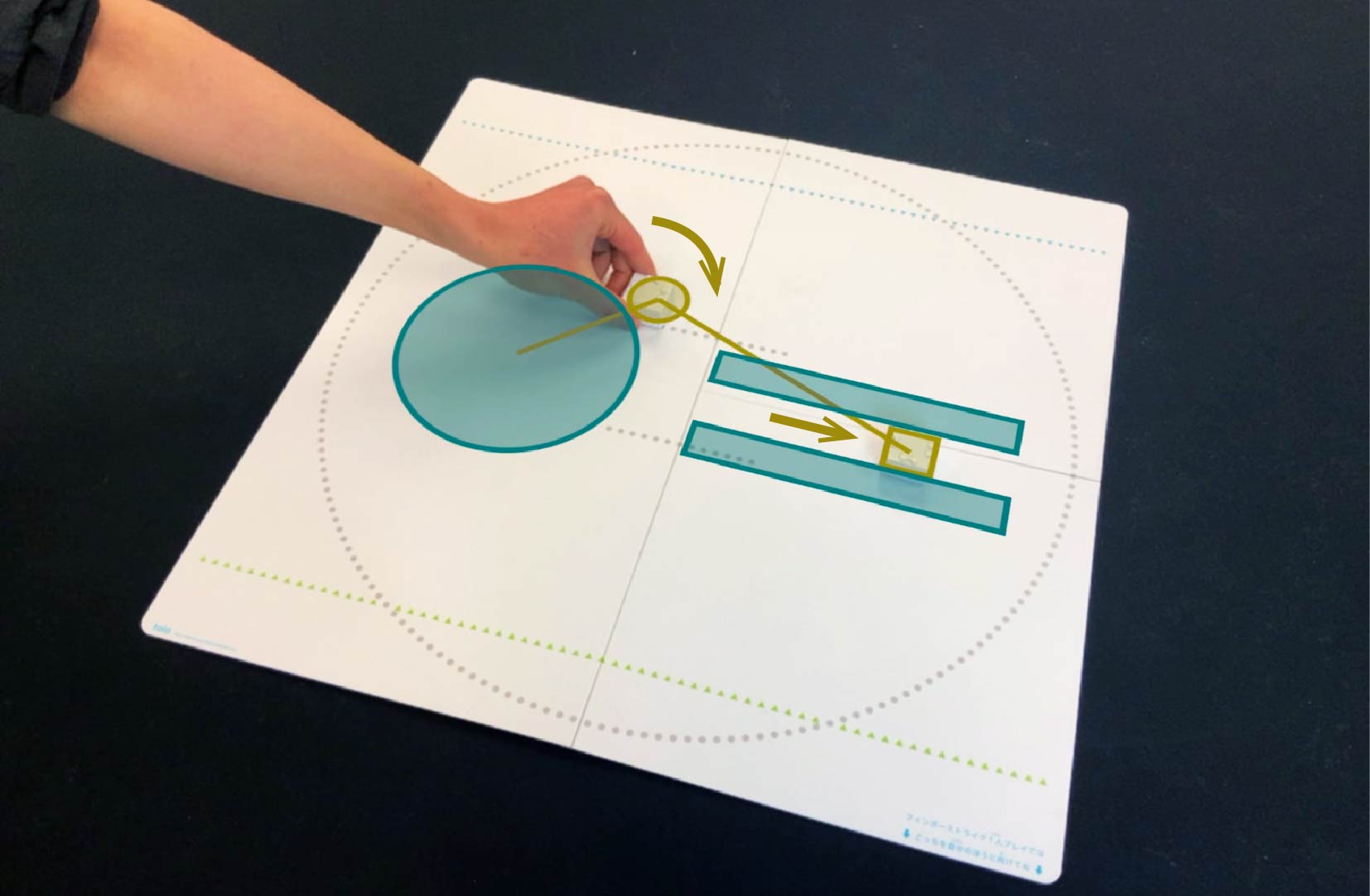}
\includegraphics[width=0.32\linewidth]{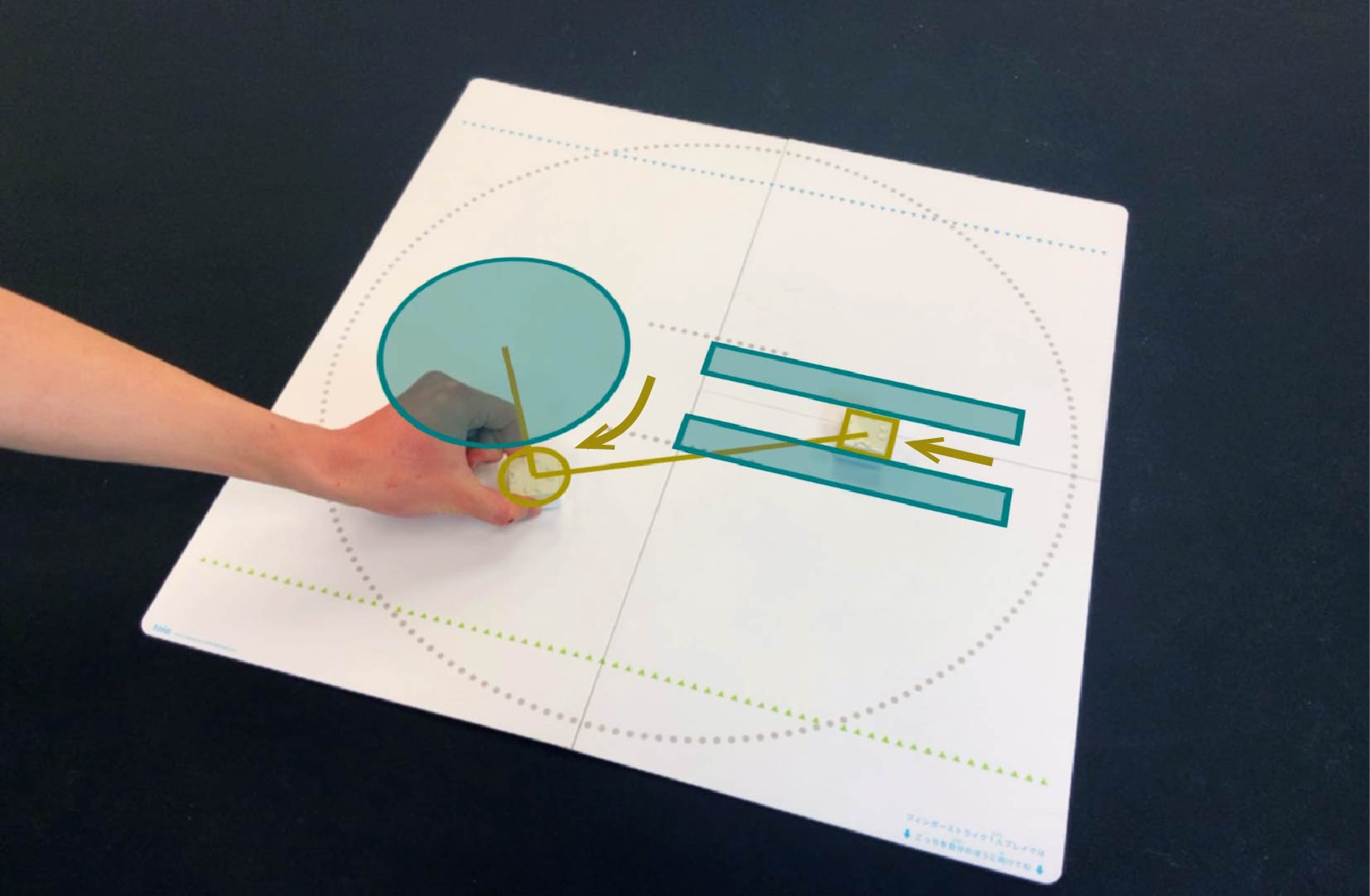}
\caption{Mechanical Exploration - Piston Mechanism: As a user sketches a set of virtual link mechanisms attached to physical robot bodies, he/she can manually move one of the kinetic elements to explore and simulate the designed kinematics. The result is employed through the other robot's physical motion.}
\label{fig:piston}
\end{figure}

Explorable Mechanism application demonstrates the use of Sketched Reality for mechanical design tasks. Similar to physics education, users can improvisationally design mechanisms where the virtual relationship is defined between actuated TUIs. Once interacted kinematic relationships are defined through AR sketches, users can grasp the Toio robots to control and examine the behavior of the mechanism. For example, as shown in Figure \ref{fig:piston}, a basic mechanism such as a piston can be verified through AR and actuated TUI, that dynamically responds to human input by grasping the actuated TUIs representing a handle of mechanism to control.

\change{Mechanism design tasks could be difficult to get a tangible sense to simulate and test in screen-based CAD tools without tangibility. For example, some research works like Mechanism Perfboard~\cite{jeong2018mechanism} argue that the use of AR simulation improves the trial and error in mechanism design.}
In Sketched Reality, mechanism design and exploration application incorporate advantages from these tools for users to quickly iterate with sketches and physically explore and test through tangible interactions.

\subsection{Tangible Gaming}

Tangible Gaming applications bring virtual and graphical information together with immersive haptic and tangible interaction via the robots for creating storytelling and entertainment. For example, similar to Angry Birds \cite{angrybirds}, users can pull the robots using a virtually sketched spring or sling-shot to aim at targets either virtual or physical (Figure \ref{fig:top}). 

\begin{figure}[h]
\includegraphics[width=0.32\linewidth]{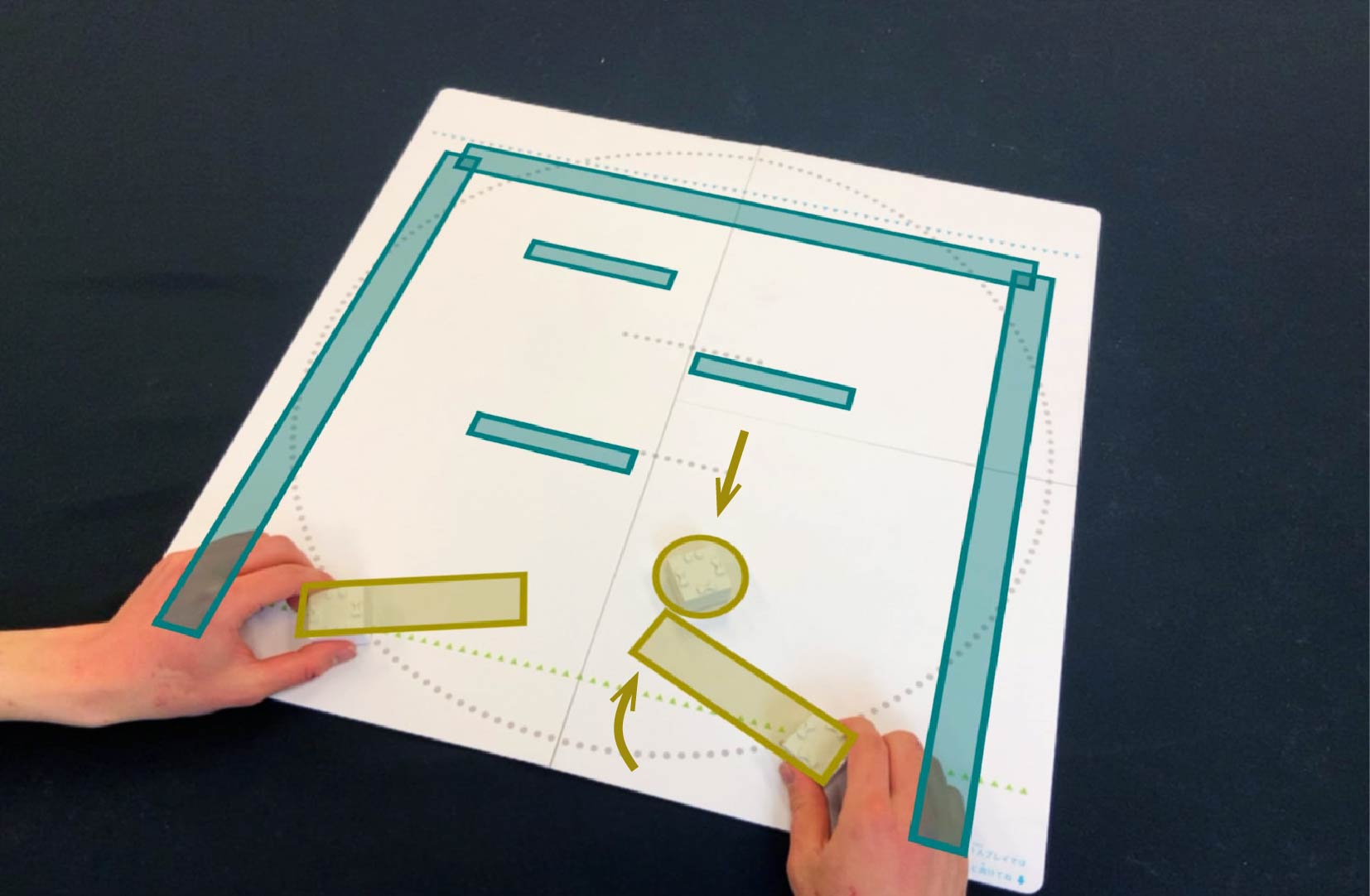}
\includegraphics[width=0.32\linewidth]{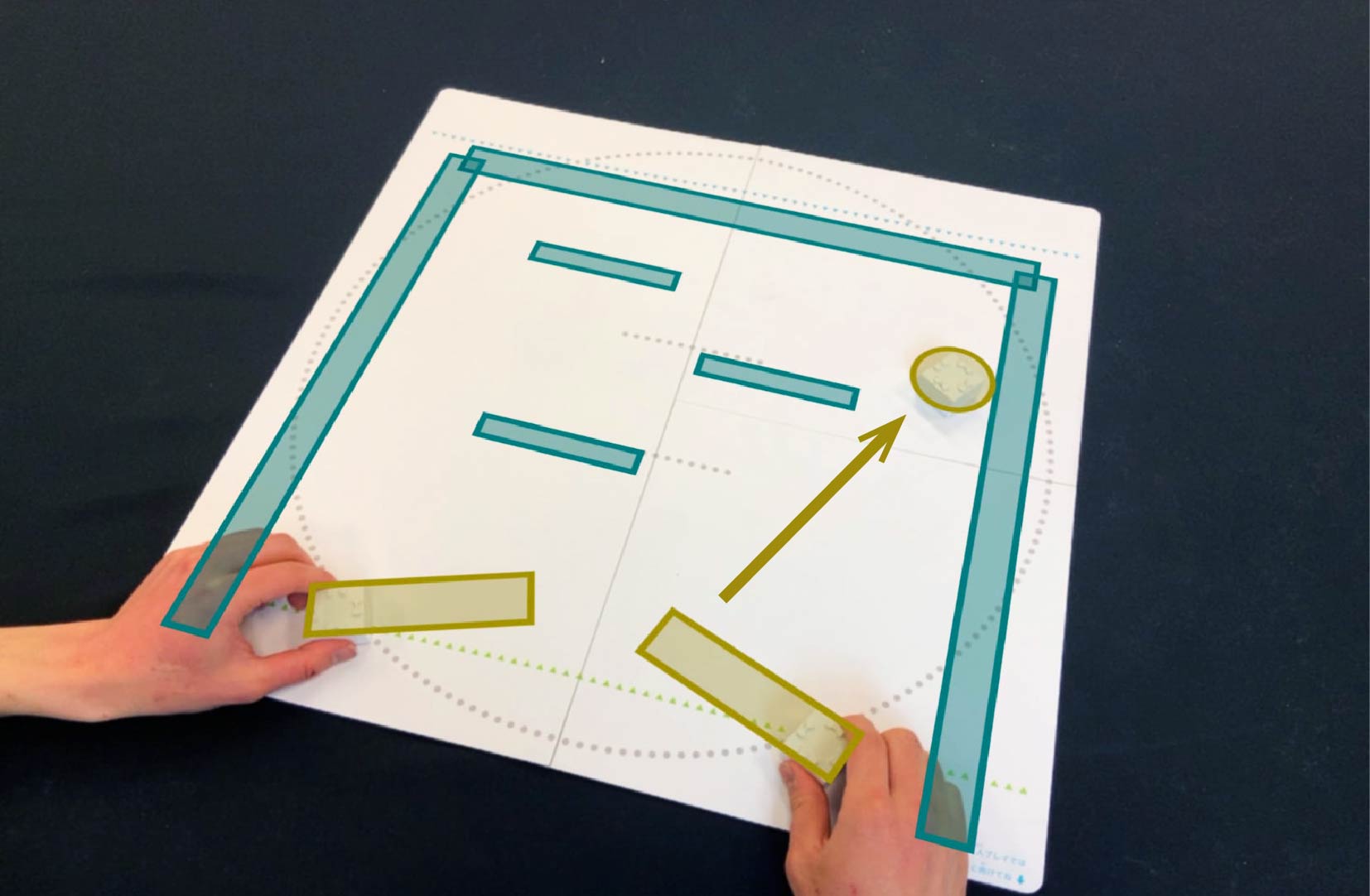}
\includegraphics[width=0.32\linewidth]{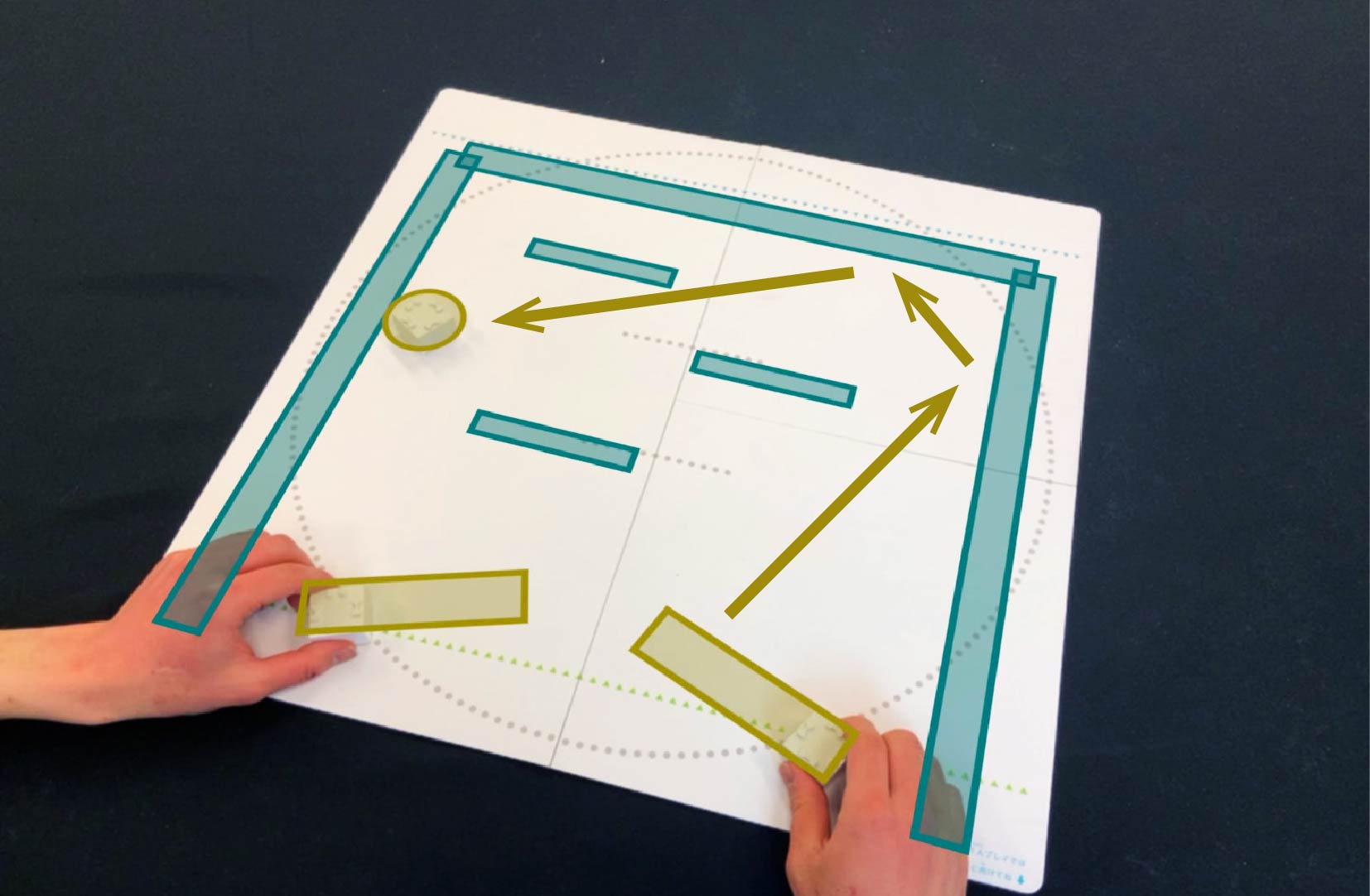}
\caption{Tangible Gaming - Pin Ball: With the combination of virtually sketched obstacles and physical robots, users can enjoy playing virtual pinball by manually moving flippers (represented by Toio robots) to hit a robot, representing a ball. Users can feel the force through the flipper robots.}
\label{fig:pin_ball}
\end{figure}

\begin{figure}[h]
\includegraphics[width=0.32\linewidth]{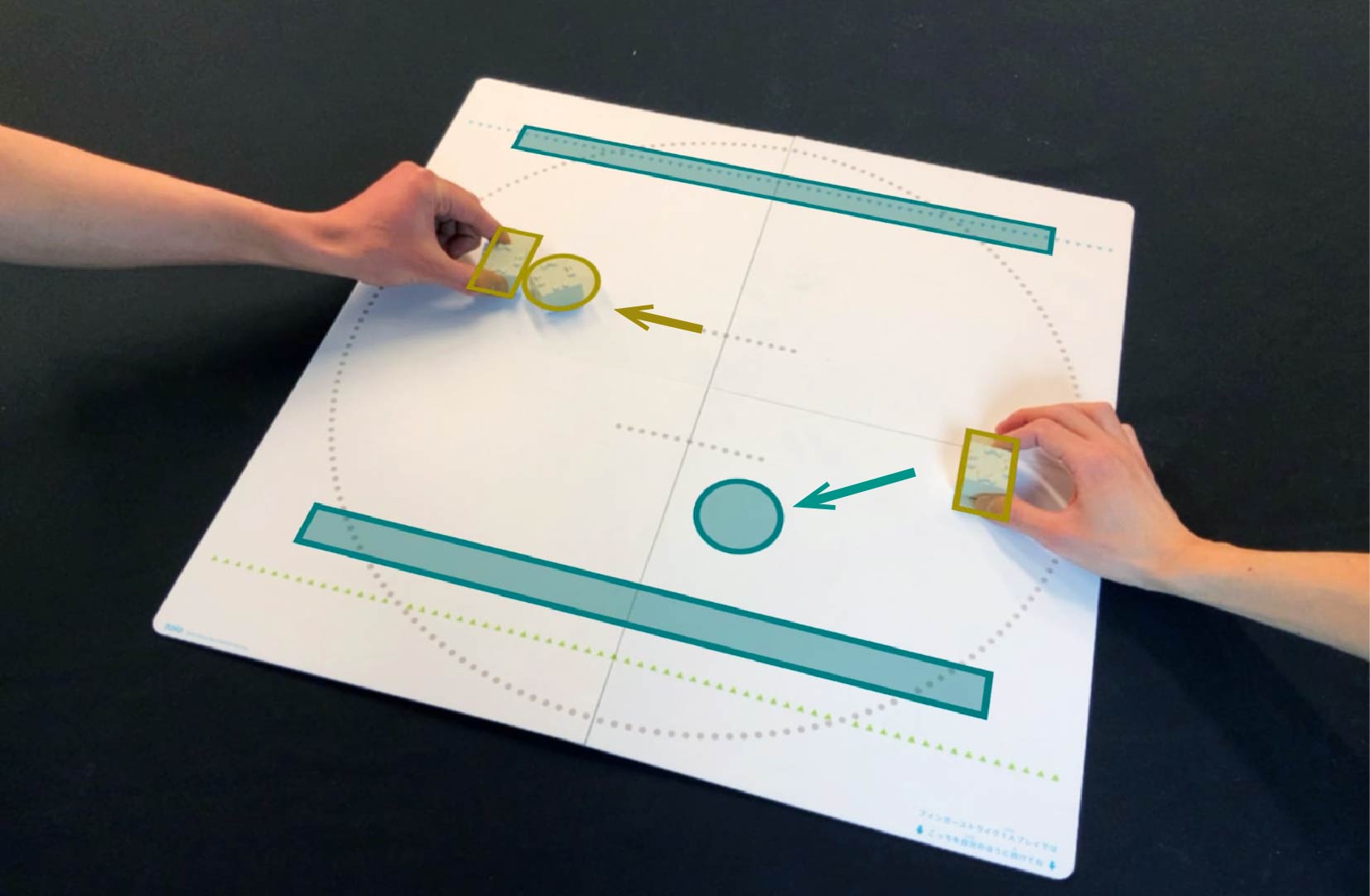}
\includegraphics[width=0.32\linewidth]{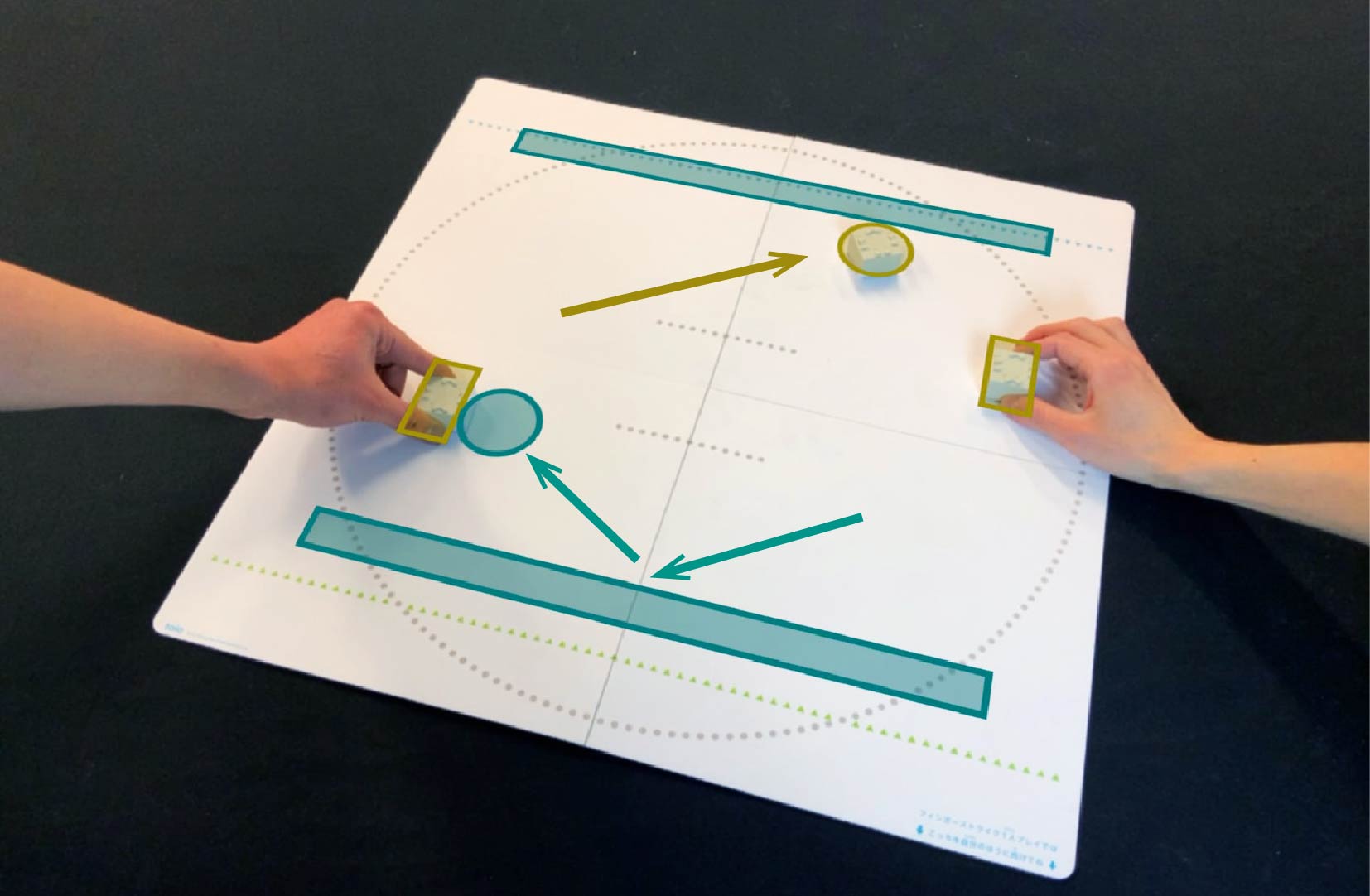}
\includegraphics[width=0.32\linewidth]{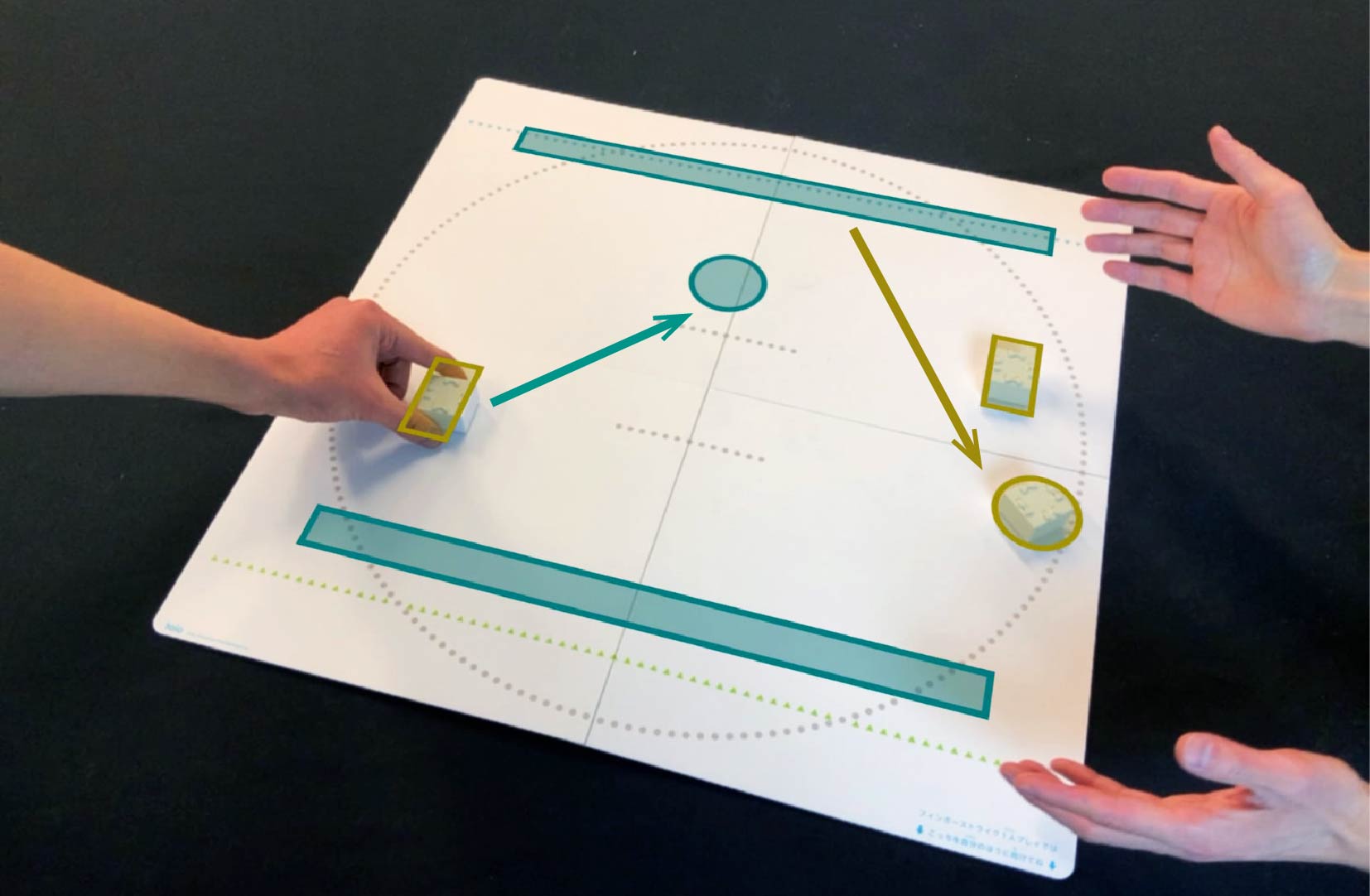}
\caption{Tangible Gaming - Pong: a pong game can have both a virtual ball and a physical ball for two competing players to hit and play. }
\label{fig:pong}
\end{figure}

Users can feel the spring-like haptic feedback as gradually pulling the robots on the spring by dynamically controlling the actuation of robots, then see the trajectory to hit targets after releasing. Users can sketch different obstacles and stage gimmicks for making interactivity and gaming. Figure \ref{fig:pin_ball} represents how AR sketching can be used for pinball gaming, allowing users to sketch obstacles and play the game with a physical controller and moving physical balls. A similar gaming experience can be created for pong (Figure \ref{fig:pong}), where virtual and physical balls are mixed together for advanced game mechanics.

\subsection{In-situ Programming}

\begin{figure}[h]
\includegraphics[width=0.32\linewidth]{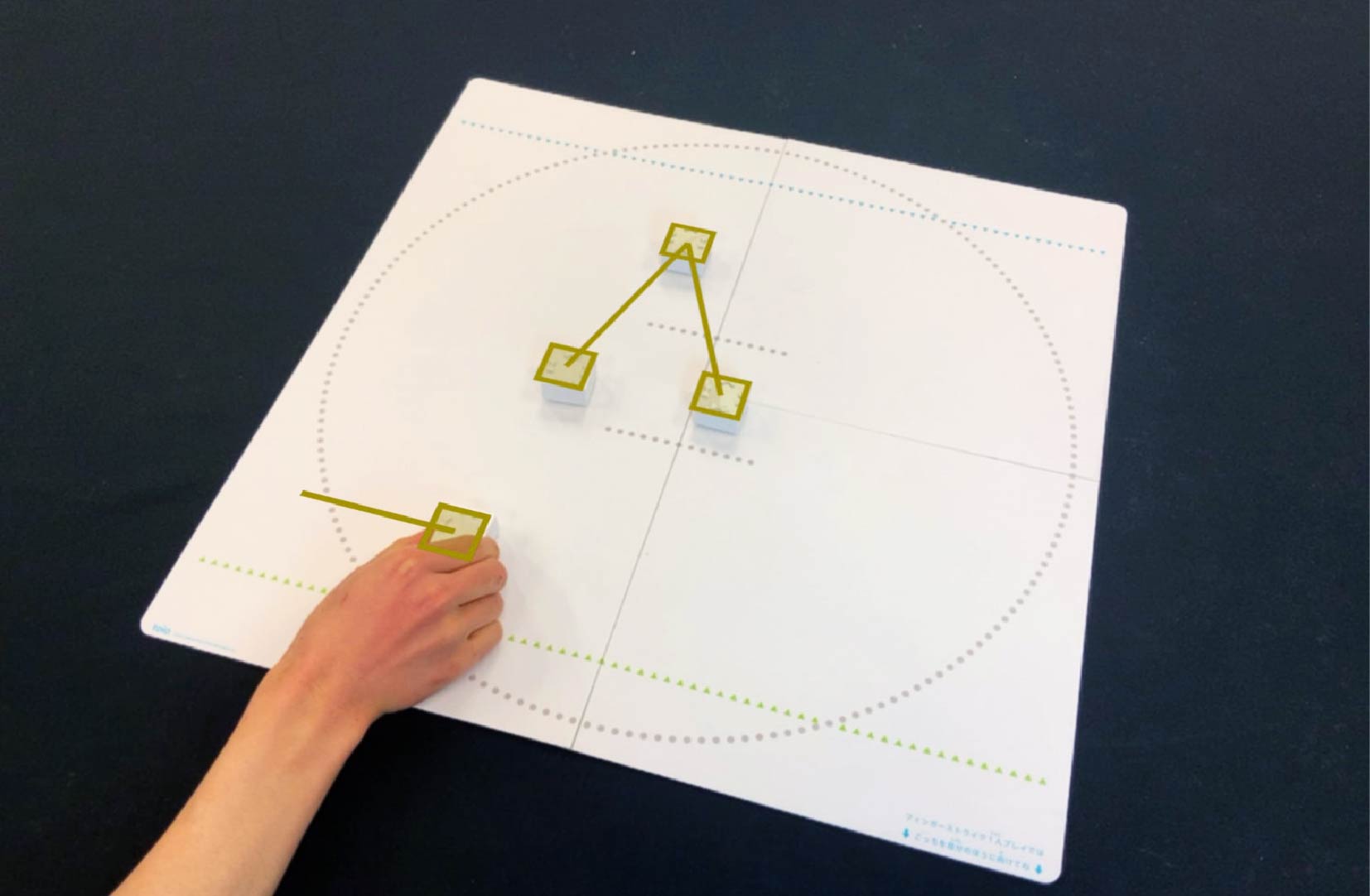}
\includegraphics[width=0.32\linewidth]{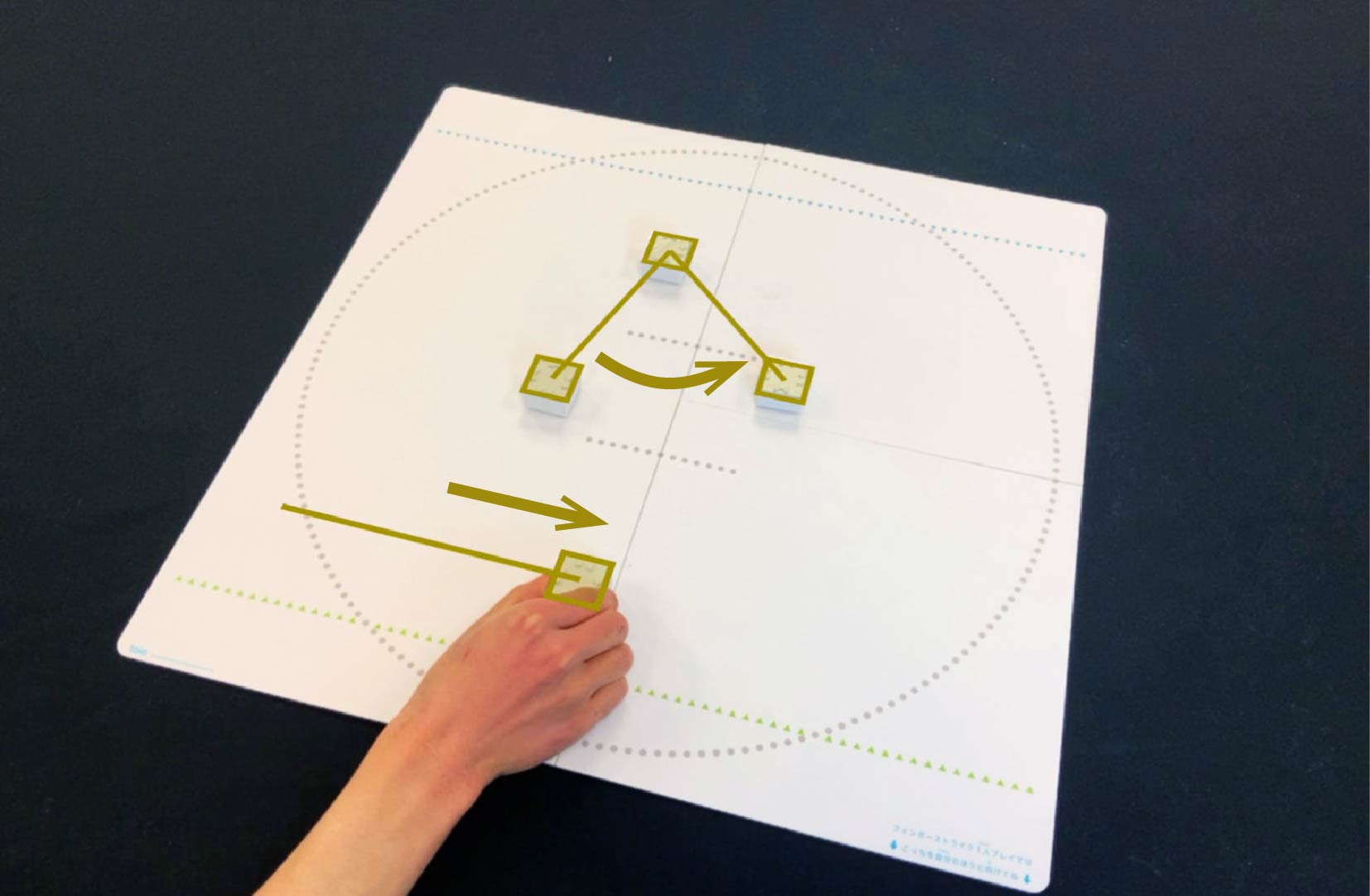}
\includegraphics[width=0.32\linewidth]{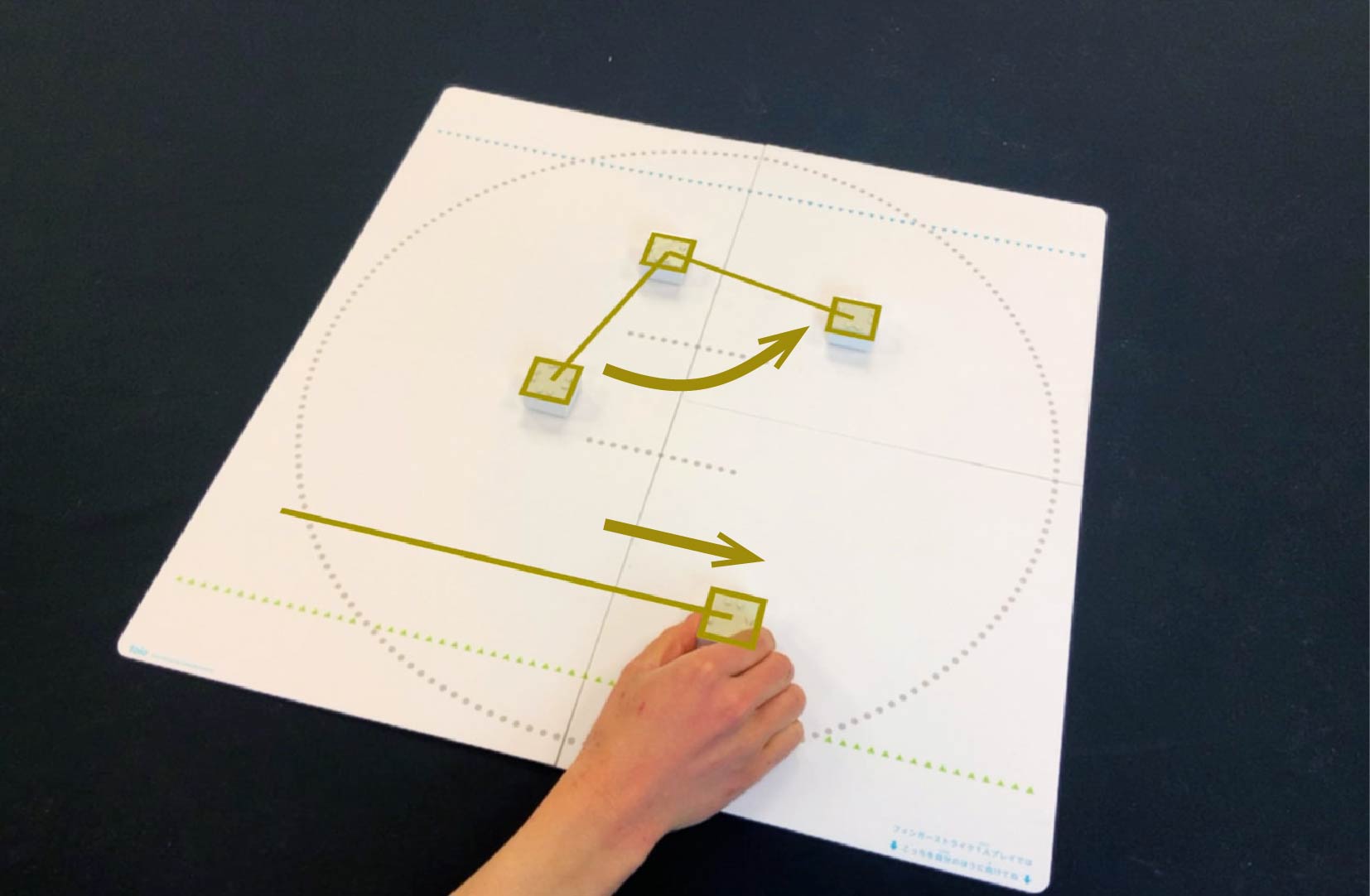}

\caption{In-situ Programming - Actuated TUI control: by drawing linkage lines between the three robots as well as an input slider for another robot, users can tangibly define the angular relationship between robots through the input slider.}
\label{fig:insitu}
\end{figure}

As AR interfaces and sketches have been used for defining and programming the behavior of IoT devices and robots \cite{heun2013reality, suzuki2022augmented}, Sketched Reality can be used to define the relationship of multiple robots as a way for users to program their behavior as everyday physical and actuated user interfaces. For example, as shown in Figure \ref{fig:insitu}, by sketching lines between robots with variable angles in-between with an input slider, users can define the relationship in-between these components to bridge the input and output via AR sketch. 

\begin{figure}[h]
\includegraphics[width=0.32\linewidth]{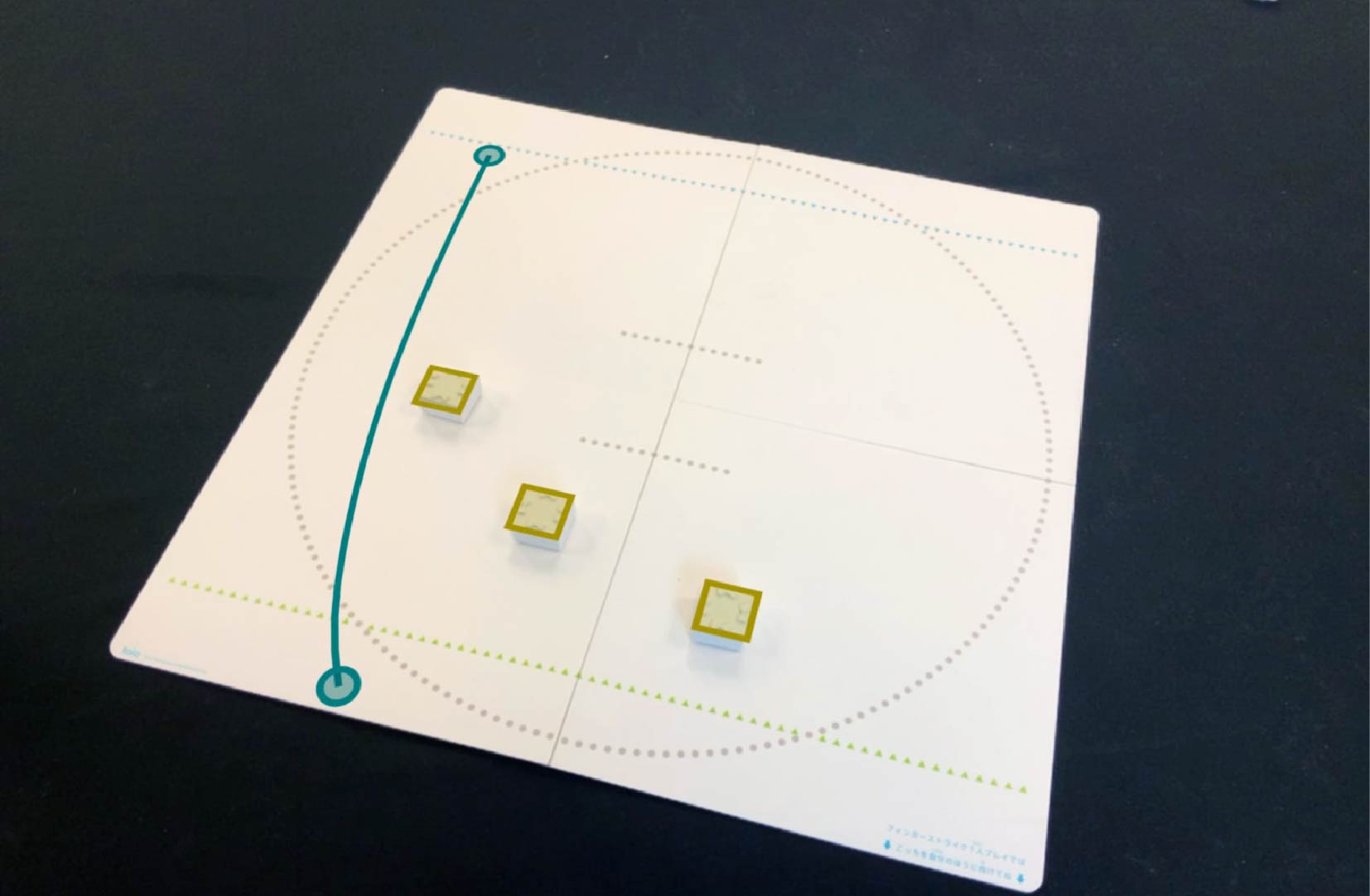}
\includegraphics[width=0.32\linewidth]{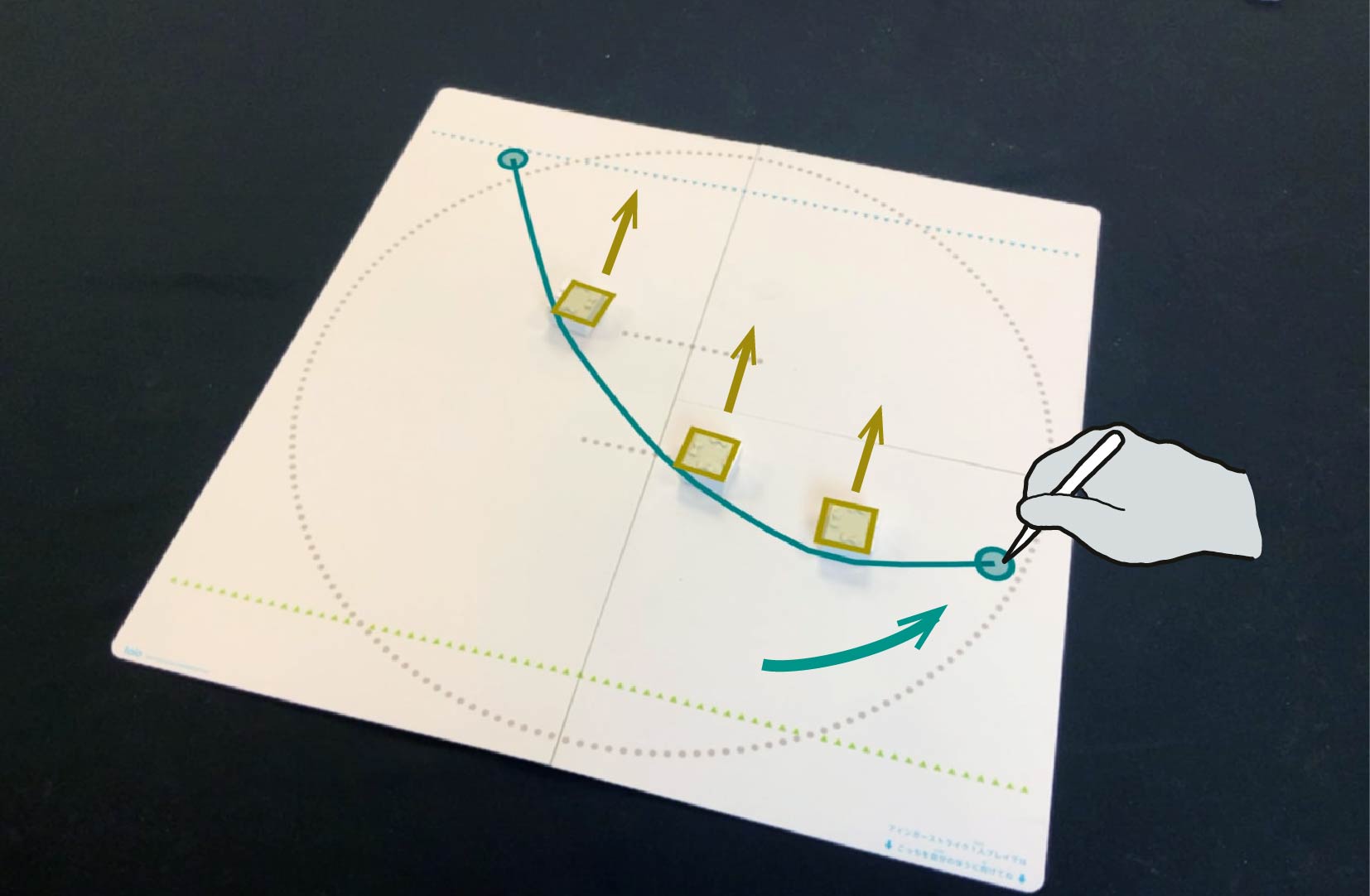}
\includegraphics[width=0.32\linewidth]{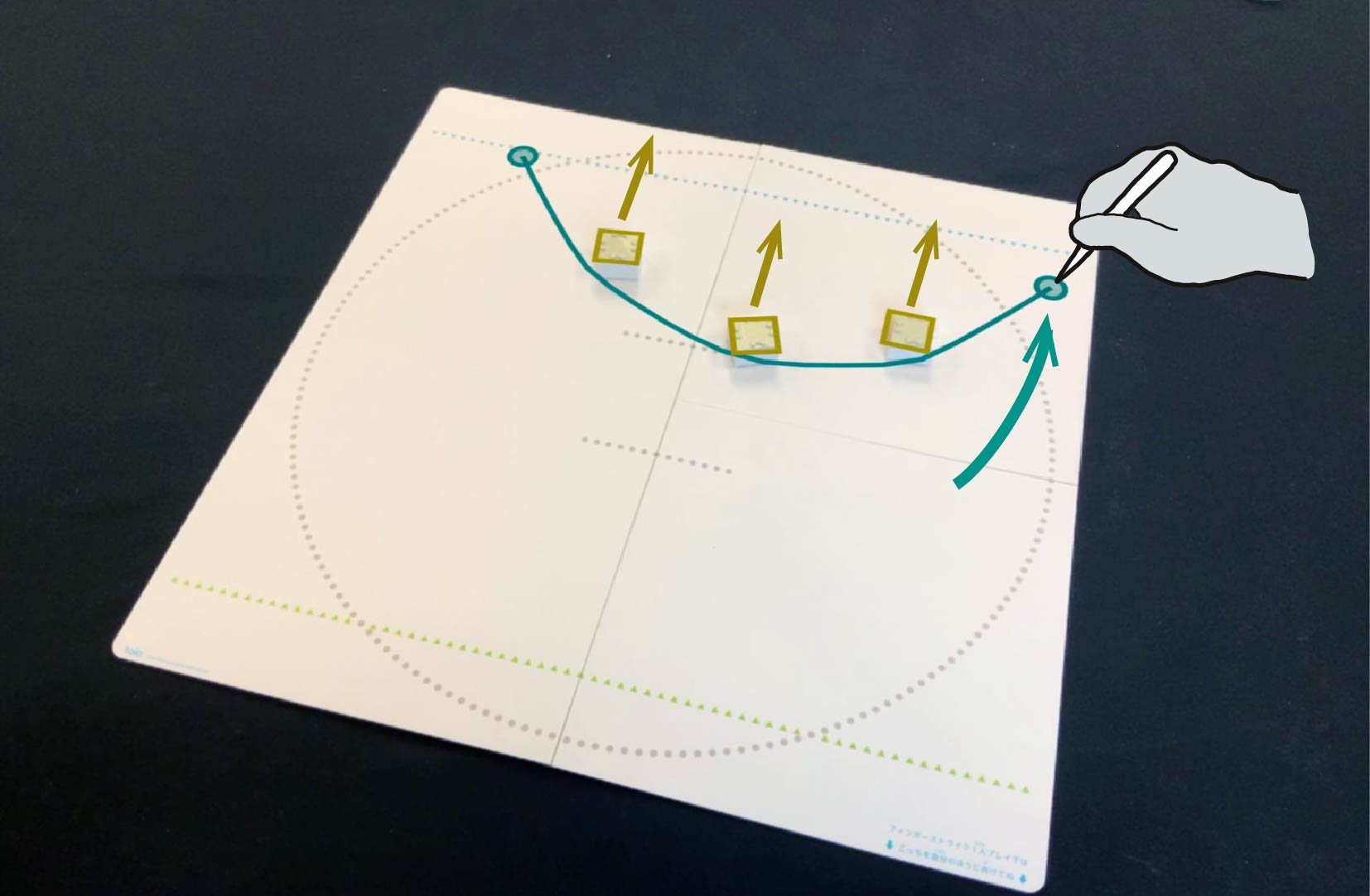}

\caption{In-situ Programming - Rope Control: With the combination of a virtually sketched rope and physical robots, users can control the movement of physical robots through the control of virtual rope action.}
\label{fig:insitu_rope}
\end{figure}

Such behavior can be dynamically used to develop bi-directional actuated interaction for users to define the relationship between actuated objects. In Figure \ref{fig:insitu_rope}, sketching allows users to create an in-situ virtual rope, that, after sketching, they can use this virtual rope to manipulate the behavior of multiple robots.

\section{Discussion and Future Work}


\subsection{Different Approaches for AR Sketching}
\change{While our implementation mostly focuses on tablet-based AR, we acknowledge that \textit{mobile AR} is just one of the possible approaches among many.
Based on AR and Robotics taxonomy~\cite{suzuki2022augmented}, there are three possible approaches: \textit{1) mobile AR, 2) head-mounted displays (HMDs), and 3) projection mapping}. In this section, we will discuss the pros and cons of each approach and give a holistic comparison of each of them.}

\subsubsection{Mobile AR}
\change{As we showed, tablet-based AR sketching tools allow an \textbf{easy setup} and implementation in a mobile setting. In addition, it allows using a pen or finger for tactile feedback while drawing. However, the biggest limitation of this approach is \textbf{non-hands-free interaction}. For example, in our setup, we used a tripod to overcome this issue, but this may not always be optimal. The use of a smartphone could alleviate this problem, but the screen size is limited.} 

\subsubsection{HMDs}
\change{On the other hand, head-mounted displays (HMDs), like Hololens, allow \textbf{hands-free interaction}. This approach can address the current limitation, as the user does not need to hold the tablet when interacting with the virtual sketches and physical TUIs.
In contrast, however, \textbf{precise sketching interaction} through finger and gestural interaction in HMD is still a challenge. In particular, the \textbf{lack of tactile feedback} of mid-air gestures makes the precise sketching interactions much harder~\cite{arora2017experimental}.
Therefore, we should consider the sketching interface for touch interaction like MRTouch~\cite{xiao2018mrtouch}.
In addition, it is not trivial to share the experience with multiple users.}

\subsubsection{Projection Mapping}
\change{A projector allows the \textbf{collaborative multi-user experience}. This approach also enables hands-free interaction for multiple users, which is especially appropriate for educational use cases. However, The setup is not mobile and always requires a \textbf{tedious calibration process}. Moreover, in this way, the interaction and implementation can become more complex. For example, the system needs to distinguish between sketching, menu selection, virtual object dragging, and physical object dragging, which is not trivial and often requires another tracking mechanism for sketching.}
 
\subsubsection{Combination}
\change{Alternatively, an exciting future direction could be to combine multiple approaches. This allows us to leverage each benefit and overcome limitations of each approach. 
For example, previous research shows the benefits and advantages of using \textbf{Mobile AR + HMD} (e.g., BISHARE~\cite{zhu2020bishare}, SymbiosisSketch~\cite{arora:2018:symbiosissketch}) or \textbf{HMD + Projector} (e.g., ShareVR~\cite{gugenheimer2017sharevr}, AAR~\cite{hartmann2020aar}).
In future work, we should also explore different approaches for AR sketching and investigate how each approach would benefit the user interaction and experiences.}

\subsection{Benefits of Bi-Directionality}
\change{In this paper, we mostly focused on the exploration and demonstration of the Sketched Reality concept, and we did not formally evaluate the benefits of bi-directionality of AR and actuated TUIs. 
However, we believe AR sketches and actuated TUIs can benefit each other in many ways. Therefore, we would like to outline such benefits in this section.}

\subsubsection{How AR Sketching Benefits from Actuated TUIs}
\change{Actuated TUIs allow a number of benefits that cannot be done solely on AR. For instance, consider that in situations like Tangible Gaming, users can \textbf{feel the elastic force} of a virtual spring while pulling the robot (Figure~\ref{fig:top}). Also, in situations like exploration of mechanical linkage examples (Figure~\ref{fig:piston}), the direct tangible manipulation allows the user to \textbf{feel the constraint} of the virtual object. In addition, if it enables the multi-user collaboration in a classroom, each user can interact with each other through virtually \textbf{inter-connected physical objects}. Such an experience can never be achieved solely with AR sketching.}

\subsubsection{How Actuated TUIs Benefit from AR Sketching}
\change{On the other hand, AR sketching allows the instantaneous creation of virtual objects, which may greatly increase the expressiveness of the actuated TUIs. For example, AR allows the scalable and flexible \textbf{object instantiation} (like virtual rope), which cannot be done with physical objects.
The user can easily change, scale, and modify such a virtual object beyond the traditional physical constraints ~\cite{patten2007mechanical}.}

\subsubsection{Beyond the Current Implementation}
\change{Due to the current limitation of mobile AR, such benefits may not be entirely clear as the experience seemingly takes place in the virtual world, not in the physical world. This is mostly because the mobile AR approach cannot fully blend the virtual and physical worlds, as it enforces the user to watch the tablet while their hands interact with the tangible. However, we envision the near future where everyone has an AR headset and the entire physical world becomes the interactive canvas. In such a fully blended reality world, we expect these benefits would become more interesting and clearer. We are interested in how our concept can be implemented in such a fully blended world in the future.}

\subsection{Exploration of Different Hardware}
\change{The concepts mentioned in this paper are not limited to mobile tabletop robots, but also can be expanded to different hardware. 
For example, larger-size robots like robotic vacuum cleaners or robot dogs could also be controlled through virtual sketched objects.
Alternatively, we are also interested in further applying our concept to other actuated TUIs or shape-changing user interfaces.
For example, the bi-directional interaction between AR and shape displays could be an interesting future exploration. 
Such interactions are partially explored in inFORM~\cite{follmer2013inform}, but we believe there should be a larger design space for bi-directional virtual-physical interactions. 
In addition, we are also interested in exploring bi-directional interactions between virtual sketches and IoT devices. 
For example, the user could turn off or change the color of light bulb with sketched AR objects.
By leveraging IoT and other robots, we could bring such an interaction to the everyday environment.
We believe the ground concept and design space proposed in this paper help the future exploration in the future.}

\subsection{Advanced Sketch and Control Properties}
In this paper, we have demonstrated the concept of Sketched Reality through preliminary sketching primitives (e.g., lines, blocks, spheres, and springs) with basic behaviors (e.g., bouncing, colliding, and linkage motions). A future implementation should incorporate more complex shapes, behaviors, and properties for users to flexibly sketch different elements and properties such as mechanical gears, friction/adhesion properties, or elastic surfaces. Increasing the library of sketch shapes and properties would enrich the versatility and adaptability of each application, for users to explore, improvise and touch AR sketches.

\subsection{Scalability of Robots, Users, and Interaction Area}
The scalability of robots, users, and interaction areas is obvious limitations in our prototype that can be further explored in the next steps. While a variety of user interaction methods with swarm robots have been proposed~\cite{kim2020user}, we believe our proposed sketching interaction will further allow users to manipulate and interact with tens and hundreds of robots with in-situ sketches. Additionally, as sketching interactions are commonly introduced for multi-user interaction setup for collaborative and cooperative sketching/drawing, such a direction should be further explored in the future physical space, where a number of robots and a number of people exist. Additionally, while the current sketching area is limited to the mat of Toio, we plan to extend the interaction area much larger to, for example, our entire living space for full-spatial interactivity taking advantage of explorable AR, combined with embedded actuation in the real world.

\subsection{User Study and Evaluation}
While our paper primarily focused on the basic concept and prototype, we hope to further explore and evaluate our system with user studies. Such studies could compare the immersion, control efficiency, as well as enjoyment of user interaction with AR systems with and without our system. Application-specific evaluations could be explored further to understand the effect of our system. 
\change{For example, for learning abstract physics or kinematics concepts, previous research like Mechanism Perfboard~\cite{jeong2018mechanism} or HoloBoard~\cite{gong2021holoboard} provides evidence of the benefits of blending virtual and physical interactions for education. 
We believe the bi-directional sketching interaction proposed in this paper can also provide similar benefits in education and entertainment.}
As sketching interaction gives full freedom for people to control and design interaction with actuated TUIs, such evaluations should further suggest novel design space and functionality for bi-directional sketching interaction.


\section{Conclusion}

In this study, we proposed \textit{Sketched Reality}, a concept of bi-directional virtual-physical interaction between AR sketches and actuated tangible user interfaces. Our design space categorized such a general interaction into four different categories, and we demonstrated this design space with a proof-of-concept prototype using tabletop mobile robots and an iPad-based AR sketching tool. With our implemented prototype, which combines AR, sketching interface, robot control, and primitive physics simulation, we have introduced a set of interaction techniques and demonstrated several applications, including tangible physics education for children, explorable mechanism, tangible gaming, and in-situ robot programming and actuated TUIs. We have discussed limitations and future work to highlight broader challenges to enable virtual-physical bi-directional interaction for the future physical environments in which digital computation and actuated robots are integrated.





\begin{acks}
This research was funded in part by the Natural Sciences and Engineering Research Council of Canada (NSERC) and Mitacs Globalink Research Award.
\end{acks}

\ifdouble
  \balance
\fi
\bibliographystyle{ACM-Reference-Format}
\bibliography{references}


\begin{thebibliography}{58}


\ifx \showCODEN    \undefined \def \showCODEN     #1{\unskip}     \fi
\ifx \showDOI      \undefined \def \showDOI       #1{#1}\fi
\ifx \showISBNx    \undefined \def \showISBNx     #1{\unskip}     \fi
\ifx \showISBNxiii \undefined \def \showISBNxiii  #1{\unskip}     \fi
\ifx \showISSN     \undefined \def \showISSN      #1{\unskip}     \fi
\ifx \showLCCN     \undefined \def \showLCCN      #1{\unskip}     \fi
\ifx \shownote     \undefined \def \shownote      #1{#1}          \fi
\ifx \showarticletitle \undefined \def \showarticletitle #1{#1}   \fi
\ifx \showURL      \undefined \def \showURL       {\relax}        \fi
\providecommand\bibfield[2]{#2}
\providecommand\bibinfo[2]{#2}
\providecommand\natexlab[1]{#1}
\providecommand\showeprint[2][]{arXiv:#2}

\bibitem[\protect\citeauthoryear{8th Wall~Inc.}{8th Wall~Inc.}{2022}]%
        {8th-wall}
\bibfield{author}{\bibinfo{person}{8th Wall~Inc.}}
  \bibinfo{year}{2022}\natexlab{}.
\newblock \bibinfo{title}{8th Wall}.
\newblock
\newblock
\urldef\tempurl%
\url{https://www.8thwall.com/}
\showURL{%
\tempurl}


\bibitem[\protect\citeauthoryear{Alexander, Roudaut, Steimle, Hornb{\ae}k,
  Bruns~Alonso, Follmer, and Merritt}{Alexander et~al\mbox{.}}{2018}]%
        {alexander2018grand}
\bibfield{author}{\bibinfo{person}{Jason Alexander}, \bibinfo{person}{Anne
  Roudaut}, \bibinfo{person}{J{\"u}rgen Steimle}, \bibinfo{person}{Kasper
  Hornb{\ae}k}, \bibinfo{person}{Miguel Bruns~Alonso}, \bibinfo{person}{Sean
  Follmer}, {and} \bibinfo{person}{Timothy Merritt}.}
  \bibinfo{year}{2018}\natexlab{}.
\newblock \showarticletitle{Grand challenges in shape-changing interface
  research}. In \bibinfo{booktitle}{\emph{Proceedings of the 2018 CHI
  conference on human factors in computing systems}}. \bibinfo{pages}{1--14}.
\newblock


\bibitem[\protect\citeauthoryear{Aoki, Matsushita, Iio, Mitake, Toyama,
  Hasegawa, Ayukawa, Ichikawa, Sato, Kuriyama, et~al\mbox{.}}{Aoki
  et~al\mbox{.}}{2005}]%
        {aoki2005kobito}
\bibfield{author}{\bibinfo{person}{Takafumi Aoki}, \bibinfo{person}{Takashi
  Matsushita}, \bibinfo{person}{Yuichiro Iio}, \bibinfo{person}{Hironori
  Mitake}, \bibinfo{person}{Takashi Toyama}, \bibinfo{person}{Shoichi
  Hasegawa}, \bibinfo{person}{Rikiya Ayukawa}, \bibinfo{person}{Hiroshi
  Ichikawa}, \bibinfo{person}{Makoto Sato}, \bibinfo{person}{Takatsugu
  Kuriyama}, {et~al\mbox{.}}} \bibinfo{year}{2005}\natexlab{}.
\newblock \showarticletitle{Kobito: virtual brownies}.
\newblock In \bibinfo{booktitle}{\emph{ACM SIGGRAPH 2005 emerging
  technologies}}. \bibinfo{pages}{11--es}.
\newblock


\bibitem[\protect\citeauthoryear{Arora, Habib~Kazi, Grossman, Fitzmaurice, and
  Singh}{Arora et~al\mbox{.}}{2018}]%
        {arora:2018:symbiosissketch}
\bibfield{author}{\bibinfo{person}{Rahul Arora}, \bibinfo{person}{Rubaiat
  Habib~Kazi}, \bibinfo{person}{Tovi Grossman}, \bibinfo{person}{George
  Fitzmaurice}, {and} \bibinfo{person}{Karan Singh}.}
  \bibinfo{year}{2018}\natexlab{}.
\newblock \showarticletitle{Symbiosissketch: Combining 2d \& 3d sketching for
  designing detailed 3d objects in situ}. In
  \bibinfo{booktitle}{\emph{Proceedings of the 2018 CHI Conference on Human
  Factors in Computing Systems}}. ACM, \bibinfo{pages}{185}.
\newblock


\bibitem[\protect\citeauthoryear{Arora, Kazi, Anderson, Grossman, Singh, and
  Fitzmaurice}{Arora et~al\mbox{.}}{2017}]%
        {arora2017experimental}
\bibfield{author}{\bibinfo{person}{Rahul Arora}, \bibinfo{person}{Rubaiat~Habib
  Kazi}, \bibinfo{person}{Fraser Anderson}, \bibinfo{person}{Tovi Grossman},
  \bibinfo{person}{Karan Singh}, {and} \bibinfo{person}{George~W Fitzmaurice}.}
  \bibinfo{year}{2017}\natexlab{}.
\newblock \showarticletitle{Experimental Evaluation of Sketching on Surfaces in
  VR.}. In \bibinfo{booktitle}{\emph{CHI}}, Vol.~\bibinfo{volume}{17}.
  \bibinfo{pages}{5643--5654}.
\newblock


\bibitem[\protect\citeauthoryear{Boniardi, Valada, Burgard, and
  Tipaldi}{Boniardi et~al\mbox{.}}{2016}]%
        {boniardi2016autonomous}
\bibfield{author}{\bibinfo{person}{Federico Boniardi}, \bibinfo{person}{Abhinav
  Valada}, \bibinfo{person}{Wolfram Burgard}, {and} \bibinfo{person}{Gian~Diego
  Tipaldi}.} \bibinfo{year}{2016}\natexlab{}.
\newblock \showarticletitle{Autonomous indoor robot navigation using a sketch
  interface for drawing maps and routes}. In \bibinfo{booktitle}{\emph{2016
  IEEE International Conference on Robotics and Automation (ICRA)}}. IEEE,
  \bibinfo{pages}{2896--2901}.
\newblock


\bibitem[\protect\citeauthoryear{Cavallo, Dholakia, Havlena, Ocheltree, and
  Podlaseck}{Cavallo et~al\mbox{.}}{2019}]%
        {cavallo2019dataspace}
\bibfield{author}{\bibinfo{person}{Marco Cavallo}, \bibinfo{person}{Mishal
  Dholakia}, \bibinfo{person}{Matous Havlena}, \bibinfo{person}{Kenneth
  Ocheltree}, {and} \bibinfo{person}{Mark Podlaseck}.}
  \bibinfo{year}{2019}\natexlab{}.
\newblock \showarticletitle{Dataspace: A reconfigurable hybrid reality
  environment for collaborative information analysis}. In
  \bibinfo{booktitle}{\emph{2019 IEEE Conference on Virtual Reality and 3D User
  Interfaces (VR)}}. IEEE, \bibinfo{pages}{145--153}.
\newblock


\bibitem[\protect\citeauthoryear{Coelho and Zigelbaum}{Coelho and
  Zigelbaum}{2011}]%
        {coelho2011shape}
\bibfield{author}{\bibinfo{person}{Marcelo Coelho} {and} \bibinfo{person}{Jamie
  Zigelbaum}.} \bibinfo{year}{2011}\natexlab{}.
\newblock \showarticletitle{Shape-changing interfaces}.
\newblock \bibinfo{journal}{\emph{Personal and Ubiquitous Computing}}
  \bibinfo{volume}{15}, \bibinfo{number}{2} (\bibinfo{year}{2011}),
  \bibinfo{pages}{161--173}.
\newblock


\bibitem[\protect\citeauthoryear{Drey, Gugenheimer, Karlbauer, Milo, and
  Rukzio}{Drey et~al\mbox{.}}{2020}]%
        {drey2020vrsketchin}
\bibfield{author}{\bibinfo{person}{Tobias Drey}, \bibinfo{person}{Jan
  Gugenheimer}, \bibinfo{person}{Julian Karlbauer}, \bibinfo{person}{Maximilian
  Milo}, {and} \bibinfo{person}{Enrico Rukzio}.}
  \bibinfo{year}{2020}\natexlab{}.
\newblock \showarticletitle{VRSketchIn: Exploring the Design Space of Pen and
  Tablet Interaction for 3D Sketching in Virtual Reality}. In
  \bibinfo{booktitle}{\emph{Proceedings of the 2020 CHI Conference on Human
  Factors in Computing Systems}}. \bibinfo{pages}{1--14}.
\newblock


\bibitem[\protect\citeauthoryear{Follmer, Leithinger, Olwal, Hogge, and
  Ishii}{Follmer et~al\mbox{.}}{2013}]%
        {follmer2013inform}
\bibfield{author}{\bibinfo{person}{Sean Follmer}, \bibinfo{person}{Daniel
  Leithinger}, \bibinfo{person}{Alex Olwal}, \bibinfo{person}{Akimitsu Hogge},
  {and} \bibinfo{person}{Hiroshi Ishii}.} \bibinfo{year}{2013}\natexlab{}.
\newblock \showarticletitle{inFORM: dynamic physical affordances and
  constraints through shape and object actuation.}. In
  \bibinfo{booktitle}{\emph{Uist}}, Vol.~\bibinfo{volume}{13}.
  \bibinfo{pages}{2501--988}.
\newblock


\bibitem[\protect\citeauthoryear{Gasques, Johnson, Sharkey, and Weibel}{Gasques
  et~al\mbox{.}}{2019}]%
        {gasques2019you}
\bibfield{author}{\bibinfo{person}{Danilo Gasques}, \bibinfo{person}{Janet~G
  Johnson}, \bibinfo{person}{Tommy Sharkey}, {and} \bibinfo{person}{Nadir
  Weibel}.} \bibinfo{year}{2019}\natexlab{}.
\newblock \showarticletitle{What you sketch is what you get: Quick and easy
  augmented reality prototyping with pintar}. In
  \bibinfo{booktitle}{\emph{Extended Abstracts of the 2019 CHI Conference on
  Human Factors in Computing Systems}}. \bibinfo{pages}{1--6}.
\newblock


\bibitem[\protect\citeauthoryear{Gong, Han, Guo, Li, Zha, Zhang, Tian, Wang,
  and Rui}{Gong et~al\mbox{.}}{2021}]%
        {gong2021holoboard}
\bibfield{author}{\bibinfo{person}{Jiangtao Gong}, \bibinfo{person}{Teng Han},
  \bibinfo{person}{Siling Guo}, \bibinfo{person}{Jiannan Li},
  \bibinfo{person}{Siyu Zha}, \bibinfo{person}{Liuxin Zhang},
  \bibinfo{person}{Feng Tian}, \bibinfo{person}{Qianying Wang}, {and}
  \bibinfo{person}{Yong Rui}.} \bibinfo{year}{2021}\natexlab{}.
\newblock \showarticletitle{HoloBoard: a Large-format Immersive Teaching Board
  based on pseudo HoloGraphics}. In \bibinfo{booktitle}{\emph{The 34th Annual
  ACM Symposium on User Interface Software and Technology}}.
  \bibinfo{pages}{441--456}.
\newblock


\bibitem[\protect\citeauthoryear{Gugenheimer, Stemasov, Frommel, and
  Rukzio}{Gugenheimer et~al\mbox{.}}{2017}]%
        {gugenheimer2017sharevr}
\bibfield{author}{\bibinfo{person}{Jan Gugenheimer}, \bibinfo{person}{Evgeny
  Stemasov}, \bibinfo{person}{Julian Frommel}, {and} \bibinfo{person}{Enrico
  Rukzio}.} \bibinfo{year}{2017}\natexlab{}.
\newblock \showarticletitle{Sharevr: Enabling co-located experiences for
  virtual reality between hmd and non-hmd users}. In
  \bibinfo{booktitle}{\emph{Proceedings of the 2017 CHI Conference on Human
  Factors in Computing Systems}}. \bibinfo{pages}{4021--4033}.
\newblock


\bibitem[\protect\citeauthoryear{Hartmann, Yeh, and Vogel}{Hartmann
  et~al\mbox{.}}{2020}]%
        {hartmann2020aar}
\bibfield{author}{\bibinfo{person}{Jeremy Hartmann}, \bibinfo{person}{Yen-Ting
  Yeh}, {and} \bibinfo{person}{Daniel Vogel}.} \bibinfo{year}{2020}\natexlab{}.
\newblock \showarticletitle{AAR: Augmenting a wearable augmented reality
  display with an actuated head-mounted projector}. In
  \bibinfo{booktitle}{\emph{Proceedings of the 33rd Annual ACM Symposium on
  User Interface Software and Technology}}. \bibinfo{pages}{445--458}.
\newblock


\bibitem[\protect\citeauthoryear{Heun, Hobin, and Maes}{Heun
  et~al\mbox{.}}{2013}]%
        {heun2013reality}
\bibfield{author}{\bibinfo{person}{Valentin Heun}, \bibinfo{person}{James
  Hobin}, {and} \bibinfo{person}{Pattie Maes}.}
  \bibinfo{year}{2013}\natexlab{}.
\newblock \showarticletitle{Reality editor: programming smarter objects}. In
  \bibinfo{booktitle}{\emph{Proceedings of the 2013 ACM conference on Pervasive
  and ubiquitous computing adjunct publication}}. \bibinfo{pages}{307--310}.
\newblock


\bibitem[\protect\citeauthoryear{Inc.}{Inc.}{[n.d.]}]%
        {just-a-line}
\bibfield{author}{\bibinfo{person}{Google Inc.}}
  \bibinfo{year}{[n.d.]}\natexlab{}.
\newblock \bibinfo{title}{Just a Line}.
\newblock
\newblock
\urldef\tempurl%
\url{https://justaline.withgoogle.com/}
\showURL{%
\tempurl}


\bibitem[\protect\citeauthoryear{Inc.}{Inc.}{2016}]%
        {tiltbrush}
\bibfield{author}{\bibinfo{person}{Google Inc.}}
  \bibinfo{year}{2016}\natexlab{}.
\newblock \bibinfo{title}{TiltBrush}.
\newblock
\newblock
\urldef\tempurl%
\url{https://www.tiltbrush.com/}
\showURL{%
\tempurl}


\bibitem[\protect\citeauthoryear{Inc.}{Inc.}{2017a}]%
        {gravity-sketch}
\bibfield{author}{\bibinfo{person}{Gravity~Sketch Inc.}}
  \bibinfo{year}{2017}\natexlab{a}.
\newblock \bibinfo{title}{Gravity Sketch}.
\newblock
\newblock
\urldef\tempurl%
\url{https://www.gravitysketch.com/}
\showURL{%
\tempurl}


\bibitem[\protect\citeauthoryear{Inc.}{Inc.}{2017b}]%
        {vuforia-chalk}
\bibfield{author}{\bibinfo{person}{PTC Inc.}} \bibinfo{year}{2017}\natexlab{b}.
\newblock \bibinfo{title}{Vuforia Chalk AR}.
\newblock
\newblock
\urldef\tempurl%
\url{https://chalk.vuforia.com/}
\showURL{%
\tempurl}


\bibitem[\protect\citeauthoryear{Ishii, Lakatos, Bonanni, and Labrune}{Ishii
  et~al\mbox{.}}{2012}]%
        {ishii2012radical}
\bibfield{author}{\bibinfo{person}{Hiroshi Ishii}, \bibinfo{person}{D{\'a}vid
  Lakatos}, \bibinfo{person}{Leonardo Bonanni}, {and}
  \bibinfo{person}{Jean-Baptiste Labrune}.} \bibinfo{year}{2012}\natexlab{}.
\newblock \showarticletitle{Radical atoms: beyond tangible bits, toward
  transformable materials}.
\newblock \bibinfo{journal}{\emph{interactions}} \bibinfo{volume}{19},
  \bibinfo{number}{1} (\bibinfo{year}{2012}), \bibinfo{pages}{38--51}.
\newblock


\bibitem[\protect\citeauthoryear{Ishii, Zhao, Inami, Igarashi, and Imai}{Ishii
  et~al\mbox{.}}{2009}]%
        {ishii2009designing}
\bibfield{author}{\bibinfo{person}{Kentaro Ishii}, \bibinfo{person}{Shengdong
  Zhao}, \bibinfo{person}{Masahiko Inami}, \bibinfo{person}{Takeo Igarashi},
  {and} \bibinfo{person}{Michita Imai}.} \bibinfo{year}{2009}\natexlab{}.
\newblock \showarticletitle{Designing laser gesture interface for robot
  control}. In \bibinfo{booktitle}{\emph{IFIP Conference on Human-Computer
  Interaction}}. Springer, \bibinfo{pages}{479--492}.
\newblock


\bibitem[\protect\citeauthoryear{Jeong, Kim, and Nam}{Jeong
  et~al\mbox{.}}{2018}]%
        {jeong2018mechanism}
\bibfield{author}{\bibinfo{person}{Yunwoo Jeong}, \bibinfo{person}{Han-Jong
  Kim}, {and} \bibinfo{person}{Tek-Jin Nam}.} \bibinfo{year}{2018}\natexlab{}.
\newblock \showarticletitle{Mechanism perfboard: An augmented reality
  environment for linkage mechanism design and fabrication}. In
  \bibinfo{booktitle}{\emph{Proceedings of the 2018 CHI Conference on Human
  Factors in Computing Systems}}. \bibinfo{pages}{1--11}.
\newblock


\bibitem[\protect\citeauthoryear{Kasahara, Niiyama, Heun, and Ishii}{Kasahara
  et~al\mbox{.}}{2013}]%
        {kasahara2013extouch}
\bibfield{author}{\bibinfo{person}{Shunichi Kasahara}, \bibinfo{person}{Ryuma
  Niiyama}, \bibinfo{person}{Valentin Heun}, {and} \bibinfo{person}{Hiroshi
  Ishii}.} \bibinfo{year}{2013}\natexlab{}.
\newblock \showarticletitle{exTouch: spatially-aware embodied manipulation of
  actuated objects mediated by augmented reality}. In
  \bibinfo{booktitle}{\emph{Proceedings of the 7th International Conference on
  Tangible, Embedded and Embodied Interaction}}. \bibinfo{pages}{223--228}.
\newblock


\bibitem[\protect\citeauthoryear{Kim, Drew, Domova, and Follmer}{Kim
  et~al\mbox{.}}{2020}]%
        {kim2020user}
\bibfield{author}{\bibinfo{person}{Lawrence~H Kim}, \bibinfo{person}{Daniel~S
  Drew}, \bibinfo{person}{Veronika Domova}, {and} \bibinfo{person}{Sean
  Follmer}.} \bibinfo{year}{2020}\natexlab{}.
\newblock \showarticletitle{User-defined swarm robot control}. In
  \bibinfo{booktitle}{\emph{Proceedings of the 2020 CHI Conference on Human
  Factors in Computing Systems}}. \bibinfo{pages}{1--13}.
\newblock


\bibitem[\protect\citeauthoryear{Kim and Follmer}{Kim and Follmer}{2019}]%
        {kim2019swarmhaptics}
\bibfield{author}{\bibinfo{person}{Lawrence~H Kim} {and} \bibinfo{person}{Sean
  Follmer}.} \bibinfo{year}{2019}\natexlab{}.
\newblock \showarticletitle{Swarmhaptics: Haptic display with swarm robots}. In
  \bibinfo{booktitle}{\emph{Proceedings of the 2019 CHI conference on human
  factors in computing systems}}. \bibinfo{pages}{1--13}.
\newblock


\bibitem[\protect\citeauthoryear{Kim and Bae}{Kim and Bae}{2016}]%
        {kim2016sketchingwithhands}
\bibfield{author}{\bibinfo{person}{Yongkwan Kim} {and}
  \bibinfo{person}{Seok-Hyung Bae}.} \bibinfo{year}{2016}\natexlab{}.
\newblock \showarticletitle{SketchingWithHands: 3D sketching handheld products
  with first-person hand posture}. In \bibinfo{booktitle}{\emph{Proceedings of
  the 29th Annual Symposium on User Interface Software and Technology}}.
  \bibinfo{pages}{797--808}.
\newblock


\bibitem[\protect\citeauthoryear{LaViola~Jr and Zeleznik}{LaViola~Jr and
  Zeleznik}{2006}]%
        {laviola2006mathpad2}
\bibfield{author}{\bibinfo{person}{Joseph~J LaViola~Jr} {and}
  \bibinfo{person}{Robert~C Zeleznik}.} \bibinfo{year}{2006}\natexlab{}.
\newblock \showarticletitle{Mathpad2: a system for the creation and exploration
  of mathematical sketches}.
\newblock In \bibinfo{booktitle}{\emph{ACM SIGGRAPH 2006 Courses}}.
  \bibinfo{pages}{33--es}.
\newblock


\bibitem[\protect\citeauthoryear{Le~Goc, Kim, Parsaei, Fekete, Dragicevic, and
  Follmer}{Le~Goc et~al\mbox{.}}{2016}]%
        {le2016zooids}
\bibfield{author}{\bibinfo{person}{Mathieu Le~Goc}, \bibinfo{person}{Lawrence~H
  Kim}, \bibinfo{person}{Ali Parsaei}, \bibinfo{person}{Jean-Daniel Fekete},
  \bibinfo{person}{Pierre Dragicevic}, {and} \bibinfo{person}{Sean Follmer}.}
  \bibinfo{year}{2016}\natexlab{}.
\newblock \showarticletitle{Zooids: Building blocks for swarm user interfaces}.
  In \bibinfo{booktitle}{\emph{Proceedings of the 29th annual symposium on user
  interface software and technology}}. \bibinfo{pages}{97--109}.
\newblock


\bibitem[\protect\citeauthoryear{Lee, Post, and Ishii}{Lee
  et~al\mbox{.}}{2011}]%
        {lee2011zeron}
\bibfield{author}{\bibinfo{person}{Jinha Lee}, \bibinfo{person}{Rehmi Post},
  {and} \bibinfo{person}{Hiroshi Ishii}.} \bibinfo{year}{2011}\natexlab{}.
\newblock \showarticletitle{ZeroN: mid-air tangible interaction enabled by
  computer controlled magnetic levitation}. In
  \bibinfo{booktitle}{\emph{Proceedings of the 24th annual ACM symposium on
  User interface software and technology}}. \bibinfo{pages}{327--336}.
\newblock


\bibitem[\protect\citeauthoryear{Lee, Kim, and Kim}{Lee et~al\mbox{.}}{2020}]%
        {lee2020rolling}
\bibfield{author}{\bibinfo{person}{Yujin Lee}, \bibinfo{person}{Myeongseong
  Kim}, {and} \bibinfo{person}{Hyunjung Kim}.} \bibinfo{year}{2020}\natexlab{}.
\newblock \showarticletitle{Rolling Pixels: Robotic Steinmetz Solids for
  Creating Physical Animations}. In \bibinfo{booktitle}{\emph{Proceedings of
  the Fourteenth International Conference on Tangible, Embedded, and Embodied
  Interaction}}. \bibinfo{pages}{557--564}.
\newblock


\bibitem[\protect\citeauthoryear{Li, Luo, Zheng, Xu, and Fu}{Li
  et~al\mbox{.}}{2017}]%
        {li:2017:sweepcanvas}
\bibfield{author}{\bibinfo{person}{Yuwei Li}, \bibinfo{person}{Xi Luo},
  \bibinfo{person}{Youyi Zheng}, \bibinfo{person}{Pengfei Xu}, {and}
  \bibinfo{person}{Hongbo Fu}.} \bibinfo{year}{2017}\natexlab{}.
\newblock \showarticletitle{SweepCanvas: Sketch-based 3D prototyping on an
  RGB-D image}. In \bibinfo{booktitle}{\emph{Proceedings of the 30th Annual ACM
  Symposium on User Interface Software and Technology}}.
  \bibinfo{pages}{387--399}.
\newblock


\bibitem[\protect\citeauthoryear{Makhataeva and Varol}{Makhataeva and
  Varol}{2020}]%
        {makhataeva2020augmented}
\bibfield{author}{\bibinfo{person}{Zhanat Makhataeva} {and}
  \bibinfo{person}{Huseyin~Atakan Varol}.} \bibinfo{year}{2020}\natexlab{}.
\newblock \showarticletitle{Augmented reality for robotics: A review}.
\newblock \bibinfo{journal}{\emph{Robotics}} \bibinfo{volume}{9},
  \bibinfo{number}{2} (\bibinfo{year}{2020}), \bibinfo{pages}{21}.
\newblock


\bibitem[\protect\citeauthoryear{Marshall, Carter, Alexander, and
  Subramanian}{Marshall et~al\mbox{.}}{2012}]%
        {marshall2012ultra}
\bibfield{author}{\bibinfo{person}{Mark Marshall}, \bibinfo{person}{Thomas
  Carter}, \bibinfo{person}{Jason Alexander}, {and} \bibinfo{person}{Sriram
  Subramanian}.} \bibinfo{year}{2012}\natexlab{}.
\newblock \showarticletitle{Ultra-tangibles: creating movable tangible objects
  on interactive tables}. In \bibinfo{booktitle}{\emph{Proceedings of the
  SIGCHI Conference on Human Factors in Computing Systems}}.
  \bibinfo{pages}{2185--2188}.
\newblock


\bibitem[\protect\citeauthoryear{Mueller, Lopes, and Baudisch}{Mueller
  et~al\mbox{.}}{2012}]%
        {mueller2012interactive}
\bibfield{author}{\bibinfo{person}{Stefanie Mueller}, \bibinfo{person}{Pedro
  Lopes}, {and} \bibinfo{person}{Patrick Baudisch}.}
  \bibinfo{year}{2012}\natexlab{}.
\newblock \showarticletitle{Interactive construction: interactive fabrication
  of functional mechanical devices}. In \bibinfo{booktitle}{\emph{Proceedings
  of the 25th annual ACM symposium on User interface software and technology}}.
  \bibinfo{pages}{599--606}.
\newblock


\bibitem[\protect\citeauthoryear{Nakagaki, Leong, Tappa, Wilbert, and
  Ishii}{Nakagaki et~al\mbox{.}}{2020}]%
        {nakagaki2020hermits}
\bibfield{author}{\bibinfo{person}{Ken Nakagaki}, \bibinfo{person}{Joanne
  Leong}, \bibinfo{person}{Jordan~L Tappa}, \bibinfo{person}{Jo{\~a}o Wilbert},
  {and} \bibinfo{person}{Hiroshi Ishii}.} \bibinfo{year}{2020}\natexlab{}.
\newblock \showarticletitle{Hermits: Dynamically reconfiguring the
  interactivity of self-propelled tuis with mechanical shell add-ons}. In
  \bibinfo{booktitle}{\emph{Proceedings of the 33rd Annual ACM Symposium on
  User Interface Software and Technology}}. \bibinfo{pages}{882--896}.
\newblock


\bibitem[\protect\citeauthoryear{Nakagaki, Umapathi, Leithinger, and
  Ishii}{Nakagaki et~al\mbox{.}}{2017}]%
        {nakagaki2017animastage}
\bibfield{author}{\bibinfo{person}{Ken Nakagaki}, \bibinfo{person}{Udayan
  Umapathi}, \bibinfo{person}{Daniel Leithinger}, {and}
  \bibinfo{person}{Hiroshi Ishii}.} \bibinfo{year}{2017}\natexlab{}.
\newblock \showarticletitle{AnimaStage: hands-on animated craft on pin-based
  shape displays}. In \bibinfo{booktitle}{\emph{Proceedings of the 2017
  Conference on Designing Interactive Systems}}. \bibinfo{pages}{1093--1097}.
\newblock


\bibitem[\protect\citeauthoryear{Nakayama, Suzuki, Nakamaru, Niiyama, Kawahara,
  and Kakehi}{Nakayama et~al\mbox{.}}{2019}]%
        {nakayama2019morphio}
\bibfield{author}{\bibinfo{person}{Ryosuke Nakayama}, \bibinfo{person}{Ryo
  Suzuki}, \bibinfo{person}{Satoshi Nakamaru}, \bibinfo{person}{Ryuma Niiyama},
  \bibinfo{person}{Yoshihiro Kawahara}, {and} \bibinfo{person}{Yasuaki
  Kakehi}.} \bibinfo{year}{2019}\natexlab{}.
\newblock \showarticletitle{Morphio: Entirely soft sensing and actuation
  modules for programming shape changes through tangible interaction}. In
  \bibinfo{booktitle}{\emph{Proceedings of the 2019 on Designing Interactive
  Systems Conference}}. \bibinfo{pages}{975--986}.
\newblock


\bibitem[\protect\citeauthoryear{Nowacka, Ladha, Hammerla, Jackson, Ladha,
  Rukzio, and Olivier}{Nowacka et~al\mbox{.}}{2013}]%
        {nowacka2013touchbugs}
\bibfield{author}{\bibinfo{person}{Diana Nowacka}, \bibinfo{person}{Karim
  Ladha}, \bibinfo{person}{Nils~Y Hammerla}, \bibinfo{person}{Daniel Jackson},
  \bibinfo{person}{Cassim Ladha}, \bibinfo{person}{Enrico Rukzio}, {and}
  \bibinfo{person}{Patrick Olivier}.} \bibinfo{year}{2013}\natexlab{}.
\newblock \showarticletitle{Touchbugs: Actuated tangibles on multi-touch
  tables}. In \bibinfo{booktitle}{\emph{Proceedings of the SIGCHI conference on
  human factors in computing systems}}. \bibinfo{pages}{759--762}.
\newblock


\bibitem[\protect\citeauthoryear{Pangaro, Maynes-Aminzade, and Ishii}{Pangaro
  et~al\mbox{.}}{2002}]%
        {pangaro2002actuated}
\bibfield{author}{\bibinfo{person}{Gian Pangaro}, \bibinfo{person}{Dan
  Maynes-Aminzade}, {and} \bibinfo{person}{Hiroshi Ishii}.}
  \bibinfo{year}{2002}\natexlab{}.
\newblock \showarticletitle{The actuated workbench: computer-controlled
  actuation in tabletop tangible interfaces}. In
  \bibinfo{booktitle}{\emph{Proceedings of the 15th annual ACM symposium on
  User interface software and technology}}. \bibinfo{pages}{181--190}.
\newblock


\bibitem[\protect\citeauthoryear{Patten and Ishii}{Patten and Ishii}{2007}]%
        {patten2007mechanical}
\bibfield{author}{\bibinfo{person}{James Patten} {and} \bibinfo{person}{Hiroshi
  Ishii}.} \bibinfo{year}{2007}\natexlab{}.
\newblock \showarticletitle{Mechanical constraints as computational constraints
  in tabletop tangible interfaces}. In \bibinfo{booktitle}{\emph{Proceedings of
  the SIGCHI conference on Human factors in computing systems}}.
  \bibinfo{pages}{809--818}.
\newblock


\bibitem[\protect\citeauthoryear{Poupyrev, Nashida, and Okabe}{Poupyrev
  et~al\mbox{.}}{2007}]%
        {poupyrev2007actuation}
\bibfield{author}{\bibinfo{person}{Ivan Poupyrev}, \bibinfo{person}{Tatsushi
  Nashida}, {and} \bibinfo{person}{Makoto Okabe}.}
  \bibinfo{year}{2007}\natexlab{}.
\newblock \showarticletitle{Actuation and tangible user interfaces: the
  Vaucanson duck, robots, and shape displays}. In
  \bibinfo{booktitle}{\emph{Proceedings of the 1st international conference on
  Tangible and embedded interaction}}. \bibinfo{pages}{205--212}.
\newblock


\bibitem[\protect\citeauthoryear{Raffle, Parkes, and Ishii}{Raffle
  et~al\mbox{.}}{2004}]%
        {raffle2004topobo}
\bibfield{author}{\bibinfo{person}{Hayes~Solos Raffle},
  \bibinfo{person}{Amanda~J Parkes}, {and} \bibinfo{person}{Hiroshi Ishii}.}
  \bibinfo{year}{2004}\natexlab{}.
\newblock \showarticletitle{Topobo: a constructive assembly system with kinetic
  memory}. In \bibinfo{booktitle}{\emph{Proceedings of the SIGCHI conference on
  Human factors in computing systems}}. \bibinfo{pages}{647--654}.
\newblock


\bibitem[\protect\citeauthoryear{Rajaram and Nebeling}{Rajaram and
  Nebeling}{2022}]%
        {rajaram2022papertrail}
\bibfield{author}{\bibinfo{person}{Shwetha Rajaram} {and}
  \bibinfo{person}{Michael Nebeling}.} \bibinfo{year}{2022}\natexlab{}.
\newblock \showarticletitle{Paper Trail: An Immersive Authoring System for
  Augmented Reality Instructional Experiences}. In
  \bibinfo{booktitle}{\emph{Proceedings of the 2022 CHI Conference on Human
  Factors in Computing Systems}}. \bibinfo{pages}{1--14}.
\newblock


\bibitem[\protect\citeauthoryear{Rasmussen, Pedersen, Petersen, and
  Hornb{\ae}k}{Rasmussen et~al\mbox{.}}{2012}]%
        {rasmussen2012shape}
\bibfield{author}{\bibinfo{person}{Majken~K Rasmussen},
  \bibinfo{person}{Esben~W Pedersen}, \bibinfo{person}{Marianne~G Petersen},
  {and} \bibinfo{person}{Kasper Hornb{\ae}k}.} \bibinfo{year}{2012}\natexlab{}.
\newblock \showarticletitle{Shape-changing interfaces: a review of the design
  space and open research questions}. In \bibinfo{booktitle}{\emph{Proceedings
  of the SIGCHI Conference on Human Factors in Computing Systems}}.
  \bibinfo{pages}{735--744}.
\newblock


\bibitem[\protect\citeauthoryear{ROVIO}{ROVIO}{2022}]%
        {angrybirds}
\bibfield{author}{\bibinfo{person}{ROVIO}.} \bibinfo{year}{2022}\natexlab{}.
\newblock \bibinfo{title}{Angry Birds}.
\newblock
\newblock
\urldef\tempurl%
\url{https://www.angrybirds.com/}
\showURL{%
\tempurl}


\bibitem[\protect\citeauthoryear{Scott and Davis}{Scott and Davis}{2013}]%
        {scott2013physink}
\bibfield{author}{\bibinfo{person}{Jeremy Scott} {and} \bibinfo{person}{Randall
  Davis}.} \bibinfo{year}{2013}\natexlab{}.
\newblock \showarticletitle{Physink: sketching physical behavior}. In
  \bibinfo{booktitle}{\emph{Proceedings of the adjunct publication of the 26th
  annual ACM symposium on User interface software and technology}}.
  \bibinfo{pages}{9--10}.
\newblock


\bibitem[\protect\citeauthoryear{Setalaphruk, Ueno, Kume, Kono, and
  Kidode}{Setalaphruk et~al\mbox{.}}{2003}]%
        {setalaphruk2003robot}
\bibfield{author}{\bibinfo{person}{Vachirasuk Setalaphruk},
  \bibinfo{person}{Atsushi Ueno}, \bibinfo{person}{Izuru Kume},
  \bibinfo{person}{Yasuyuki Kono}, {and} \bibinfo{person}{Masatsugu Kidode}.}
  \bibinfo{year}{2003}\natexlab{}.
\newblock \showarticletitle{Robot navigation in corridor environments using a
  sketch floor map}. In \bibinfo{booktitle}{\emph{Proceedings 2003 IEEE
  International Symposium on Computational Intelligence in Robotics and
  Automation. Computational Intelligence in Robotics and Automation for the New
  Millennium (Cat. No. 03EX694)}}, Vol.~\bibinfo{volume}{2}. IEEE,
  \bibinfo{pages}{552--557}.
\newblock


\bibitem[\protect\citeauthoryear{Shah, Schneider, and Campbell}{Shah
  et~al\mbox{.}}{2010}]%
        {shah2010robust}
\bibfield{author}{\bibinfo{person}{Danelle Shah}, \bibinfo{person}{Joseph
  Schneider}, {and} \bibinfo{person}{Mark Campbell}.}
  \bibinfo{year}{2010}\natexlab{}.
\newblock \showarticletitle{A robust sketch interface for natural robot
  control}. In \bibinfo{booktitle}{\emph{2010 IEEE/RSJ International Conference
  on Intelligent Robots and Systems}}. IEEE, \bibinfo{pages}{4458--4463}.
\newblock


\bibitem[\protect\citeauthoryear{Sugimoto, Kagotani, Kojima, Nii, Nakamura, and
  Inami}{Sugimoto et~al\mbox{.}}{2005}]%
        {sugimoto2005augmented}
\bibfield{author}{\bibinfo{person}{Maki Sugimoto}, \bibinfo{person}{Georges
  Kagotani}, \bibinfo{person}{Minoru Kojima}, \bibinfo{person}{Hideaki Nii},
  \bibinfo{person}{Akihiro Nakamura}, {and} \bibinfo{person}{Masahiko Inami}.}
  \bibinfo{year}{2005}\natexlab{}.
\newblock \showarticletitle{Augmented coliseum: display-based computing for
  augmented reality inspiration computing robot}.
\newblock In \bibinfo{booktitle}{\emph{ACM SIGGRAPH 2005 Emerging
  technologies}}. \bibinfo{pages}{1--es}.
\newblock


\bibitem[\protect\citeauthoryear{Suzuki, Karim, Xia, Hedayati, and
  Marquardt}{Suzuki et~al\mbox{.}}{2022}]%
        {suzuki2022augmented}
\bibfield{author}{\bibinfo{person}{Ryo Suzuki}, \bibinfo{person}{Adnan Karim},
  \bibinfo{person}{Tian Xia}, \bibinfo{person}{Hooman Hedayati}, {and}
  \bibinfo{person}{Nicolai Marquardt}.} \bibinfo{year}{2022}\natexlab{}.
\newblock \showarticletitle{Augmented Reality and Robotics: A Survey and
  Taxonomy for AR-enhanced Human-Robot Interaction and Robotic Interfaces}. In
  \bibinfo{booktitle}{\emph{Proceedings of the 2022 CHI Conference on Human
  Factors in Computing Systems}}. \bibinfo{pages}{1--32}.
\newblock
\urldef\tempurl%
\url{https://doi.org/10.1145/1122445.1122456}
\showURL{%
\tempurl}


\bibitem[\protect\citeauthoryear{Suzuki, Kato, Gross, and Yeh}{Suzuki
  et~al\mbox{.}}{2018}]%
        {suzuki2018reactile}
\bibfield{author}{\bibinfo{person}{Ryo Suzuki}, \bibinfo{person}{Jun Kato},
  \bibinfo{person}{Mark~D Gross}, {and} \bibinfo{person}{Tom Yeh}.}
  \bibinfo{year}{2018}\natexlab{}.
\newblock \showarticletitle{Reactile: Programming swarm user interfaces through
  direct physical manipulation}. In \bibinfo{booktitle}{\emph{Proceedings of
  the 2018 CHI Conference on Human Factors in Computing Systems}}.
  \bibinfo{pages}{1--13}.
\newblock


\bibitem[\protect\citeauthoryear{Suzuki, Kazi, Wei, DiVerdi, Li, and
  Leithinger}{Suzuki et~al\mbox{.}}{2020}]%
        {suzuki2020realitysketch}
\bibfield{author}{\bibinfo{person}{Ryo Suzuki}, \bibinfo{person}{Rubaiat~Habib
  Kazi}, \bibinfo{person}{Li-Yi Wei}, \bibinfo{person}{Stephen DiVerdi},
  \bibinfo{person}{Wilmot Li}, {and} \bibinfo{person}{Daniel Leithinger}.}
  \bibinfo{year}{2020}\natexlab{}.
\newblock \showarticletitle{Realitysketch: Embedding responsive graphics and
  visualizations in AR through dynamic sketching}. In
  \bibinfo{booktitle}{\emph{Proceedings of the 33rd Annual ACM Symposium on
  User Interface Software and Technology}}. \bibinfo{pages}{166--181}.
\newblock


\bibitem[\protect\citeauthoryear{Suzuki, Ofek, Sinclair, Leithinger, and
  Gonzalez-Franco}{Suzuki et~al\mbox{.}}{2021}]%
        {suzuki2021hapticbots}
\bibfield{author}{\bibinfo{person}{Ryo Suzuki}, \bibinfo{person}{Eyal Ofek},
  \bibinfo{person}{Mike Sinclair}, \bibinfo{person}{Daniel Leithinger}, {and}
  \bibinfo{person}{Mar Gonzalez-Franco}.} \bibinfo{year}{2021}\natexlab{}.
\newblock \showarticletitle{HapticBots: Distributed Encountered-type Haptics
  for VR with Multiple Shape-changing Mobile Robots}. In
  \bibinfo{booktitle}{\emph{The 34th Annual ACM Symposium on User Interface
  Software and Technology}}. \bibinfo{pages}{1269--1281}.
\newblock


\bibitem[\protect\citeauthoryear{Suzuki, Zheng, Kakehi, Yeh, Do, Gross, and
  Leithinger}{Suzuki et~al\mbox{.}}{2019}]%
        {suzuki2019shapebots}
\bibfield{author}{\bibinfo{person}{Ryo Suzuki}, \bibinfo{person}{Clement
  Zheng}, \bibinfo{person}{Yasuaki Kakehi}, \bibinfo{person}{Tom Yeh},
  \bibinfo{person}{Ellen Yi-Luen Do}, \bibinfo{person}{Mark~D Gross}, {and}
  \bibinfo{person}{Daniel Leithinger}.} \bibinfo{year}{2019}\natexlab{}.
\newblock \showarticletitle{Shapebots: Shape-changing swarm robots}. In
  \bibinfo{booktitle}{\emph{Proceedings of the 32nd annual ACM symposium on
  user interface software and technology}}. \bibinfo{pages}{493--505}.
\newblock


\bibitem[\protect\citeauthoryear{Walker, Hedayati, Lee, and Szafir}{Walker
  et~al\mbox{.}}{2018}]%
        {walker2018communicating}
\bibfield{author}{\bibinfo{person}{Michael Walker}, \bibinfo{person}{Hooman
  Hedayati}, \bibinfo{person}{Jennifer Lee}, {and} \bibinfo{person}{Daniel
  Szafir}.} \bibinfo{year}{2018}\natexlab{}.
\newblock \showarticletitle{Communicating robot motion intent with augmented
  reality}. In \bibinfo{booktitle}{\emph{Proceedings of the 2018 ACM/IEEE
  International Conference on Human-Robot Interaction}}.
  \bibinfo{pages}{316--324}.
\newblock


\bibitem[\protect\citeauthoryear{Xiao, Schwarz, Throm, Wilson, and Benko}{Xiao
  et~al\mbox{.}}{2018}]%
        {xiao2018mrtouch}
\bibfield{author}{\bibinfo{person}{Robert Xiao}, \bibinfo{person}{Julia
  Schwarz}, \bibinfo{person}{Nick Throm}, \bibinfo{person}{Andrew~D Wilson},
  {and} \bibinfo{person}{Hrvoje Benko}.} \bibinfo{year}{2018}\natexlab{}.
\newblock \showarticletitle{MRTouch: Adding touch input to head-mounted mixed
  reality}.
\newblock \bibinfo{journal}{\emph{IEEE transactions on visualization and
  computer graphics}} \bibinfo{volume}{24}, \bibinfo{number}{4}
  (\bibinfo{year}{2018}), \bibinfo{pages}{1653--1660}.
\newblock


\bibitem[\protect\citeauthoryear{Yu, Arora, Stanko, B{\ae}rentzen, Singh, and
  Bousseau}{Yu et~al\mbox{.}}{2021}]%
        {yu2021cassie}
\bibfield{author}{\bibinfo{person}{Emilie Yu}, \bibinfo{person}{Rahul Arora},
  \bibinfo{person}{Tibor Stanko}, \bibinfo{person}{J~Andreas B{\ae}rentzen},
  \bibinfo{person}{Karan Singh}, {and} \bibinfo{person}{Adrien Bousseau}.}
  \bibinfo{year}{2021}\natexlab{}.
\newblock \showarticletitle{Cassie: Curve and surface sketching in immersive
  environments}. In \bibinfo{booktitle}{\emph{Proceedings of the 2021 CHI
  Conference on Human Factors in Computing Systems}}. \bibinfo{pages}{1--14}.
\newblock


\bibitem[\protect\citeauthoryear{Zhu and Grossman}{Zhu and Grossman}{2020}]%
        {zhu2020bishare}
\bibfield{author}{\bibinfo{person}{Fengyuan Zhu} {and} \bibinfo{person}{Tovi
  Grossman}.} \bibinfo{year}{2020}\natexlab{}.
\newblock \showarticletitle{Bishare: Exploring bidirectional interactions
  between smartphones and head-mounted augmented reality}. In
  \bibinfo{booktitle}{\emph{Proceedings of the 2020 CHI Conference on Human
  Factors in Computing Systems}}. \bibinfo{pages}{1--14}.
\newblock


\end{thebibliography}

\end{document}